\documentclass{aa}
\usepackage{txfonts}
\usepackage{epsfig}
\usepackage{rotating}
\usepackage{times}
\usepackage{booktabs}
\usepackage{longtable}
\usepackage{color}
\usepackage{natbib}
\usepackage{amsmath}
\usepackage[utf8]{inputenc}
\usepackage{caption}
\usepackage{float}
\usepackage{pdflscape}

\bibpunct{(}{)}{;}{a}{}{,}

\begin{document}

   \title{An XMM-Newton view of FeK$\alpha$ in HMXBs}

   \author{A. Gim\'{e}nez-Garc\'{i}a\inst{1,4,6}
   	   \and
	   J. M. Torrej\'{o}n\inst{1,2}
	   \and
	   W. Eikmann\inst{3}
	   \and 
	   S. Mart\'{i}nez-N\'{u}\~{n}ez\inst{2}  
	   \and
	   L. M. Oskinova\inst{4}
	   \and
	   J. J. Rodes-Roca\inst{1,2,5}
	   \and
	   G. Bernab\'{e}u\inst{1,2}
	   }
	   
   \authorrunning{A. Gim\'{e}nez-Garc\'{i}a\inst{1}}
   \titlerunning{Title}

   \offprints{A. Gim\'{e}nez-Garc\'{i}a}

   \institute{ \footnotesize University Institute of Physics Applied to Sciences and Technologies, University of Alicante, P.O. Box 99, E03080 Alicante, Spain \\ e-mail: angelgimenez@ua.es
   \and
   X-ray Astronomy Group. Departamento de F\'{\i}sica, Ingenier\'{\i}a de Sistemas y Teor\'{\i}a de la Se\~{n}al, University of Alicante, P.O. Box 99, E03080 Alicante, Spain
   \and 
   Dr. Karl Remeis-Sternwarte FAU Erlangen-N\"{u}rnberg, D96049 Bamberg, Germany
   \and
   Institute for Physics and Astronomy, University of Potsdam, D-14476 Potsdam, Germany
   \and
   MAXI team, Institute of Physical and Chemical Research (RIKEN), 2-1 Hirosawa, Wako, Saitama, 351-0198, Japan
   \and
   School of Physics, Faculty of Science, Monash University, Clayton, VIC 3800, Australia }
   
   \date{Accepted date: 25th December 2014}

\abstract{We present a comprehensive analysis of the whole sample of available XMM-Newton observations of High Mass X-ray Binaries (HMXBs) until August, 2013, focusing on the FeK$\alpha$ emission line. This line is a key tool to better understand the physical properties of the material surrounding the X-ray source within a few stellar radii (the circumstellar medium). We have collected observations from 46 HMXBs, detecting FeK$\alpha$ in 21 of them. We have used the standard classification of HMXBs to divide the sample in different groups. We find that: (1)~different classes of HMXBs  display different qualitative behaviour in the FeK$\alpha$ spectral region. This is specially visible in SGXBs (showing ubiquitous Fe fluorescence but not recombination Fe lines), and $\gamma$~Cass~analogs (showing both fluorescent and recombination Fe lines). (2)~FeK$\alpha$ is centred at a mean value of 6.42~keV. Considering the instrumental and fits uncertainties, this value is compatible with ionization states lower than Fe~\textsc{xviii}. (3)~The flux of the continuum is well correlated with the flux of the line, as expected. Eclipse observations show that the Fe fluorescence emission comes from an extended region surrounding the X-ray source. (4) We observe an inverse correlation between the X-ray luminosity and the equivalent width of FeK$\alpha$~(EW). This phenomenon is known as \textit{X-rays Baldwin effect}. (5)~FeK$\alpha$ is narrow ($\sigma_{line}<0.15$~keV), reflecting that the reprocessing material does not move at high speeds. We attempt to explain the broadness of the line in terms of three possible broadening phenomena: line blending, Compton scattering and Doppler shifts (with velocities of the reprocessing material $V \sim 1000$~km/s). (6)~The equivalent hydrogen column~($N_H$) directly correlates with the EW of FeK$\alpha$, displaying clear similarities to numerical simulations. It highlights the strong link between the absorbing and the fluorescent matter. (7)~The observed $N_H$ in Supergiant X-ray Binaries (SGXBs) is in general higher than in Supergiant Fast X-ray Transients (SFXTs). We suggest two possible explanations: different orbital configurations, or a different interaction \textit{compact object - wind}. (8)~Finally, we analysed in more detail the sources IGR~J16320-4751 and 4U~1700-37, covering several orbital phases. The observed variation of $N_H$ between phases is compatible with the absorption produced by the wind of their optical companions. The obtained results clearly point to a very important contribution of the donor's wind in the FeK$\alpha$ emission and the absorption when the donor is a supergiant massive star. 
}


   \maketitle

\setlongtables

\section{Introduction}
Since the early stages of X-ray astronomy, Fe lines in the spectral region of $\sim$6-7~keV (the Fe complex) have been studied in a large number of X-ray sources given its fruitfulness as a tool for plasma diagnostics. They were reported for the first time in the supernova remnant Cas~A \citep{Serlemitsos1973}, and only two years later in a High Mass X-ray Binary (HMXB) using the Ariel~5 satellite \citep{Sanford1975}. The most recent X-ray space missions (Swift, Suzaku, Chandra and XMM-Newton) have triggered a notable improvement in the attainable spectral resolution and effective area, permitting to distinguish between different emission features in the Fe complex: narrow and broad fluorescence lines (FeK$\alpha$ and FeK$\beta$), Compton shoulders and recombination lines (Fe~\textsc{xxv} and Fe~\textsc{xxvi}) \citep{Torrejon2010}. This improvement has given a remarkable impetus in the study of the Fe complex, and motivates a comprehensive analysis in HMXBs. 

In particular, FeK$\alpha$ has been proven as a fundamental tool in the study of HMXBs \citep{Martinez-Nunez2014,Rodes-Roca2011,Van_der_Meer2005}. The origin of the fluorescence emitting region has been discussed by many authors in the past. \cite{Nagase1989} considered accretion disks and the matter stagnated in the accretion and ionization wakes in the stellar wind as plausible areas of FeK$\alpha$ production. \cite{Watanabe2006} analysed the classical HMXB Vela~X-1, and proposed the extended stellar wind, reflection off the stellar photosphere and an accretion wake, as the most likely candidates for fluorescence reprocessing regions. In any case, FeK$\alpha$ is very sensitive to the physical conditions of the vicinity of the X-ray source,  and provides remarkable information that must be analysed.   

Fluorescence is produced as a consequence of the X-ray illumination of matter. When an Fe atom absorbs a photon carrying sufficient energy to remove an electron from its K-shell (E$>$7.2~keV), the vacancy can be occupied by another electron from an outer shell. If the electron comes from the L-shell, the transition produces FeK$\alpha$ emission. FeK$\beta$ emission is produced when the vacancy is filled by a former M-shell electron. When Fe is more ionized than Fe~\textsc{xix}, the fluorescence yield starts to decrease with the ionization state \citep{Kallman2004}. Therefore, FeK$\alpha$ is a footprint of not extremely ionized Fe (less than Fe~\textsc{xx}). On the other hand, recombination lines Fe~\textsc{xxv} and Fe~\textsc{xxvi} unveil the presence of very hot gas, where Fe atoms are almost completely stripped. 

Previous comprehensive surveys of the Fe complex in HMXBs were carried out by \cite{Gottwald1995} using EXOSAT and \cite{Torrejon2010} using the High Energy Transmission Gratings (HETGS) on board Chandra. The high spectral resolution provided by Chandra gratings proved to be instrumental in disentangling the different ionization species present in the Fe complex. However, the relatively low throughput of the instrument allowed to study only the brightest binaries. In this study we increase significantly the previous sample by using the high throughput of XMM-Newton EPN. This has allowed us to include fainter systems (like Be X-ray binaries (BeXBs) or SFXTs in quiescence) while the moderate resolution of the EPN CCDs has allowed us to test previous correlations based on a small sample.

HMXBs are specially susceptible to be studied using the Fe complex, on account of the significance of the circumstellar medium in the observable phenomena. These systems consists of a compact object, either a neutron star (NS) or a black hole (BH), accreting matter from a massive OB star (usually called optical or normal star of the system). In HMXBs the observed luminosity is commonly powered via accretion. Consequently, the way that matter is accreted from the donor directly defines the observable luminosity features of every source. 

When the optical star is a Be star, the system is a BeXB. Be stars are fast-rotating BIII-V stars, which have shown spectral emission lines at some point of their lives. They also show an excess of infrared emission, when they are compared to non Be stars of the same spectral type. These observables are explained appealing to an extended circumstellar decretion disk. BeXBs are usually transient in the X-rays, although some systems exhibit a persistent quiescence emission (L$\leq 10^{34-35}$~erg/s). The outbursts have been traditionally classified in two types. Type I outburst (L$\leq 10^{37}$~erg/s) are related to periastron passages. Type II outburst are not related to the orbital phase and imply an even higher increase in luminosity than Type I outbursts, reaching the Eddington luminosity (for a review on BeXBs see \cite{Reig2011}).

In the case of classical Supergiant X-ray Binaries (SGXBs), the compact object is embedded in the dense and powerful wind of a OB supergiant companion, swallowing everything that enters its gravitational domain. The mass loss rate of the donor is of the order of $\gtrsim 10^{-7}\,M_{\odot}\,yr^{-1}$, and the compact object is usually found at a close distance of a~$\sim 1.5-2\,R_{\star}$. In such a close orbit, the captured matter is able to fuel a persistent X-ray emission of $\sim 10^{33-39}$~erg/s. Flares and off-states are often observed in SGXBs, indicating an abrupt transition in the accretion rate. They might be produced either by sudden variations of density in the medium transited by the compact object \citep{Martinez-Nunez2014,Kreykenbohm2008}, or either by instabilities above the magnetosphere of the neutron star, as proposed in the quasi-spherical accretion theory by \cite{Shakura2012}. 

The medium transited by the compact object through the extended atmosphere of an OB supergiant star is not smooth because of, at least, two phenomena. First, density inhomogeneities (clumps) are present as an intrinsic feature of the radiatively driven winds of hot stars \citep{Lucy1980,Oskinova2012}. Second, hydrodynamical simulations show that the X-ray radiation and the gravity field of the compact object disturb the wind of the donor, inducing the formation of denser structures like filaments, bow shocks and wakes \citep{Blondin1990,Blondin1991}. 

In the last decade and a half, new discoveries have led to the addition of new groups to the previous picture of HMXBs, stressing the value of grasping the different features of the sources such as geometry, compact object properties, optical star peculiarities and wind clumpiness. The new groups are Supergiant Fast X-ray Transient systems (SFXTs), $\gamma$ Cassiopeae analogs and $\gamma$-ray Binaries. 

SFXTs are systems with a supergiant optical star, as in SGXBs, but defined by an extremely transient behaviour. During quiescence they exhibit low luminosity ($\sim 10^{32}$~erg/s), but they spend most of their time in an intermediate level of emission ($\sim 10^{33-34}$~erg/s). They display short outbursts ($\sim$few hours), reaching luminosities up to $10^{36-37}$ erg/s \citep{Sidoli2009}. It is likely that the clumpiness of the wind plays a main role in the variability of these sources. Other mechanisms involving centrifugal and magnetic barriers could enhance the observed luminosity swings, relaxing the needed variation amplitudes in the physical conditions of the wind \citep{Bozzo2008}. Nevertheless, other authors explain the variability appealing to the quasi-spherical accretion model \citep{Drave2013,Paizis2014}.

$\gamma$ Cassiopeae analogs are characterized by the thermal nature of the X-ray emission, with plasma temperatures of $\sim 10^8\;K$ ($\sim 10$~keV), an X-ray luminosity of $10^{32-33}$~erg/s, and high flux variability on various time scales. However, they do not display giant outbursts as observed in BeXBs \citep{Lopes_de_Oliveira2010}. Presently, it is not clear that the X-ray emission is emitted by accretion processes (onto a neutron star or a white dwarf), or alternatively generated from the interaction between the surface of the star, the circumstellar disk and its magnetic field. 

High~Mass~$\gamma$-ray Binary systems (HMGBs) are HMXBs where the emission peaks above 1~MeV. Nowadays, it is thought that the emission is caused by accelerated particles in the shock that is produced when the pulsar wind collides the massive star wind. Therefore, they are powered by the rotational energy of the neutron star, in opposition to the rest of HMXBs, which are accretion fed. There are currently five confirmed HMGBs, all of them with a main sequence optical star (for a review on HMGBs see \cite{Dubus2013}).

Finally, there are sources which, following a number of reasons, cannot be classified in any of the already mentioned classes of HMXBs. Particularly, among the set of sources studied in this paper, they are 4U~2206+54, Centaurus~X-3 and Cygnus~X-1. The optical star in 4U~2206+54 is a O9.5V \citep{Blay2006}, neither a supergiant nor a Be star. The system may be part of a new group of wind-fed HMXBs with a main sequence donor \citep{Ribo2006}. Centaurus~X-3 and Cygnus~X-1 are the only systems here collected where accretion is persistently driven by an accretion disk \citep{Tjemkes1986,Shapiro1976}, what is reflected in the spectra of both sources. 

In this paper, we study the FeK$\alpha$ line for the whole sample of HMXBs available with XMM-Newton until August, 2013. In Section~\ref{sec:obs} we present the set of observations, the reduction process and the more important details concerning the spectral fits. In Section~\ref{sec:results} we show the obtained results: an spectral atlas including every fit and different plots relating fit parameters. In Section~\ref{sec:discussion} we interpret the obtained results and we summarize the most important conclusions in Section~\ref{sec:conclusions}. In Appendix~A we present a set of tables describing the obtained parameters from the spectral fits. In Appendix~B we show the spectral atlas, containing the plot of every spectrum that we have analysed in this survey. We show the observations and the models, together with the ratio between them.  

\section{\label{sec:obs}Observations and data treatment}
The XMM-Newton observatory \citep{XMM_Observatory2012} is fitted with three X-ray telescopes of 1500~$cm^2$ and a coalignated optical telescope.  Spectroscopy and photometry is done by the 6 instruments on board: three X-ray imaging cameras EPIC (European Photon Imaging Camera), two grating X ray spectrometers RGS (Reflection Grating Spectrometer) and an optical monitor (OM). EPIC cameras (0.1-15~keV) are the only instruments at XMM-Newton covering the energy range of the Fe complex. Among EPIC, one camera uses PN CCDs, and the other two use MOS CCDs. EPIC PN cameras (EPN) surpass by a factor $\sim$3 the effective area of the MOS cameras at 6-7~keV, making EPN more suitable for our purposes. Compared to other missions, the High Energy Transmission Grating Spectrometer (HETGS) on board of Chandra provides better energy resolution in the energy range of the Fe complex, but the effective area available with EPN is significantly higher. EPN provides the adequate conditions to perform the study here presented, on account of the moderate (but sufficient) spectral resolution ($\Delta E/E\sim40$), and great effective area ($\sim1000 \; cm^2$), enabling us to analyze a large amount of sources in an homogeneous and consistent way.

Since HMXBs are usually variable, we often observe in the same observation a dramatic change in luminosity, remarkably affecting the spectral parameters. In these cases, an averaged spectrum does not reproduce the actual emission of the source, and it is advisable to split the observation in more than one time interval. We have considered five different states\footnote{The states considered in this work and the also called states in black hole binary systems must not be mislead.} of the systems in order to define the time intervals: dips, quiescence, flares, eclipse ingress/egress, and eclipse. We have used the following criteria. When luminosity drops a factor $\gtrsim$2 in the timescale of $\lesssim$1~hour, we have tagged the time interval as a dip. Analogously, when luminosity rises $\gtrsim$2 in the timescale of $\lesssim$1~hour, we have labelled the time interval as a flare. For observations covering eclipsing phases, we have defined time intervals for eclipse ingress/egress and eclipse. The rest of time intervals are tagged as quiescent states. 

\begin{figure} 
\begin{center}
\includegraphics[width=0.5\textwidth]{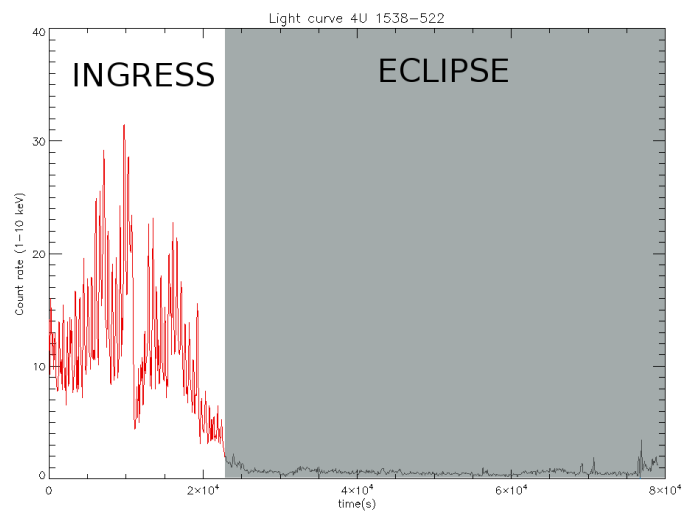}
\end{center}
\caption{\footnotesize Light curve of the observation of 4U~1538-522 (ObsID:0152780201). We have split the observation in two parts, one for the ingress in the eclipse, and another one for the eclipse.  }
\label{states}
\end{figure}

In Figure~\ref{states} we see the light curve of an observation of 4U~1538-522, as an example of how we have split the time intervals in the observations. The source was observed during the ingress in an X-ray eclipse, which is clearly noticeable in the light curve. We have separated the observation in two time intervals, one covering the ingress in eclipse, and another one covering the eclipse. 

Summarizing the sample of observations, we have collected data from 46 HMXBs. 21 of them exhibit FeK$\alpha$ emission. We note that some sources have more than one available observation. Taking everything into account (46 sources, temporal splitting depending on the state of the source, and more than one observation per source in some cases), we end up with a total number of 108 spectra that we have analysed. 

We have followed the catalogue of \cite{Liu2006}, in addition to later discoveries or confirmations, to identify the currently known HMXBs, and used every available XMM-Newton public observation\footnote{http://xmm.esac.esa.int/xsa/}. The sources not included in the Liu catalogue, but here considered, are: HD~119682, SS~397, IGR~J16328-4726, HD~45314, HD~157832, Swift~J045106.8-694803, IGR~J16207-5129 and XTE~J1743-363.

\subsection{Data reduction}
We have reduced the data using Science Analysis System (SAS), version 12.0.1. Since the sample of observations contain an heterogenous group of HMXBs, we found different observation modes (timing and imaging) to account for the different properties of the sources. In the brightest systems, the observations were usually performed using the timing mode, while the faintest sources were observed using imaging modes.

Timing modes permit to process the arrival of photons at high rate, since only one CCD operates and the information is collapsed into one dimension, allowing a fast read out. The time resolution is as high as 30 $\mu s$ (7 $\mu s$ in burst mode, \cite{Kirsch2006}). Even though the high timing resolution reached with these observation modes, pile-up is still present in several cases, specially when the count rate is $\gtrsim$ 800 counts s$^{-1}$. We have checked in every observation if pile-up is affecting the data, using the SAS task \textit{epatplot}, and we have excised the core of the source's point spread function in the pertinent cases. The size of the excised region has been chosen wide enough to remove the unwanted pile-up effects (see examples of the use of \textit{epatplot} in \cite{Ng2010}). 

Background subtraction process is also dependent on the brightness of the source. In the EPN timing mode, the PSF of the sources displaying $\gtrsim$ 200 counts s$^{-1}$ will span across the whole CCD. Therefore, any area selected as a background region will be contaminated by source photons. Since this effect is strongly energy dependent, for the brightest sources we have chosen a method of background subtraction similar to the one performed in the analysis of Vela X-1 by \cite{Martinez-Nunez2014}, where a blank sky spectrum taken in timing mode is used as the real background for energies below 2.5 keV, while the rest of the spectrum corresponds to the outermost pixels of the CCD. Meanwhile, for \textit{common} observations, we have used source-free regions to extract a background spectrum and subtract it from the former source plus background energy distribution. 

Ancillary response files were generated using the SAS task \textit{arfgen}. For observations taken in timing mode affected by pile-up, we have followed the recommendations of the \textit{XMM-Newton SAS User Guide} in order to generate the appropriate ancillary response files. Response matrices were created using the SAS task \textit{rmfgen}.

\subsection{\label{sec:fitting}Spectral fitting}
For the spectral analysis we have used XSPEC, version 12.8.0\footnote{http://heasarc.nasa.gov/xanadu/xspec/}. We have rebinned the spectra to have a minimum of 20 counts per bin, and a bin size of at least 1/3 of the FWHM of the intrinsic energy resolution, in order to be allowed to apply $\chi^2$ statistics in the fitting of a set of poissonian data \citep{Cash1979}. 

In Table~\ref{tab:models} we present the sample of models employed for the continuum in the fits. Every model is a combination of additive and multiplicative models. An additive model stands for a source of X-rays (e.g. bremsstrahlung radiation), and a multiplicative model represents a energy-dependent change of an additive model (e.g. photoelectric absorption).

\begin{figure} 
\begin{center}
\includegraphics[width=0.5\textwidth]{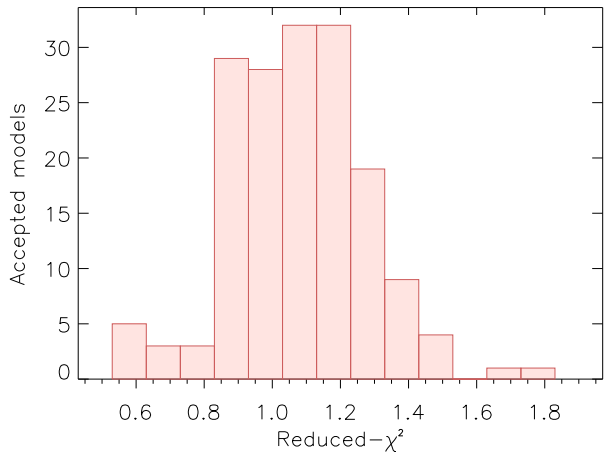}
\end{center}
\caption{\footnotesize Number of accepted models depending on the Reduced-$\chi^2$ value. } 
\label{histo_chis}
\end{figure}

The models presented in Table~\ref{tab:models} have been tested in every observation, and accepted depending on the Reduced-$\chi^2$ ($\frac{\chi^2}{n-m}$, with \textit{n} the number of bins and \textit{m} the number of fitted parameters). Every observation has particular characteristics, and therefore the decision of which Reduced-$\chi^2$ value is acceptable has been taken one by one. In Figure~\ref{histo_chis} we can see that most of the fits result in a Reduced-$\chi^2 \simeq 1$, as expected for a suitable fit. The highest value of Reduced-$\chi^2$ for an accepted model has been 1.82. The parameters arising from the fits are listed in Tables~\ref{tab:cont} and \ref{tab:FeKa}. 

We can classify the additive components of the models as thermal or non-thermal. A component is called \textit{thermal} when radiation is produced as a consequence of the thermal motion of the plasma particles (e.g. blackbody radiation). Otherwise, the emitted radiation is non-thermal (e.g. non-thermal Inverse-Compton emission). If all the additive components of a model are thermal, we classify the model as thermal (analogously for non-thermal). We have also used hybrid models, combining thermal and non-thermal components. The thermal components used in this work are the following: 
\begin{itemize}
\item \textit{bbody}: blackbody emission.
\item \textit{diskbb}: model of an accretion disk emission made of multiple blackbody components.
\item \textit{bremss}: thermal bremsstrahlung emission (electrons distributed according to the Maxwell–Boltzmann distribution).
\item \textit{mekal}: emission from optically thin hot gas, including spectral lines from several elements \citep{Mewe1985}.  
\item \textit{cemekl}: built from the mekal model, incorporating multi-temperature emission.
\end{itemize}

On the other hand, the only non-thermal component used in this work is: 
\begin{itemize}
\item \textit{powerlaw}: phenomenological model consisting of a simple inverse power law profile ($\propto E^{-\Gamma}$). This profile is a footprint of Inverse-Compton scattering by hot electrons (non-thermally distributed) of a seed radiation field.
\end{itemize}

For the photoelectric absorption, we have used \textit{tbnew}\footnote{http://pulsar.sternwarte.uni-erlangen.de/wilms/research/tbabs/}, the improved version of the \cite{Wilms2000} model \textit{tbabs}, setting the cross sections to the \cite{Verner1996} ones and the abundances according to \cite{Wilms2000}. The most important parameter of this model is the total equivalent hydrogen column $N_H$, which is the integrated amount of hydrogen atoms in the line of sight from the observer to the source, per $cm^{2}$. We have also added the model \textit{cabs} to account for the Compton scattering, which is not comprised in the \textit{tbnew} model and is especially significative for $N_H \gtrsim 10^{24} \; cm^{-2}$. 

The emission lines are fitted using Gaussian profiles. We have categorized as FeK$\alpha$ any emission line that fulfils the following conditions: \begin{itemize} 
\item[1)] The centroid energy of the Gaussian component lies in the interval $[6.3,6.65]$~keV. The interval includes the expected energy of FeK$\alpha$ emission from Fe~\textsc{ii} ($\sim 6.395$~keV) to Fe~\textsc{xxiii} ($\sim 6.63$~keV) \citep{Kallman2004}. This condition excludes the detection of any hypothetical fluorescent emission from Fe~\textsc{xxiv-xxv} at $\sim 6.67-6.7$~keV, thereby excluding any confusion between FeK$\alpha$ and the recombination line Fe~\textsc{xxv} at similar wavelength. The fluorescence yields of Fe~\textsc{xxiv-xxv} are low compared to lower ionization states.
\item[2)] The statistical significance ($\sigma_{sign}$) of the Gaussian component is greater than 2$\sigma$. We have calculated $\sigma_{sign}$ from $\chi_{k_1}^2-\chi_{k_2}^2$, assuming $\chi_{k_1}^2-\chi_{k_2}^2 \sim \chi_{k_1-k_2}^2$ \footnote{This assumption is not strictly true, since $\chi_{k_1}^2$ and $\chi_{k_2}^2$ are not independent. However, it provides an estimation on the impact of the Gaussian component in the model. }; where $\chi_{k_1}^2$ arises from a fit using certain model with the Gaussian component included, and $\chi_{k_2}^2$ arising from a fit using the same model without the Gaussian component. 
\end{itemize} 
 
In some cases, FeK$\alpha$ line is clearly noticeable, but FeK$\beta$ is not prominent enough to permit error estimation of its parameters. In these cases, we have constrained the centroid energy and the norm of FeK$\beta$ according to \cite{Kallman2004}:
\begin{itemize}
\item Energy(FeK$\beta$) = Energy(FeK$\alpha$) + 0.652~keV
\item Norm(FeK$\beta$) = Norm(FeK$\alpha$) $\times$ 0.13~photons/cm$^2$/s
\end{itemize}
The estimated parameters, like the EW, are very sensitive to the fit of the continuum. Therefore, although the Fe complex appears in the $\sim$6-7~keV energies, we broadened the spectral scope to an energy range of 1-10~keV to perform the analysis. It also allows us to consider possible calibration inaccuracies in the charge transfer inefficiency (CTI) and the X-ray loading (XRL), an issue reported in previous analysis of EPN observations (see \cite{Martinez-Nunez2014} and \cite{Furst2011}). In the few cases of possible CTI or XRL, we applied an artificial gain $E_{new} = \frac{E_{old}}{slope} + offset$ (see Table~\ref{tab:cont}).  

The estimation of the parameter confidence regions (at 90$\%$ level) have been calculated with a Markov Chain Monte Carlo (MCMC) technique, implemented in XSPEC, where \textit{N} generations of the set of free parameters are used to determine the best-fit values and the confidence regions. We have set $N=1.5\times10^4$ in our calculations. These chains are also valid to estimate fluxes and equivalent widths.
\begin{center}
\begin{table*}
\begin{tabular*}{\textwidth}{ccc}
 & Model & Continuum models \\
\toprule
 & $N_1$  & $powerlaw \times tbnew \times cabs$ \\
Non-thermal & $N_2$   & $powerlaw_1 \times tbnew_1 \times cabs_1 + powerlaw_2 \times tbnew_2 \times cabs_2$ \\
 & $N_3$   & $powerlaw_1 \times tbnew_1 \times cabs_1 + powerlaw_2 \times tbnew_2 \times cabs_2 + powerlaw_3 \times tbnew_3 \times cabs_3$ \\
\midrule
 & $T_1$  & $mekal \times tbnew \times cabs$ \\
 & $T_2$   & $(mekal+mekal) \times tbnew \times cabs$ \\
 & $T_3$   & $(mekal+mekal+mekal) \times tbnew \times cabs$ \\
 & $T_4$   & $(cemekl) \times tbnew \times cabs$ \\
 & $T_5$   & $bbody \times tbnew \times cabs$ \\
 & $T_6$   & $(bbody+bbody) \times tbnew \times cabs$ \\
Thermal & $T_7$   & $(bbody_1 + bbody_2) \times tbnew \times cabs$ \\
 & $T_8$   & $bbody_1 \times tbnew_1 \times cabs_1 + bbody_2 \times tbnew_2 \times cabs_2$  \\
 & $T_9$   & $diskbb \times tbnew \times cabs$ \\
 & $T_{10}$  & $(diskbb+bbody) \times tbnew \times cabs$ \\
 & $T_{11}$  & $bremss \times tbnew \times cabs$ \\
 & $T_{12}$  & $bbody \times tbnew \times cabs + bremss \times tbnew \times cabs$ \\
 & $T_{13}$  & $(bremss+bbody) \times tbnew \times cabs$ \\
\midrule
 & $TN_1$   & $(powerlaw+bbody) \times tbnew \times cabs$ \\
 & $TN_2$   & $powerlaw \times tbnew_1 \times cabs_1 + bbody \times tbnew_2 \times cabs_2$   \\
Both & $TN_3$   & $(powerlaw+diskbb) \times tbnew \times cabs$ \\
 & $TN_4$   & $powerlaw \times tbnew_1 \times cabs_1 + diskbb \times tbnew_2 \times cabs_2$  \\
 & $TN_5$   & $(powerlaw+mekal) \times tbnew \times cabs$ \\
\bottomrule
\end{tabular*}
\caption{\footnotesize \label{tab:models}List of models used to fit the continuum, described in XSPEC notation. The basic components are \textit{powerlaw, bbody, diskbb, bremss, mekal} and \textit{cemekl}, together with \textit{tbnew} to account for the absorption and \textit{cabs} for the non-relativistic Compton scattering. The employed models are a combination of these components, in addition to Gaussian profiles modelling emission lines. We divide the models in three types: $N_{\#}$ for non-thermal, $T_{\#}$ for thermal and $TN_{\#}$ for models containing both thermal and non-thermal components. }
\end{table*}
\end{center}
\section{\label{sec:results}Results}

\subsection{Spectral Atlas}
In Appendix~\ref{Atlas}, we present the full sample of analysed spectra. The figures in Appendix~\ref{Atlas} show the set of analysed observations (cross points), the model employed (solid line), the components of the model (dotted line) and the ratio between observation and model (lower box in each spectrum plot). 

\begin{figure*} 
\begin{center}
\includegraphics[width=0.32\textwidth]{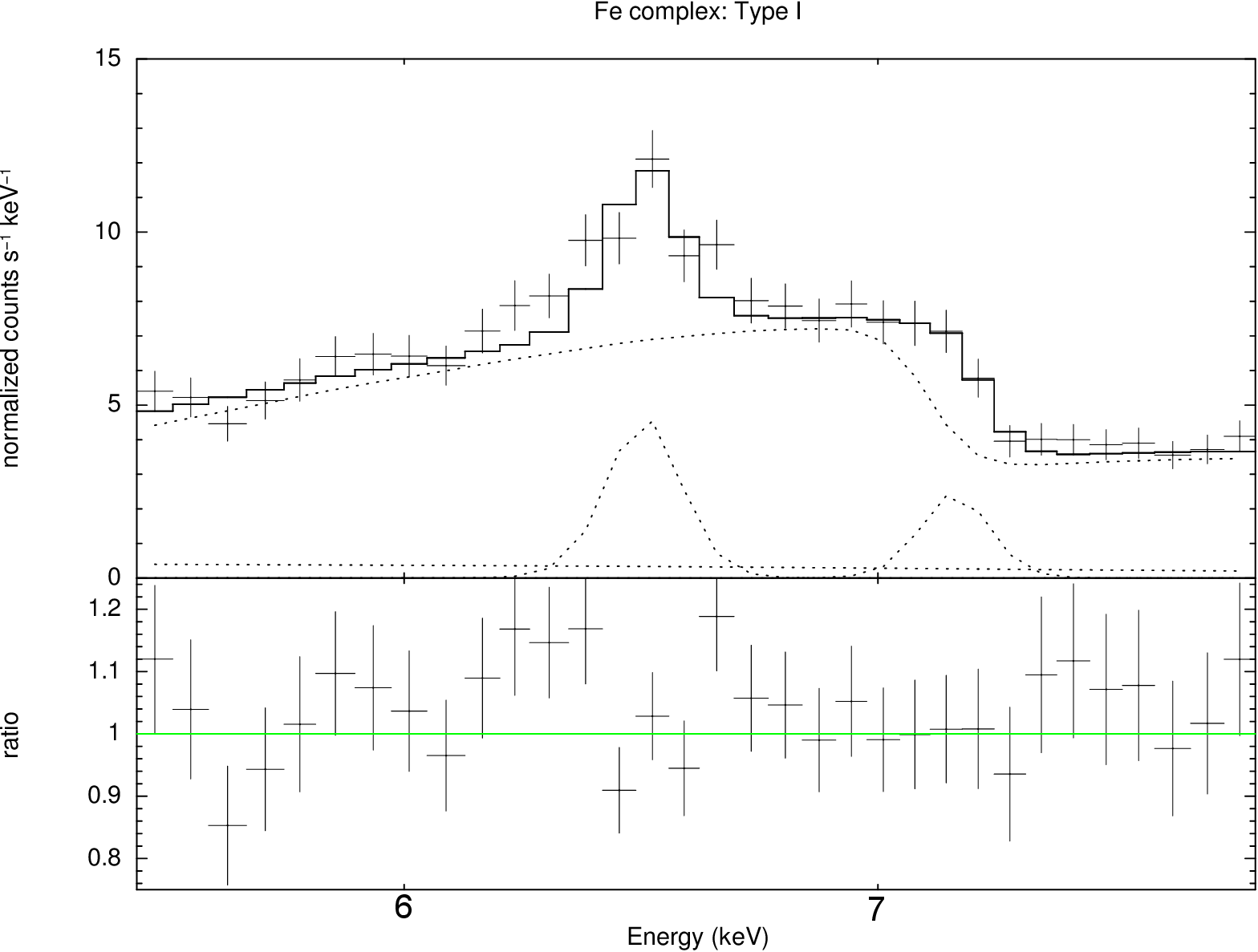}
\includegraphics[width=0.32\textwidth]{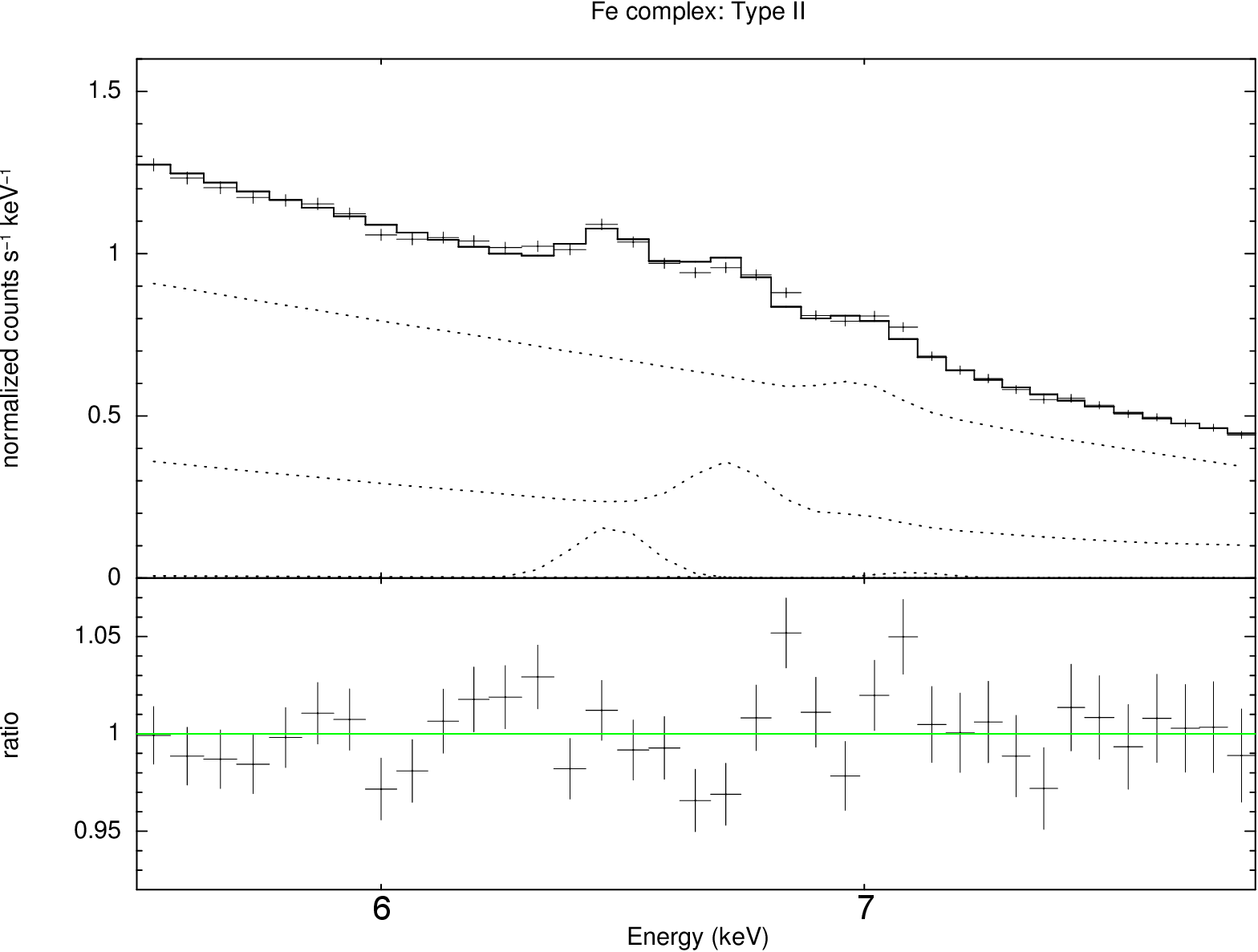}
\includegraphics[width=0.32\textwidth]{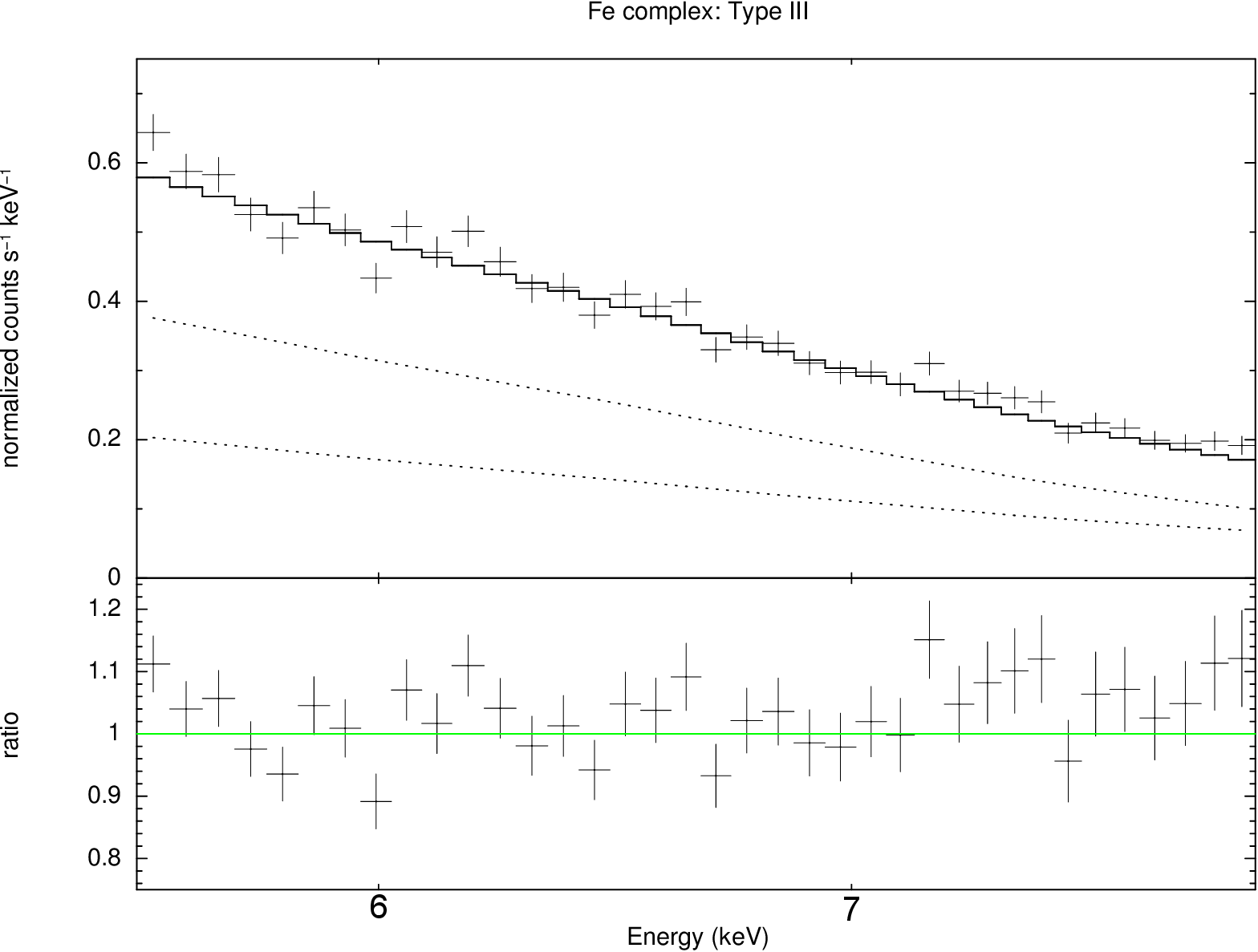}
\caption{\footnotesize Patterns found in the Fe complex of HMXBs: Type~I (left panel), Type~II (central panel), and Type~III (right panel).} 
\label{FeComplex}
\end{center}
\end{figure*}

We show a list of the sources in Table~\ref{tab:sources}, giving the class where we have grouped them and the reference for such a classification. We can see that the different classes of HMXBs behave qualitatively different in the region of the Fe complex ($\sim$6-7~keV), reflecting the distinct accretion regimes that characterize them. We have observed three patterns in the Fe complex, that we define as Type~I, II and III (see Figure~\ref{FeComplex}). We define Type~I, when fluorescence lines FeK$\alpha$ and FeK$\beta$ are observed, but not recombination lines Fe~\textsc{xxv} and Fe~\textsc{xxvi}. We define Type~II, when fluorescence lines are detected, together with recombination lines Fe~\textsc{xxv} and Fe~\textsc{xxvi}. Finally, we define Type~III, when Fe lines are not detected.

\begin{center}
\begin{table}
\begin{tabular}{c|c|c|c|c}
\toprule
 Group & $\#$ Sources & Fe complex & Models & $N_H$ \\
\hline
  & & & & \\
 BeXB               & 10 & Type III & T, TN    & Low  \\
  & & & & \\
 SGXB               & 12 & Type I   & N       & High \\
  & & & & \\
 SFXT               & 10 & Type III & T, N, TN & High \\
                    & & (in quiescence) &      &      \\
  & & & & \\
 $\gamma$~Cass~like & 8  & Type II  & T        & Low  \\
  & & & & \\
 HMGB               & 2  & Type III & T, N, TN & Low  \\  
\bottomrule
\end{tabular}
\caption{\footnotesize \label{tab:atlas} Description of the features observed in this work, for the different groups of HMXBs. We have analyzed data of 46 sources. However, those classified as peculiars (3 sources) or non classified (AX~J1749.1-2733) are not included in this table. We define $N_H$ of a group as \textit{high}, when the typical value observed is well over the estimations of the interstellar $N_H$ in the line of sight of the sources following \cite{Willingale2013}. That is, we say that the $N_H$ of a group is \textit{high} when the absorption is typically intrinsic of the systems.}
\end{table}
\end{center}
The general features observed in this work, for the different groups of HMXB, are summarized in Table~\ref{tab:atlas}, and explained below in more details:
\begin{itemize}
\item BeXBs. We have collected data from 10 sources. All the observations were performed in quiescence. We have detected FeK$\alpha$ emission in only one BeXB (SAX~J2103.5+4545). The upper limit of the FeK$\alpha$ EW in the rest of BeXBs is in general higher than the observed value in SAX~J2103.5+4545, implying that the lack of detections might be due to a poor signal-to-noise. The spectra can be modeled by thermal or a combination of thermal and non-thermal components, except for Swift J045106.8-694803 (fitted using an absorbed power law). Seven sources accept a thermal model, and 6 a combination model (4 of them accept both). 
\item SGXBs. We have gathered data from 12 sources. Ten of them show detectable Fe fluorescence emission. The only exceptions are IGR~J16465-4507 and SAX~J1802.7-2017, the most distant SGXBs included in this work, at 12.5 and 12.4~kpc respectively. The EW upper limits in these two sources is high, implying that their faintness is very likely the reason why we do not detect FeK$\alpha$. The 12 SGXBs can be well fitted using non-thermal models, although thermal components are also plausible in some sources. In general, SGXBs are characterized by high absorption and the presence of Fe fluorescence emission lines. 
\item SFXTs. We have collected data from 10 sources. Three of them show FeK$\alpha$: AX~J1841.0-0536, IGR~J11215-5952 and IGR~J16479-4514. The EW upper limit in the rest of sources is high. Therefore, FeK$\alpha$ would be probably detectable with a better signal-to-noise. The models employed for fitting the SFXT systems are very heterogeneous, with no preference of thermal, non-thermal, neither a combination of both kinds of models. 
\item $\gamma$~Cassiopeae~analogs. We have gathered observations from 8 sources. Five of them exhibit FeK$\alpha$. The EW upper limit in the other 3 sources is very high. Again, it implies a very likely presence of fluorescence in the case of better signal-to-noise. In addition, recombination lines of Fe~\textsc{xxv} and Fe~\textsc{xxvi}, are always present in the set of $\gamma$~Cassiopeae~analogs. These lines are included in the XSPEC model \textit{mekal}. For most of the observations we have achieved a good fit using a combination of \textit{mekal} components. In a few cases we have used other components: \textit{diskbb} and \textit{powerlaw}, but \textit{mekal} is by far the most employed one in $\gamma$~Cassiopeae~like systems, in agreement with previous X-ray analysis \citep{Lopes_de_Oliveira2010,Lopes_de_Oliveira2006}. 
\item HMGBs. We have collected data from two HMGBs: LS~I+61~303 and LS~5039. None of them show Fe features. However, the signal-to-noise in these observations is poor and the upper limits of the FeK$\alpha$ EW are high enough to do not rule out the presence of the line. We have used both thermal and non-thermal components in the fits. 
\item Peculiars. Set of sources that do not accommodate in any of the aforementioned classes of HMXBs, as explained in the introduction. We have collected data of 3 such systems: 4U~2206+54, Centaurus~X-3 and Cygnus~X-1: 
\begin{itemize}
\item 4U~2206+54 does not show any detectable Fe emission line, and the upper limit in the EW of FeK$\alpha$ is low (it is comparable to the upper limits in the BeXBs). It can be fitted by means of an hybrid model (thermal plus non-thermal components).
\item Centaurus~X-3 presents a rich emission lines spectrum. Concretely, in the Fe complex we are able to identify FeK$\alpha$, FeK$\beta$, Fe~\textsc{xxv} and Fe~\textsc{xxvi}. We have used either an hybrid model either a non-thermal model.
\item Cygnus~X-1 exhibits a broad Fe feature, sometimes combined with a faint and narrow, but statistically significant, FeK$\alpha$ line. We have mostly used non-thermal models, occasionally combined with a \textit{bbody} or \textit{diskbb} component. 
\end{itemize} 
\item AX~J1749.1-2733. In this system, the optical member has been classified as a B1-2 \citep{Karasev2010}, but the luminosity class remains unknown, preventing us to incorporate the source in a defined group. Although it does not exhibit detectable Fe emission, the high absorption clearly points to a supergiant companion. In addition, the EW upper limit of FeK$\alpha$ is compatible with the values observed in SGXBs and SFXTs. It can be well fitted using an absorbed \textit{powerlaw} or a \textit{blackbody}. 

In summary, it is very likely that FeK$\alpha$ is an ubiquitous feature in HMXBs, and its detection highly depends on the quality of the observations. In this regard, the EW of the line is very affected by the level of intrinsic absorption present on the sources (see also Section~\ref{sec:curve_growth}). SGXBs and SFXTs, which show higher absorption, tend to exhibit a more prominent Fe fluorescence.
\end{itemize}

\subsection{FeK$\alpha$ width}
\begin{figure} 
\begin{center}
\includegraphics[width=0.5\textwidth]{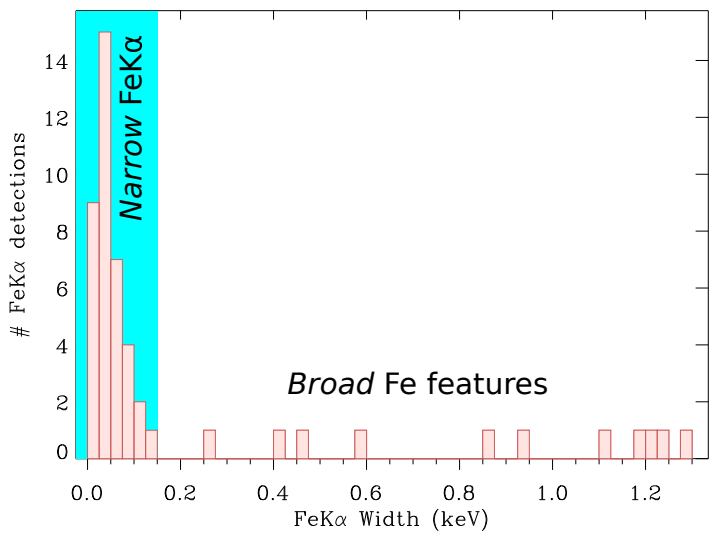}
\end{center}
\caption{\footnotesize Histogram of the FeK$\alpha$ width. The bulk of the detections are grouped showing $\sigma_{line}<0.15$~keV. We define them as \textit{narrow} FeK$\alpha$. The rest are defined as \textit{broad} Fe features. Even though we have detected 60 \textit{narrow} FeK$\alpha$, this plot only includes 38. The reason is that 22 of them are very narrow (or the signal-to-noise very low) and they have been treated in the fits as delta functions (in Table~\ref{tab:FeKa} we present their width as $\sigma_{line}=0$).  }
\label{histo_widths}
\end{figure}

In Table~\ref{tab:FeKa} we show the parameters of every detected FeK$\alpha$, including the width of the line ($\sigma_{line}$). We have made a distinction between \textit{narrow} and \textit{broad} FeK$\alpha$. We have defined \textit{narrow}~FeK$\alpha$ when $\sigma_{line}<0.15$~keV, and \textit{broad}~FeK$\alpha$ when $\sigma_{line}>0.15$~keV. This separation is both physically and observationally motivated:

The origin of broad Fe features in X-ray Binaries is still an opened question, but the most likely alternatives are related to the presence of an accretion disk (see e.g. \cite{Hanke2009,Ng2010,Duro2011}). However, narrow features are not compatible with material rotating at high velocities or relativistically broadened. Given that broad and narrow FeK$\alpha$ have a clearly different origin, they must be analysed in a separated way. Then, it raises the question of how to define the separation mark between them. In Figure~\ref{histo_widths} we can see that in our sample the number of detected sources decreases when increasing $\sigma_{line}$. Moreover, most of the detections are grouped at $\sigma_{line}<0.15$~keV. Hence, $\sigma_{line}<0.15$~keV seems a natural threshold for the definition of \textit{narrow} lines in the sample. In addition, we must pay attention to the plausible contamination of FeK$\alpha$ with Fe~\textsc{xxv}, which is located at $\sim 6.7$~keV. The chosen criterion separates the sources where it is very unlikely that FeK$\alpha$ is contaminated by Fe~\textsc{xxv} (\textit{narrow} lines), from the sources which probably suffer from this issue (\textit{broad} lines). A more detailed analysis of broad Fe features in HMXBs is out of the scope of this paper and its discussion will be fully developed in a forthcoming work. 

\emph{Hereafter, when FeK$\alpha$ is mentioned, we refer to the narrow feature.} From the total number of 108 analysed observations we have detected (narrow)~FeK$\alpha$ in 60 of them.

\subsection{Centroid energy}
\begin{figure} 
\begin{center}
\includegraphics[width=0.5\textwidth]{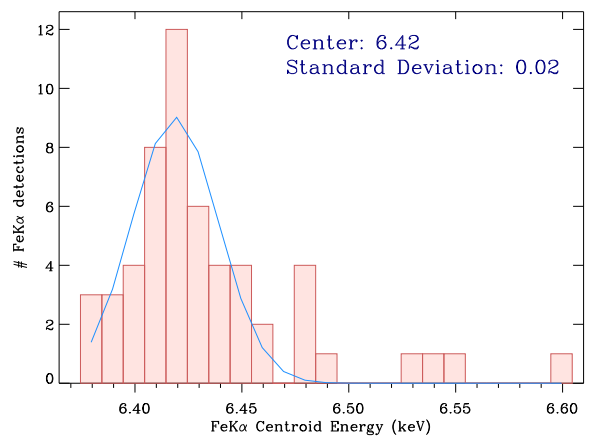}
\end{center}
\caption{\footnotesize Centroid energy of FeK$\alpha$, with a Gaussian fit overplotted (blue profile). The mean value is $6.42 \pm 0.02$~keV, compatible with Fe~\textsc{i-xviii}. Even though we have 60 detections of FeK$\alpha$, in this plot we only see 55. Five cases fulfil the requirements of detection, but the low signal-to-noise do not permit to find an accurate centroid energy. They have not been included in this plot. In these five cases we have set the centroid energy of FeK$\alpha$ to $\sim$6.4~keV. }
\label{histo energies}
\end{figure}

In Figure~\ref{histo energies} we can see an histogram presenting the centroid energy of FeK$\alpha$. A Gaussian fit of the data reveals a mean value for the centroid energy of 6.42~keV. There are not significant differences in the averaged values obtained neither for different classes of HMXBs, neither for different states. The standard deviation is 0.02~keV, comparable to the the error that we typically obtain in the estimation of the centroid energy in the fits (see Table~\ref{tab:FeKa}). Taking into account the standard deviation and the uncertainties in the CTI corrections in EPN\footnote{Please find more information about long-term CTI correction in the release note \textit{EPIC-pn Long-Term CTI}, by M.J.S~Smith et al.~(2014), at http://xmm2.esac.esa.int/docs/documents/CAL-SRN-0309-1-0.ps.gz; and \textit{EPIC status of calibration and data analysis} by Guainazzi~(2008), at http://xmm2.esac.esa.int/external/xmm\_sw\_cal/calib/CAL-TN-0018.pdf.}, the centroid energy of FeK$\alpha$ constrains the ionization state of Fe to less ionized than Fe~\textsc{xviii} \citep{Kallman2004}, in agreement with previous studies in HMXBs \citep{Torrejon2010,Gottwald1995,Nagase1989}. In this regard, the study of \cite{Torrejon2010} using HETGS (more accurate in wavelength than EPN), gives a narrower constrain in the ionization state (Fe~\textsc{i-x}). Our present work support this result adding more sources to the sample.

In the right side of Figure~\ref{histo energies} we can see seven FeK$\alpha$ detections emerging out of the Gaussian profile. Four of them are unlikely to be described by such a Gaussian profile, since they lie more than three times the standard deviation away from the mean energy. All four belong to Cygnus~X-1.

\subsection{Correlated parameters}
One of the goals of this work is to study plausible correlations involving the parameters of FeK$\alpha$ (position, width, intensity and EW, and other parameters in the fits, such as the absorbing column and the intensity of the continuum. Note that, even when a good fit is reached, the confidence region of a parameter might be occasionally difficult to find due to the dependence of the parameter with other parameters of the model. We specify in each of the following subsections the number of cases where a successful estimation of the 90$\%$ confidence region has been done. 

\subsubsection{Continuum Flux vs FeK$\alpha$ Flux}
\begin{figure} 
\begin{center}
\includegraphics[width=0.5\textwidth]{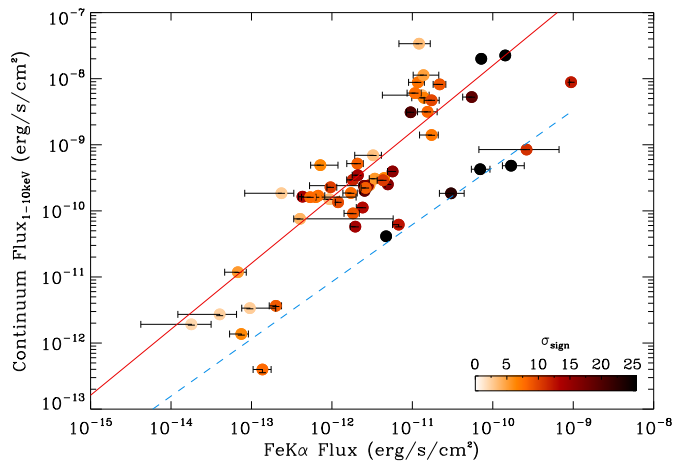}
\end{center}
\caption{\footnotesize $F_{1-10keV}$ versus $F_{FeK\alpha}$. Blue dashed line marks the correlation observed for IGR~J16318-4848 jointly with eclipse observations, and the red solid line follow the bulk of the observations. The color map indicates the $\sigma_{sign}$ of the line (defined in Section~\ref{sec:fitting}).}
\label{fluxes corr}
\end{figure}

In Figure~\ref{fluxes corr} we have represented the unabsorbed flux of the continuum between 1-10~keV cancelling FeK$\alpha$ emission ($F_{1-10keV}$), against the flux of FeK$\alpha$ ($F_{FeK\alpha}$). We have successfully found a 90$\%$ confidence region of the flux of FeK$\alpha$ in 56 cases.

In a logarithmic scale, we identify two different patterns of correlation. First, for a subset including all the eclipse observations and IGR~J16318-4848, we find a correlation with Pearson Coefficient (PC) of 0.98 (blue dashed line). Second, for the rest of observations, we find a correlation with PC=0.89 (red solid line).

A linear fit of the parameters (in logarithmic scale) in the out-of-eclipse observations (red line in Figure~\ref{fluxes corr}), gives the following dependence:
\begin{equation}
\;\;\;F_{1-10keV}(erg/s/cm^2) = F_{FeK\alpha}^{1.00 \pm 0.08} (erg/s/cm^2) \times 10^{2.18 \pm 0.87}
\end{equation}   
The errors show the standard deviation of the parameters in the linear fit. 

The observed divergence among eclipse (plus IGR~J16318-4848) and out-of-eclipse observations suggests that the companion star blocks in a different proportion the continuum and the FeK$\alpha$ emission. Therefore, an important contribution of the fluorescence emission is produced in an extended region of $R \gtrsim R_{\star}$. This is consistent with previous analysis of eclipse observations of HMXBs (e.g. \cite{Rodes-Roca2011} and \cite{Audley2006}). In particular, \cite{Audley2006} estimates that $20\%$ of FeK$\alpha$ in OAO~1657-415 is emitted from 19~light-seconds off the X-ray source.

\begin{figure} 
\begin{center}
\includegraphics[width=0.5\textwidth]{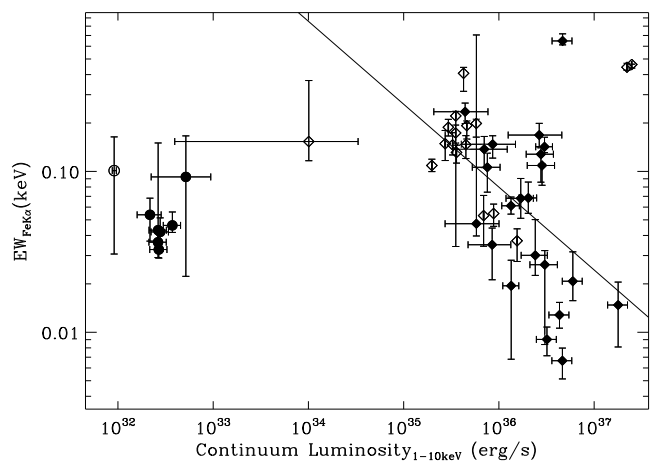}
\end{center}
\caption{\footnotesize EW of FeK$\alpha$ against $L_{1-10\,keV}$. $\gamma$~Cassiopeae~analogs (circles) lie at $L_{1-10\,keV}<10^{33}$~erg/s. Open symbols indicate that either the distance either the error in the estimation of the distance is unknown. The solid line corresponds to a linear fit in logarithmic scale of the filled diamonds, that is, the sources with available distance with error estimation and $L_{1-10\,keV}>10^{33}$~erg/s. }
\label{Bald_eff_lum}
\end{figure}

We have also the luminosity of the continuum, in order to face it with the EW of FeK$\alpha$. For the flux-to-luminosity conversion we have used the estimations of the distance shown in Table~\ref{tab:sources}. We have excluded eclipse and IGR~J16318-4848 from this analysis, given that the EW is strongly affected by the high obscuration of the continuum that they suffer from. In Figure~\ref{Bald_eff_lum} we plot the EW of FeK$\alpha$ against the unabsorbed luminosity of the continuum between 1-10~keV cancelling FeK$\alpha$ emission ($L_{1-10keV}$). We observe two different groups of sources: 1)~$\gamma$~Cassiopeae~analogs lie at low luminosities ($L_{1-10keV} < 10^{33}$~erg/s); 2)~the rest of sources which exhibit FeK$\alpha$. $\gamma$~Cassiopeae~analogs don't show any evident correlation (they are very few points), while the rest present a moderate inverse correlation (PC=-0.25, and PC=-0.39 using only the sources with an available estimation of distance with error, marked as filled diamonds in Figure~\ref{Bald_eff_lum}). A linear fit using the filled diamonds in Figure~\ref{Bald_eff_lum} gives:
\begin{equation}
\label{eq:Baldw_eff_lum}
\;\;\;EW = L_{1-10keV}^{-0.52 \pm 0.27} (erg/s) \times 10^{17.45 \pm 9.83}
\end{equation}

\cite{Baldwin1977} observed an inverse correlation in the EW of CIV and the UV luminosity in AGNs. Analogously, the decrease of the EW of FeK$\alpha$ when increasing the X-ray luminosity is called \textit{X-rays Baldwin effect}. The dependence that we observe is compatible within the error with the one observed by \cite{Torrejon2010} in X-ray Binaries using a narrower energy range: $EW \varpropto L_{1.6-2.5\AA}^{-0.29}$. 

\subsubsection{FeK$\alpha$ Width vs Centroid Energy}
\begin{figure} 
\begin{center}
\includegraphics[width=0.5\textwidth]{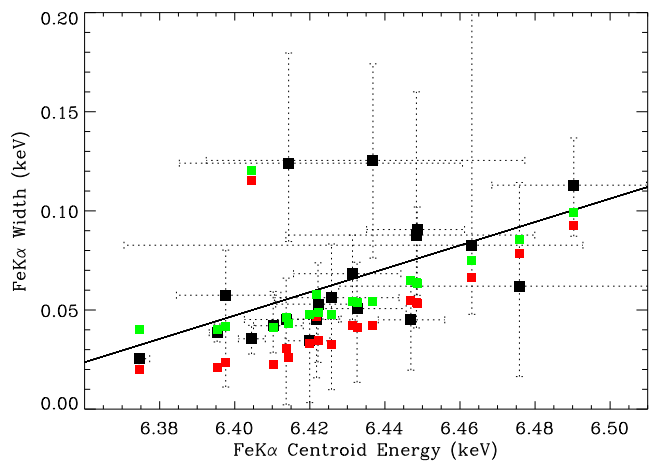}
\end{center}
\caption{\footnotesize Width of FeK$\alpha$ ($\sigma_{line}$) versus the centroid energy (black squares). Black solid line traces a linear fit. We have marked in colour the expected width from the contribution of three different broadening phenomena: line blending, Compton broadening and Doppler shifts, considering velocities of V(km/s)~=~1000~(red) and V(km/s)~=~2000~(green). Every single black square has associated a single red square and a single green square corresponding to the expected values of $\sigma_{line}$ for that observation (see Section~\ref{sec:discussion}). }
\label{Energy Sigma}
\end{figure}

In Figure~\ref{Energy Sigma} we present the centroid energy of this feature versus its width ($\sigma_{line}$). We have successfully found a 90$\%$ confidence region of $\sigma_{line}$ in 20 cases. We can see a moderate correlation (black line in Figure~\ref{Energy Sigma}, PC=0.55), indicating a possible blending of lines. Two observations (uppermost side of Fig.~\ref{Energy Sigma}) do not follow the correlation. They correspond to observations of 4U~1700-37 (Obs.ID~0083280201) and EXO~1722-363 (Obs.ID~0405640201) where the Fe complex is hardly resolved, and therefore it is very likely that a contribution of Fe~\textsc{xxv} and Fe~\textsc{xxvi} in the model of FeK$\alpha$ is increasing the measured width of the FeK$\alpha$ line. 

Coloured squares correspond to the expected width from the contribution of three different broadening phenomena: line blending, Doppler shifts and Compton broadening. A discussion on the different broadening contributions is given in Section~\ref{sec:discussion}. 

\subsubsection{Curve of growth \label{sec:curve_growth}}
\begin{figure} 
\begin{center}
\includegraphics[width=0.5\textwidth]{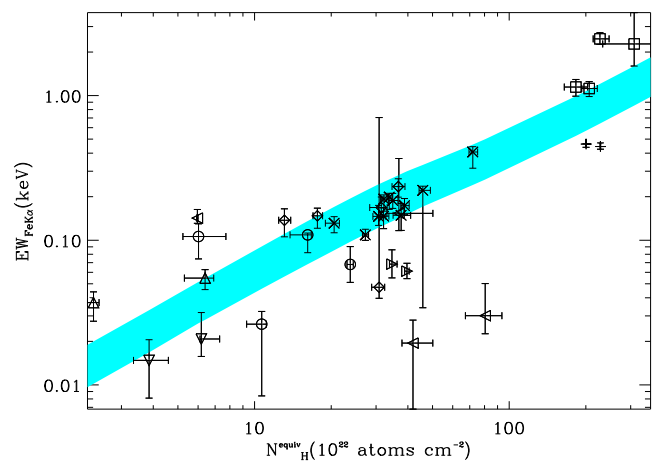}
\end{center}
\caption{\footnotesize Curve of growth observed for FeK$\alpha$, that is, EW against $N_H$. The turquoise band marks the expected correlation using numerical simulations. The sources are identified by different symbols when more than one observation is included: 4U~1700-37 (open circle), 4U~1907+09 (open upward triangle), Cygnus~X-1 (open downward triangle),  EXO1722-363 (open diamond), IGR~J16318-4848 (open square) and IGR~J16320-4751 (plus). Only one observation for Centaurus~X-3, GX~301-2, Vela~X-1 and XTE~J0421+560 (all four a star symbol). }
\label{CurveGrowth}
\end{figure}

In Figure~\ref{CurveGrowth} we show, for out of eclipse observations, the $N_H$ versus the EW of FeK$\alpha$ (what is generally known as the \textit{curve of growth}). We have successfully found a 90$\%$ confidence region of both $N_H$ and EW in 46 cases. We want to take into account observations were $N_H$ reflects the intrinsic absorption of the system. Hence, we have set $N_H > 2$ as a condition to safely exceed the typical $N_H$ of the interstellar medium for the sources here studied (checked using the online application following \cite{Willingale2013}). The use of this criterion excludes the BeXB SAX~J2103.5+4545, the $\gamma$~Cassiopeae~analogs: $\gamma$~Cassiopeae and HD~110432; and the SFXT IGR~J11215-5952. Moreover, eclipse observations show higher EW and they are not comparable to out-of-eclipse observations. Therefore, eclipse observations have not been plotted in Figure~\ref{CurveGrowth}. As a consequence of the chosen criteria, we end up with a set of 36 observations, where all the donors are supergiants. 

$N_H$ and the EW of FeK$\alpha$ are expected to correlate in HMXBs, as shown by \cite{Torrejon2010}, since the spectral lines are usually stronger when the optical depth increases. Our sample confirms these expectations, showing a notable correlation (PC=0.85). 

We have determined the theoretical curve of growth using numerical simulations. In this simulations there is an input of X-ray radiation with a power law profile, that is transmitted through a cloud of spherically distributed neutral matter \citep{Eikmann2012}. 

We have taken into account the power law index ($\Gamma$) in the simulations, since steeper profiles entail less photons available above the Fe K-shell threshold energy, decreasing the EW. That is, for the same $N_H$, the higher is $\Gamma$, the lower is the EW of FeK$\alpha$. In Figure~\ref{CurveGrowth} the turquoise band traces the results from the simulations with $\Gamma \in [0.5,2]$, which is the typical range where we find $\Gamma$ in our fits.  

\subsection{$N_H$: SGXBs and SFXTs}
\begin{figure} 
\begin{center}
\includegraphics[width=0.5\textwidth]{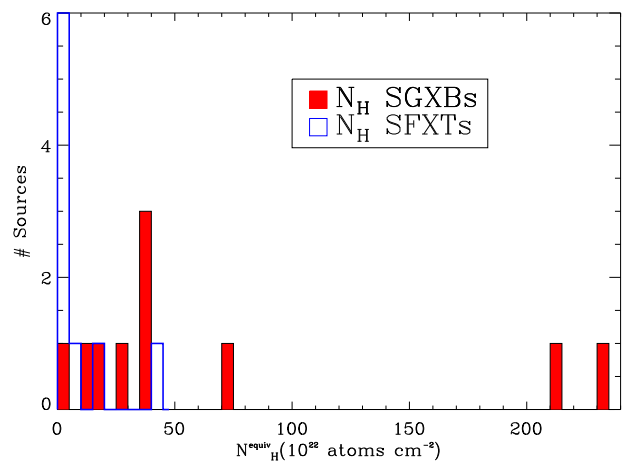}
\end{center}
\caption{\footnotesize Histograms showing a comparison of the $N_H$ values observed in SGXBs (filled red) and SFXTs (empty blue). }
\label{histo_NHs}
\end{figure}

In Figure~\ref{histo_NHs} we have plotted histograms for the $N_H$ values observed in SGXBs (red) and SFXTs (blue). In the cases where we have more than one observation for the same source, we have averaged the values in order to obtain one $N_H$ representative for each system. The orbital phase critically affects the observed $N_H$, and therefore ingress/egress and eclipse phases have not been taken into account.

We find that SGXBs are in general more absorbed than SFXTs. We have performed a permutation test in order to quantify if the observed disparity in the $N_H$ is a random effect. We have 10 $N_H$ values for SGXBs and 9 for SFXTs. We have merged them in a set of 19 elements, and considered every possible combination of 2 groups of 10 and 9 elements (92378 possibilities). We have compared the median of the two subsets, and calculated the absolute difference. $99.7\%$ of the cases have produced a lower absolute difference than the observed one. Using the mean instead of the median the percentage is also very high ($98.8\%$). In conclusion, it is very likely that the discrepancies in the observed $N_H$ values for SGXBs and SFXTs are produced by physical reasons rather than they arise by chance.

\subsection{Individual sources analysis: IGR~J16320-4751 and 4U~1700-37}
\subsubsection{IGR~J16320-4751}
IGR~J16320-4751 was detected by \textit{ASCA} in 1994 and 1997 (corresponding to AX~J1631.9-4752), and by \textit{INTEGRAL} in 2003 \citep{Tomsick2003}. It is a HMXB composed by an O8I optical star and a neutron star \citep{Rahoui2008b}. It shows a modulation of 8.96 days, that it is considered its orbital period \citep{Corbet2005}, and a pulsation period of $\sim$1300s \citep{Lutovinov2005}. The ESA archives permits to collect eleven observations of IGR~J16320-4751, enabling us to study in more detail the curve of growth, as well as tracking the absorption variation along the orbital phase. 

\begin{figure} 
\begin{center}
\includegraphics[width=0.5\textwidth]{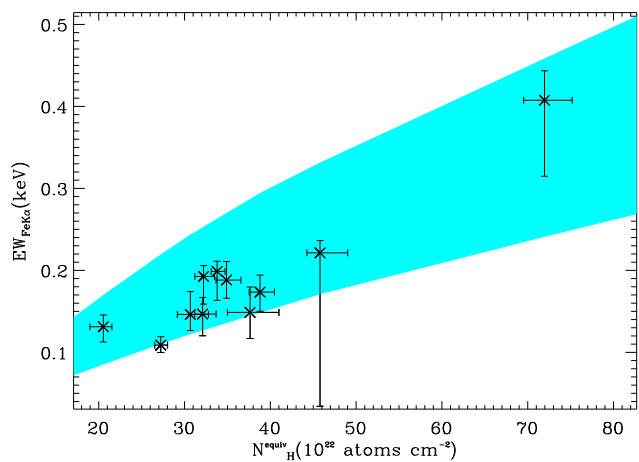}
\end{center}
\caption{\footnotesize EW of FeK$\alpha$ against $N_H$, in IGR~J16320-4751. The turquoise band marks the numerical simulations results, with $\Gamma \in [0.5,2]$.}
\label{CG_16320}
\end{figure}

In Fig.~\ref{CG_16320} we can see the curve of growth, as shown in Fig.~\ref{CurveGrowth}, restricted to IGR~J16320-4751. We clearly see the dependence between $N_H$ and EW of FeK$\alpha$, as stated for the bulk of the sources in Section~\ref{sec:curve_growth}, and the general trend following the numerical simulations marked with the turquoise band. However, the agreement with the simulations is not completely fulfilled, given that the spectral fits of IGR~J16320-4751 have brought out a power law index $\Gamma \sim 0.5$ (see Table~\ref{tab:cont}). Since from a lower power law index we expect more EW of FeK$\alpha$, the points for IGR~J16320-4751 are expected to be located in the upper edge of the turquoise band, corresponding to the simulations with $\Gamma=0.5$. We consider that the general trend is correct, but there are still some uncertainties in the fits and/or the theoretical hypothesis (spherical geometry and neutral matter).

\begin{figure} 
\begin{center}
\includegraphics[width=0.5\textwidth]{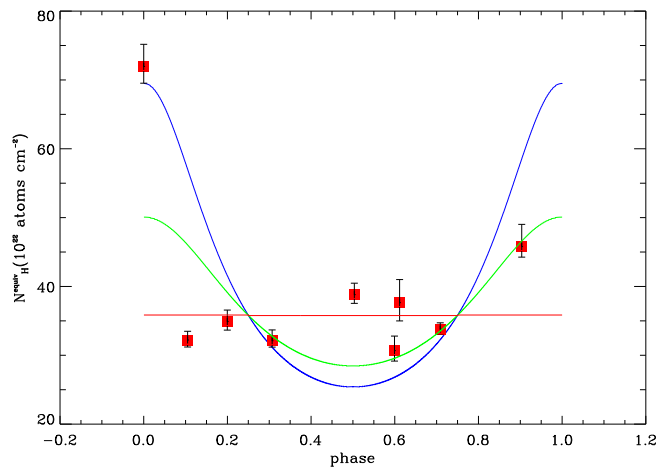}
\end{center}
\caption{\footnotesize Orbital modulation of $N_H$ in the system IGR~J16320-4751. Solid lines correspond to the expected absorption from a smooth wind and a non eccentric orbit, assuming an orbital separation $a=1.6\,R_{\star}$, $R_{\star}=20R_{\odot}$, $\dot{M}=10^{-5}\,M_{\odot}/yr$, $\varv_{\infty}=700km/s$, $\beta=0.8$; and different orbital inclinations $i=0,\pi/10, \, \pi/6$~rad (red, green and blue).  }
\label{ph_nh_16320}
\end{figure}

From 14th August to 17th September, in 2008, a campaign consisting of nine observations of IGR~J16320-4751 was performed by XMM-Newton. We have used this set of data to plot the $N_H$ modulation depending on the orbital phase (Fig.~\ref{ph_nh_16320}). We have set $\phi=0$ at the $N_H$ maximum. We have also calculated the theoretical absorption expected from a smooth wind in a non eccentric orbit using a $\beta$ velocity law \citep{Castor1975} and the motion equation, taking into account variations in the orbital inclination \textit{i}, orbital separation \textit{a}, star radius $R_{\star}$, mass loss rate $\dot{M}$, parameter $\beta$ and the terminal velocity of the wind $\varv_{\infty}$. 

Indeed, $\dot{M}$ and $\varv_{\infty}$ cannot be disentangled in this simple model, and the actual parameter used is $\dot{M}$/$\varv_{\infty}$. However, hereafter we give values of $\dot{M}$ and $\varv_{\infty}$ as if they were free variables, since they are much more commonly used than $\dot{M}$/$\varv_{\infty}$ in the literature. This way, we constrain our parameters to the observed and predicted range of values in O supergiants: $\dot{M}=10^{-7}-10^{-5}\,M_{\odot}$/yr and $\varv_{\infty}=500-3000$~km/s \citep{Kudritzki2000,Vink2001}.

For a null orbital inclination, we obviously obtain a flat $N_H$ modulation (red line in Fig.~\ref{ph_nh_16320}), that describes the observed $N_H$ (except at $\phi=0$), assuming $a=1.6\,R_{\star}$, $R_{\star}=20R_{\odot}$, $\dot{M}=10^{-5}\,M_{\odot}$/yr, $\varv_{\infty}=700$~km/s and $\beta=0.8$. Gradually increasing the orbital inclination ($i=\pi/10, \, \pi/6$~rad; green and blue lines respectively), we are able to describe a high $N_H$ at $\phi=0$, but loosing similarities around other orbital phases. Anyhow, given the simplicity of the model, the obtained parameters are certainly just indicatives.  

\subsubsection{4U~1700-37}
4U~1700-37 was detected for the first time by \textit{Uhuru} in 1970 \citep{Jones1973}. The optical star is HD~153919, a O6.5Iaf located at a distance of 1.9~kpc \citep{Ankay2001}. The orbital period is 3.41~days. Since X-ray pulsations have not been detected so far, the compact object can be either a neutron star or either a black hole. The database of ESA contains five observations of 4U~1700-37, that we split in nine spectra in order to distinguish different states of the source.

\begin{figure} 
\begin{center}
\includegraphics[width=0.5\textwidth]{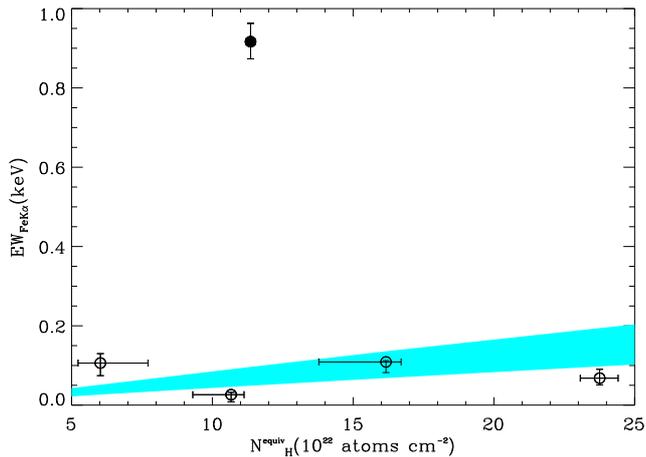}
\end{center}
\caption{\footnotesize EW of FeK$\alpha$ against $N_H$, in 4U~1700-37. The filled circle corresponds to an eclipse observation. The turquoise band traces the numerical calculations with $\Gamma \in [0.5,2]$.}
\label{CG_1700}
\end{figure}

In Figure~\ref{CG_1700} we can see the curve of growth, for 4U~1700-37. Although seven of the nine spectra show FeK$\alpha$, we were able to constrain the boundaries of $N_H$ in only five of the analysis. One of them corresponds to an eclipse observation (filled circle). It shows much more EW because the continuum flux is blocked by the optical star, whereas FeK$\alpha$ comes from a more extended region that is not completely hidden during eclipse. We do not see an obvious dependence between $N_H$ and EW, although the points lie in a region close to the expected values (turquoise band). Anyway, a set of four observations (excluding the eclipse), is too small to perform an statistical analysis.

\begin{figure} 
\begin{center}
\includegraphics[width=0.5\textwidth]{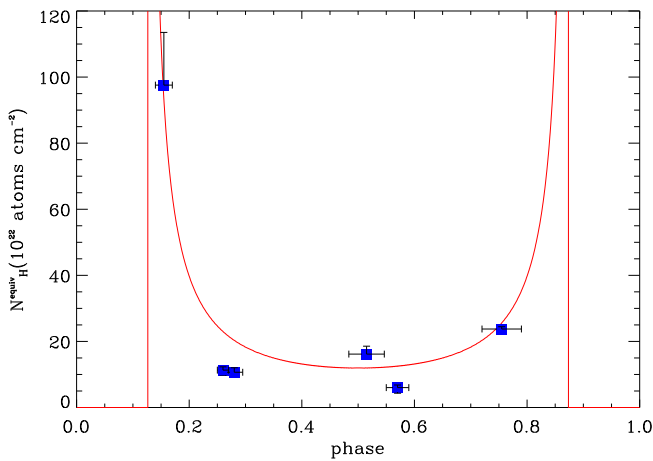}
\end{center}
\caption{\footnotesize Orbital modulation of $N_H$ in 4U~1700-37. Solid line correspond to the expected absorption from a smooth wind and a non eccentric orbit, assuming the stellar values obtained in \cite{Clark2002}, an orbital inclination $i \sim \pi/2$~rad, and an orbital separation $a=1.4\,R_{\star}$. }
\label{ph_nh_1700}
\end{figure}

From 17th to 20th of February 2001, 4U~1700-37 was observed by XMM-Newton four times, in a campaign covering different orbital phases. Therefore we can study the orbital modulation of $N_H$ in the same way as we have done with IGR~J16320-4751, but including more constrains coming from the non-LTE analysis of \cite{Clark2002}, where the following parameters are derived: $R_{\star}=21.9R_{\odot}$, $\dot{M}=9.5 \times 10^{-6}\,M_{\odot}/yr$, $\varv_{\infty}=1750km/s$ and $\beta=1.3$. Taking into account that it is an eclipsing binary, we assume $i \sim \pi/2$. Therefore the only free parameter in our toy model is the orbital separation \textit{a}. The best agreement is achieved when $a=1.4\,R_{\star}$ (see Figure~\ref{ph_nh_1700}). This orbital separation is consistent (in absolute units) to previous estimations of \cite{Conti1975} ($R_{\star}=20 \, R_{\odot}$, $a=1.35\,R_{\star}$) and \cite{Heap1992} ($R_{\star}=18 \pm 3 \, R_{\odot}$, $a=2.0 \pm 0.4 \,R_{\star}$).

\section{Discussion \label{sec:discussion}}
In Figure~\ref{histo energies} we have shown the centroid energies of FeK$\alpha$. The distribution of the histogram is roughly Gaussian, with a standard deviation that reflects the uncertainties in the fits. However, four values are too high to be compatible with this distribution, and all of them belong to Cygnus~X-1. It can be caused either by a not adequate fit or either by a physical reason. As stated before, in Cygnus~X-1 we detect a broad Fe feature, interpreted as a relativistically broadened fluorescence line. However, we have modelled the relativistically broadened feature with a Gaussian profile, that gives an acceptable fit, but might be inadequate, affecting the parameters of the narrow FeK$\alpha$ arising from the fits. Alternatively, as a plausible physical explanation, in Cygnus~X-1 the matter is accreted via an accretion disk, in opposition to the wind-fed accretion of most of the sources showing FeK$\alpha$ in this study. Therefore, the physical properties of the region emitting fluorescence might be different in Cygnus~X-1 and the rest of systems. If this region is hotter in Cygnus~X-1, the centroid energy of FeK$\alpha$ would be shifted to higher energies as we observe.

Regarding IGR~J16318-4848, it is one of the most absorbed systems of the sample, and presents a special configuration of matter in its surroundings,  where dust and cold gas distribute in a non-spherical manner, forming a disk-like structure of matter up to $\sim100R_\star$ \citep{Chaty2012}. A likely high inclination of the system would produce the extreme X-ray absorption and the eclipse-like correlation between FeK$\alpha$ and continuum fluxes. 

In Figure~\ref{Energy Sigma} we have seen that the centroid energy of FeK$\alpha$ is higher when the line is broader. When more ionized Fe goes along with the presence of more variety of Fe ions involved in the total emission, the width resulting from the blending of lines must depend on the centroid energy of the line, as observed. We have estimated the broadening produced in the lines by line blending by: $\sigma_{B} \approx E-6.4$~(keV), with \textit{E} the centroid energy of the line in keV. We note that it is also plausible that not resolved Fe~\textsc{xxv} and Fe~\textsc{xxvi} actually shift and broaden FeK$\alpha$, producing an equivalent effect.

\begin{figure} 
\begin{center}
\includegraphics[width=0.5\textwidth]{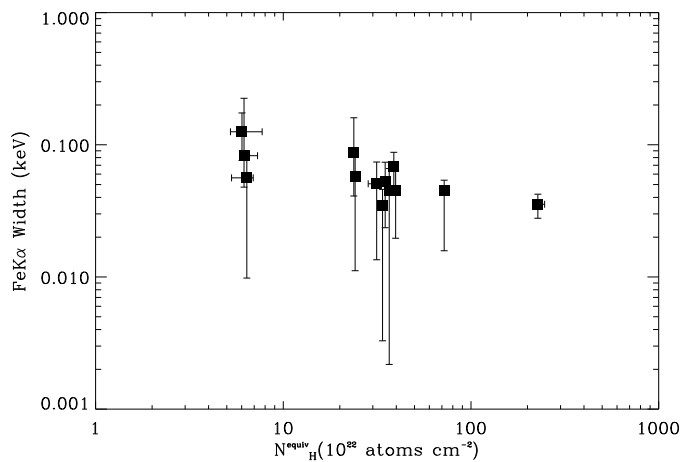}
\end{center}
\caption{\footnotesize Total equivalent hydrogen column ($N_H$) against the width of FeK$\alpha$. }
\label{nH Sigma}
\end{figure}

More processes are also able to significantly broaden FeK$\alpha$. We have considered Compton broadening and Doppler shifts as plausible candidates. Compton scattering has been proposed as a possible broadening mechanism of emission lines in neutron star LMXBs (\cite{DiazTrigo2012}, for GX~13+1). For HMXBs, Compton broadening might also be significant, given the high $N_H$ values observed (and the consequent high number of free electrons). However, if this process is determining the width of the lines, we should observe a direct correlation between the absorption column and the line width. In Figure~\ref{nH Sigma} we can see that such a direct correlation is not present. Moreover, an inverse correlation is plausible. Noticeably, \cite{Cackett2013} analysed three neutron star LMXBs and arrived to a similar result. Therefore, Compton broadening cannot be considered as the main responsible for the observed width in HMXBs, although it is not ruled out as a modest contributor. We name $\sigma_C$ to the contribution of Compton scattering in the line broadening. 

To estimate $\sigma_{C}$, we have used an empirical formula accurate to within 30$\%$, derived from \cite{Kallman1989} and corrected in \cite{Brandt1994}: $$\sigma_{C}=0.019\,E_K\,\tau_{Th}\,(1+0.78kT_e) \simeq 0.12\,\tau_{Th}\,(1+0.78kT_e)$$ where $E_K\simeq 6.4$~keV is the energy of FeK$\alpha$, $\tau_{Th}$ is the Thomson optical depth and $kT_e$ the electronic temperature in keV. We use $\tau_{Th}= \sigma_{Th} \int \, n_e(s) \, ds = \sigma_{Th} \, N_e$, where $\sigma_{Th}$ is the Thomson cross section, $n_e$ the electron number density and the integral is calculated along the line of sight. Assuming solar abundances, a temperature $kT_e \ll 1$~keV and an almost completely ionized matter, as reasonable for galactic massive stars atmospheres, we obtain (see e.g. equation 3.61 in \cite{Novotny1973}): $$N_e = \int \frac{\rho(s) \, ds}{2m_H}(1+X) = \frac{X\,(1+X)}{2} \, N_H \approx 0.7\,N_H \Rightarrow$$ $$\Rightarrow \sigma_{C} \approx 0.5 \, N_H(10^{22}\,cm^{2})\,eV$$ 

Regarding Doppler shifts, they must be taken into account, since a velocity of more than 500~km/s (a very feasible speed, either in the wind either in the accretion flow), would broaden the lines by more than 10~eV. We name $\sigma_D$ to any broadening produced by Doppler effects.  

Line blending, Compton scattering and Doppler shifts produce a resultant width of: $$\sigma_{total}=\sqrt{\sigma_{B}^2 + \sigma_{C}^2 + \sigma_{D}^2}$$ 
We have adopted $\sigma_{D}=20,40$~(eV) corresponding to velocities of $V \approx 1000,2000$~km/s, which are very plausible either in the wind of the supergiant or either in the accretion flow. We have overplotted the corresponding values of $\sigma_{total}$ in Figure~\ref{Energy Sigma}. For every observation (black squares) we have computed the expected value (red with $\sigma_{D}=20$~(eV) and green using $\sigma_{D}=40$~(eV)). A vast majority of the line widths can be described in this way. 

In IGR~J16318-4848, the high absorption measured (above $2\times 10^{24}$~cm$^2$) and the consequent expected Compton broadening of more than 100~eV, is not congruent with the measured width of $\sim 35$~eV. This is another indication that the absorbing matter in this system is cold and not ionized, as already stated by \cite{Chaty2012}. Hence, the employed expression for describing $\sigma_c$ cannot apply here, since there are not sufficient free electrons to broaden the line by means of Compton scattering.  

\begin{figure} 
\begin{center}
\includegraphics[width=0.5\textwidth]{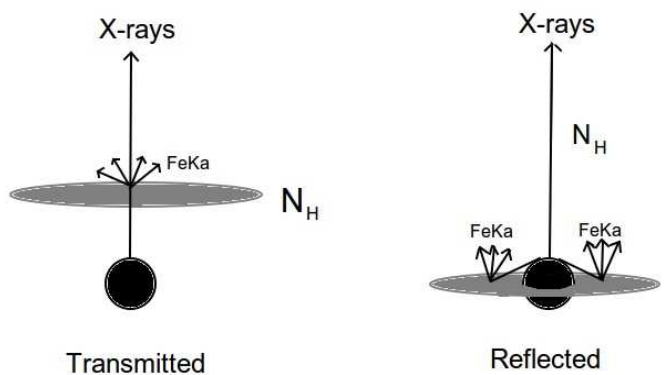}
\end{center}
\caption{\footnotesize Simple sketch of two plausible configurations of circumstellar matter in HMXBs. In the left side, X-rays are transmitted through a dense medium (e.g. the strong wind of the donor), producing high $N_H$ directly correlated with the EW of FeK$\alpha$. In the right side, X-rays are reflected in an accretion disk producing fluorescence, and also transmitted through a more diffuse medium. In this case $N_H$ is not necessarily correlated with the EW of FeK$\alpha$.}
\label{sketch}
\end{figure}

In Figure~\ref{CurveGrowth}, we show the curve of growth of FeK$\alpha$. We require that $N_H>2$ (intrinsic absorption rather than interstellar). In our sample, this criterion constrains the systems in Figure~\ref{CurveGrowth} to those with supergiant donors only. We observe a direct correlation between $N_H$ and the EW. This correlation highlights that the X-ray absorption is strongly linked to the matter that emits FeK$\alpha$, being produced by matter in the line of sight, where the X-rays are absorbed, and not in other plausible regions such as an accretion disk. (see a sketch in Figure~\ref{sketch}). Namely, in the systems included in Figure~\ref{CurveGrowth} (all with a supergiant optical star), FeK$\alpha$ is produced from the transmission of X-rays through the circumstellar medium, that is, either through the strong wind of the supergiant donor either through any structure in the line of sight such as ionization or accretion wakes. The hypothetical reflection of X-rays in an independent medium might produce an additional amount of FeK$\alpha$, as observed in the BeXB GRO~J1008$-$57 by \cite{Kuehnel2013}, which is not appreciated in the systems shown in Figure~\ref{CurveGrowth}. 

As stated before, the region where fluorescence is emitted must be more extended than $R>R_{\star}$, and consequently the wind of the companion star, being illuminated by the X-ray source, is an obvious contributor to both the absorption and the FeK$\alpha$ emission. The orbital modulation of $N_H$ shown in Figures~\ref{ph_nh_16320} and \ref{ph_nh_1700} also support such an interpretation. 

Moreover, most of the observations track the numerical simulations assuming a characteristic range of $\Gamma$ values, indicating that an isotropic distribution of absorbing (and FeK$\alpha$ emitting) matter is not far from reality. However, we do not ignore the variability and heterogeneous properties of the HMXBs environment, which is very plausible to change between different sources and even between different snapshots of the same source. We think that this variability is reflected in the observed dispersion of the curve of growth and the moderate discrepancies regarding the simplified view of spherically distributed neutral matter.

In Figure~\ref{histo_NHs} we have compared the observed values of $N_H$ for SGXBs and SFXTs. We have observed that SGXBs are in general more absorbed sources than SFXTs. It implies that, in SGXBs, either the compact object orbits a denser region of the donor wind, or either the interaction \textit{compact object - wind} modifies the environment producing an enhancement of density in its surroundings. 

Having a look to the orbital parameters of SFXTs (see e.g. Table~2 in \cite{Romano2014}), their orbital periods lie in a wide range of values, from $\sim$3~days for IGR~J16479-4514, up to 164~days for IGR~J11215-5952. Some of them show high eccentricity. However, currently there is not a complete description of the orbital parameters in SFXTs. Therefore, we cannot rule out the possibility that, in this sample, SFXTs are less absorbed than SGXBs because of the distance of the compact object to the donor star. In this regard, further studies of orbital parameters of SFXTs will be useful. 

Regarding the interaction \textit{compact object - wind}, hydrodynamic simulations show that the gravitational potential of the compact object, and the X-ray radiation field, can significantly modify the observed value of $N_H$ \citep{Manousakis2011,Manousakis2012}. In SGXBs, where the X-ray emission is more persistent, these effects might be stronger than in SFXTs, notably increasing the absorption.

\section{Conclusions \label{sec:conclusions}}
We have performed the spectral analysis of the whole sample of publicly available EPN XMM-Newton observations of HMXBs until August, 2013, in order to describe its FeK$\alpha$ emission. In total, the study involves 46 HMXBs, 21 of them showing significant FeK$\alpha$ emission. As expected, we dealed with a very heterogenous set of objects and states of the sources, that must be properly organized. We have classified the systems in the following groups: BeXBs, SGXBs, SFXTs, $\gamma$~Cass~analogs, HMGBs and peculiar sources. Furthermore, we have divided the observations depending on the source behaviour in the following states: quiescence, flare, eclipse ingress/egress, and eclipse. With these criteria, we finally had a set of 108 spectra for our analysis, that produced the following conclusions: 
\begin{itemize}
\item The spectral atlas gives a qualitative description of the different groups of HMXBs, specially recognizable for SGXBs (fluorescence but not recombination Fe lines), and $\gamma$~Cass~analogs (modeled by \textit{mekal} models and presenting fluorescence and recombination Fe lines). FeK$\alpha$ is very likely an ubiquitous feature in HMXBs, but its detection strongly depends on the quality of the observations. SGXBs and SFXTs, which show the higher $N_H$ among the HMXBs, tend to exhibit a more prominent Fe fluorescence.
\item The value of the centroid energy of FeK$\alpha$ constrains the ionization state of the reprocessing material to be below Fe~\textsc{xviii}. 
\item The FeK$\alpha$ and continuum fluxes are well correlated, as expected for the fluorescence emission of matter illuminated from an X-ray source. The different coefficients of correlation for eclipse and out-of-eclipse observations is in agreement with previous eclipse observations of HMXBs, in the sense of showing that the FeK$\alpha$ is produced in a region that ranges from the vicinity of the X-ray source to distances of the same order or greater than the stellar radius.
\item We confirm an inverse correlation between the X-ray luminosity and the EW of FeK$\alpha$ \textit{X-rays Baldwin effect}. $\gamma$~Cass~analogs do not follow this correlation. This suggests that the Fe~K$\alpha$ reprocessing scenario is fundamentally different in SGXBs and in $\gamma$~Cass~analogs.
\item The width of FeK$\alpha$ is predominantly below 0.15~keV, and can be widely explained appealing to line blending, Compton broadening and moderate Doppler shifts ($\sim$1000~km/s). 
\item The curve of growth in SGXBs shows a clear correlation of FeK$\alpha$ EW and $N_H$, indicating a strong link between the absorbing and the fluorescent matter. From numerical simulations, the assumption of spherically distributed absorbing matter is roughly correct for most of the SGXBs. 
\item The $N_H$ values observed in SGXBs are higher than in SFXTs. The disparity is hardly produced by chance, as shown by a permutation test of the sample, denoting a fundamental physical reason beneath. Systematic differences in the orbital parameters or different interaction \textit{compact object - stellar wind} are plausible candidates to explain such a discrepancy.
\item The orbital modulation of $N_H$ in IGR~J16320-4751 and 4U~1700-37, together with the aforementioned results, points to the stellar wind as the main contributor to both continuum absorption and FeK$\alpha$ emission in the case of supergiant donors.
\end{itemize}

In summary, we present the most comprehensive study of FeK$\alpha$ in HMXBs to date, complementing previous surveys at high resolution \citep{Torrejon2010}. We have increased significantly the number of sources and extended the study to all major classes of massive  binaries, leading to the already stated conclusions.  

\vspace*{0.5cm}    
{\footnotesize \textit{Acknowledgments.} Based on observations obtained with XMM-Newton, an ESA science mission with instruments and contributions directly funded by ESA member states and the USA (NASA). This research has made use of software obtained from NASA's High Energy Astrophysics Science Archive Research Center (HEASARC), a service of Goddard Space Flight Center and the Smithsonian Astrophysical Observatory. The work of AGG has been supported by the Spanish MICINN under FPI Fellowship BES-2011-050874 associated to the project AYA2010-15431. Part of this research was possible thanks to a travel grant from the Deutscher Akademischer Austauschdienst. The authors acknowledge the help of the International Space Science Institute at Bern, Switzerland, and the Faculty of the European Space Astronomy Centre. This work was supported partially by the Generalitat Valenciana project number GV2014/088 and by the Vicerectorat d'Investigació, Desenvolupament i Innovació de la Universitat d'Alacant under grant GRE12-35. JJRR acknowledges the support by the Matsumae International Foundation fellowship No14G04. SMN thanks the support of the Spanish Unemployment Agency, allowing her to continue her scientific collaborations during the critical situation of the Spanish Research System. }

\bibliographystyle{hapj}
\bibliography{bib}

\newpage
\appendix
\onecolumn
\pagestyle{empty}
\topmargin=0.5cm
\section{Tables}

\begin{table}[ht]
\begin{tabular}{c|cccc}
\toprule
Class & Source & Distance & Reference & Reference \\
 & & (kpc) & (class) & (distance) \\ 
\midrule
 & 1A 0535+26 & $2.00 \pm 0.70$ & \cite{Shenavrin2011} & \cite{Steele1998} \\
 & 2S 1845-024 & 10.00 & \cite{Bodaghee2007} & \cite{Grimm2002} \\
 & X Persei & $0.80 \pm 0.14$ & \cite{Liu2006} & \cite{Megier2009} \\
 & AX J1820.5-1434 & $8.20 \pm 3.50$ & \cite{Liu2006} & \cite{Kinugasa1998} \\
 & RX J0146.9+6121 & 1.95 & \cite{Liu2006} & \cite{Kharchenko2005} \\
BeXB & RX J0440.9+4431 & $3.30 \pm 0.50$ & \cite{Liu2006} & \cite{Reig2005} \\
 & RX J1037.5-5647 & 5.00 & \cite{Liu2006} & \cite{Grimm2002} \\
 & SAX J2103.5+4545 & $3.20 \pm 0.80$ & \cite{Liu2006} & \cite{Baykal2002} \\
 & Swift J045106.8-694803 & $50.60 \pm 2.10$ & \cite{Beardmore2009} & \cite{Bartlett2013} \\
 & V0332+53 & $7.50 \pm 1.50$ & \cite{Liu2006} & \cite{Negueruela1999} \\
\midrule
 & 4U 1538-522 & $6.40 \pm 1.00$ & \cite{Liu2006} & \cite{Reynolds1992} \\
 & 4U 1700-37 & $2.12 \pm 0.34$ & \cite{Liu2006} & \cite{Megier2009} \\
 & 4U 1907+09 & 5.00 & \cite{Liu2006} & \cite{Cox2005} \\
 & EXO1722-363 & $8.00 \pm 2.50$ & \cite{Liu2006} & \cite{Mason2009} \\
 & GX 301-2 & 3.04 & \cite{Liu2006} & \cite{Kaper2006} \\
 & IGR J16207-5129 & 6.10 & \cite{Nespoli2008} & \cite{Nespoli2008} \\
SGXB & IGR J16318-4848 & $3.60 \pm 2.60$ & \cite{Filliatre2004} & \cite{Filliatre2004} \\
 & IGR J16320-4751 & $3.50$ & \cite{Coleiro2013} & \cite{Rahoui2008b} \\
 & IGR J16465-4507 & 12.50 & \cite{Heras2007} & \cite{Smith2004ATel} \\
 & SAX J1802.7-2017 & 12.40 & \cite{Torrejon2010b} & \cite{Torrejon2010b} \\
 & Vela X-1 & $1.90 \pm 0.20$ & \cite{Liu2006} & \cite{Sadakane1985} \\
 & XTE J0421+560 & $5.00 \pm 4.00$ & \cite{Liu2006} & - \\
\midrule
 & AXJ1841.0-0536 & $7.80 \pm 0.74$ & \cite{Romano2011a} & \cite{Nespoli2007} \\
 & IGR J00370+6122 & 3.00 & \cite{Gonzalez-Galan2014} & \cite{Negueruela2004} \\
 & IGR J11215-5952 & 6.20 & \cite{Sguera2006} & \cite{Masetti2006} \\
 & IGR J16328-4726 & $6.50 \pm 3.50$ & \cite{Fiocchi2013} & \cite{Fiocchi2013} \\
 & IGR J16418-4532 & 13.00 & \cite{Sguera2006} & \cite{Drave2013} \\
SFXT & IGRJ16479-4514 & $7.50 \pm 2.50$ & \cite{Romano2011b} & \cite{Chaty2008} \\
 & XTE J1739-302 & $2.30 \pm 0.60$ & \cite{Smith2006} & \cite{Negueruela2006} \\
 & IGR J17544-2619 & $3.20 \pm 1.00$ & \cite{Pellizza2006} & \cite{Pellizza2006} \\
 & IGR J18450-0435 & 3.60 & \cite{Sguera2007} & \cite{Coe1996} \\
 & IGR J18483-0311 & $3.50 \pm 0.50$ & \cite{Rahoui2008} & \cite{Rahoui2008} \\
\midrule
 & $\gamma$~Cassiopeiae & $0.12 \pm 0.01$ & \cite{Lopes_de_Oliveira2010} & \cite{Megier2009} \\
 & HD 110432 & $0.39 \pm 0.05$ & \cite{MASmith2006} & \cite{Megier2009} \\
 & HD 119682 & 1.11 & \cite{Rakowski2006} & \cite{Kharchenko2005} \\
 & HD 157832 & 0.53 & \cite{Lopes_de_Oliveira2011} & \cite{Lopes_de_Oliveira2011} \\
$\gamma$-Cass-like & HD 161103 & $1.50 \pm 0.50$ & \cite{Lopes_de_Oliveira2006} & \cite{Lopes_de_Oliveira2006} \\
 & HD 45314 & $1.50 \pm 0.26$ & \cite{Rauw2013} & \cite{Megier2009} \\
 & SAO 49725 & $2.20 \pm 0.60$ & \cite{Lopes_de_Oliveira2010} & \cite{Lopes_de_Oliveira2006} \\
 & SS397 & $1.50 \pm 0.20$ & \cite{Lopes_de_OliveiraPhDT} & \cite{Riquelme2012} \\
\midrule
 & LS I +61 303 & 2.50 & \cite{Dubus2013} & \cite{Grimm2002} \\
HMGB & LS 5039 & $2.50 \pm 0.10$ & \cite{Dubus2013} & \cite{Casares2005} \\
\midrule	
O9.5V+NS & 4U 2206+54 & $2.90 \pm 0.20$ & \cite{Blay2006} & \cite{Riquelme2012} \\
disk-fed & Cen X-3 & $10.00 \pm 1.00$ & \cite{Liu2006} & \cite{Hutchings1979} \\
BH-SGXB & Cygnus X-1 & $2.10 \pm 0.25$ & \cite{Liu2006} & \cite{Ziolkowski2005} \\
B1-2 & AX J1749.1-2733 & $16.00 \pm 3.50$ & \cite{Karasev2010} & \cite{Karasev2010} \\
\bottomrule
\end{tabular}
\caption{Table of sources included in the sample of HMXBs.}
\label{tab:sources}
\end{table}

\begin{landscape}
\begin{longtable}{ccccccccccccc}
\caption{Parameters of the continuum.} \\
\toprule
Source & Obs ID & State & Model & Gain & $\chi_{Red}^2$ & $P_0$ & $L_{1-10\,keV}$ & $\sum N_H$ & $\Gamma$ & $kT_1$ & $kT_2$ & $kT_3$ \\
 &  &  &  &  &  &  & $(10^{33}erg/s)$ & $(10^{22}cm^{-2})$ &  & (keV) & (keV) & (keV) \\
\endfirsthead
\caption{Parameters of the continuum (continued).} \\
\endhead
\\
\caption{Parameters of the continuum, including observation ID, state of the source, model (see Table~\ref{tab:models}), artificial gain (offset or slope), reduced $\chi^2$, null hypothesis probability ($P_0$), luminosity between 1-10~keV ($L_{1-10\,keV}$), total $N_H$ (adding every absorption component), photon index ($\Gamma$), and temperature of thermal components ($kT_i$): bbody, diskbb, bremss, mekal or cemekl (depending on the model). Parameters frozen or unbounded are included without an error estimation. Therefore they are not used in the plots and the subsequent discussion. The possible states are: quiescence(Q), flare(F), eclipse ingress/egress(I/E), eclipse(E) and dip(D). } \\
\endlastfoot
\midrule
1A 0535+26 & 0674180101 & Q & $TN_1$ & - & 0.87 & 8.70E-01 & $11.03_{-6.39}^{+9.20}$ & $0.72_{-0.07}^{+0.06}$ & $1.40_{-0.08}^{+0.06}$ & $1.24_{-0.06}^{+0.05}$ & - & - \\ 
2S 1845-024 & 0302970801 & Q & $T_5$ & - & 0.73 & 7.67E-01 & $15.93_{-0.98}^{+2.54}$ & $8.82_{-4.11}^{+2.80}$ & - & $4.72_{-0.92}^{+10.76}$ & - & - \\ 
X Persei & 0151380101 & Q & $T_{13}$ & - & 1.34 & 3.68E-03 & $72.42_{-23.16}^{+27.82}$ & $0.50_{-0.04}^{+0.04}$ & - & $1.67_{-0.07}^{+0.05}$ & $10.87_{-3.41}^{+5.49}$ & - \\ 
X Persei & 0151380101 & Q & $TN_1$ & - & 1.41 & 7.44E-04 & $72.49_{-23.17}^{+27.81}$ & $0.58_{-0.06}^{+0.08}$ & $1.63_{-0.11}^{+0.14}$ & $1.59_{-0.04}^{+0.04}$ & - & - \\ 
X Persei & 0600980101 & Q & $TN_1$ & - & 1.28 & 1.06E-02 & $86.94_{-27.62}^{+33.07}$ & $0.39_{-0.11}^{+0.09}$ & $1.34_{-0.14}^{+0.13}$ & $1.08_{-0.15}^{+0.12}$ & - & - \\ 
AX J1820.5-1434 & 0511010101 & Q & $T_5$ & - & 0.53 & 9.92E-01 & $12.05_{-8.39}^{+14.36}$ & $7.85_{-1.78}^{+2.64}$ & - & $2.05_{-0.23}^{+0.29}$ & - & - \\ 
AX J1820.5-1434 & 0511010101 & Q & $N_1$ & - & 0.60 & 9.77E-01 & $12.37_{-8.63}^{+14.36}$ & $14.21_{-2.03}^{+3.60}$ & $1.34_{-0.22}^{+0.35}$ & - & - & - \\ 
RX J0146.9+6121 & 0201160101 & Q & $TN_1$ & - & 1.31 & 8.58E-03 & $7.65_{-0.12}^{+0.11}$ & 0.90 & $1.61_{-0.08}^{+0.10}$ & $1.34_{-0.09}^{+0.08}$ & - & - \\ 
RX J0440.9+4431 & 0653660101 & Q & $T_6$ & - & 1.16 & 8.55E-02 & $65.96_{-18.86}^{+22.21}$ & $0.78_{-0.07}^{+0.06}$ & - & $0.91_{-0.06}^{+0.08}$ & $2.28_{-0.12}^{+0.23}$ & - \\ 
RX J0440.9+4431 & 0653660101 & Q & $T_{10}$ & - & 1.21 & 4.14E-02 & $66.00_{-18.91}^{+22.13}$ & $1.17_{-0.05}^{+0.05}$ & - & $1.30_{-0.05}^{+0.05}$ & $5.82_{-0.35}^{+0.35}$ & - \\ 
RX J0440.9+4431 & 0653660101 & Q & $TN_1$ & - & 1.24 & 2.54E-02 & $66.06_{-18.89}^{+22.26}$ & $1.22_{-0.13}^{+0.12}$ & $0.95_{-0.13}^{+0.12}$ & $1.38_{-0.07}^{+0.09}$ & - & - \\ 
RX J1037.5-5647 & 0550560101 & Q & $T_9$ & - & 1.02 & 4.16E-01 & $11.58_{-0.43}^{+0.33}$ & 1.04 & - & $4.42_{-0.45}^{+0.37}$ & - & - \\ 
RX J1037.5-5647 & 0550560101 & Q & $TN_1$ & - & 1.00 & 4.94E-01 & $11.53_{-0.38}^{+0.38}$ & 1.04 & $1.09_{-0.12}^{+0.38}$ & $1.55_{-0.22}^{+0.32}$ & - & - \\ 
SAX J2103.5+4545 & 0149550401 & Q & $T_{10}$ & - & 1.26 & 1.77E-02 & $805.54_{-355.73}^{+459.66}$ & $0.79_{-0.04}^{+0.06}$ & - & $2.46_{-0.06}^{+0.06}$ & $1.53_{-0.15}^{+0.08}$ & - \\ 
SAX J2103.5+4545 & 0149550401 & Q & $TN_1$ & - & 1.24 & 2.69E-02 & $806.39_{-355.79}^{+462.07}$ & $0.83_{-0.11}^{+0.11}$ & $0.96_{-0.22}^{+0.24}$ & $2.15_{-0.21}^{+0.12}$ & - & - \\ 
Swift J045106.8-694803 & 0679381401 & Q & $N_1$ & - & 1.06 & 3.28E-01 & $860.26_{-112.25}^{+118.45}$ & $0.29_{-0.15}^{+0.21}$ & $1.07_{-0.10}^{+0.13}$ & - & - & - \\ 
V0332+53 & 0506190101 & Q & $T_5$ & - & 0.60 & 7.58E-01 & $0.19_{-0.09}^{+0.14}$ & 0.83 & - & $0.40_{-0.07}^{+0.08}$ & - & - \\ 
\midrule
4U 1538-522 & 0152780201 & EC & $N_1$ & - & 1.33 & 7.12E-03 & $20.04_{-6.07}^{+7.24}$ & $0.97_{-0.03}^{+0.03}$ & $0.37_{-0.05}^{+0.04}$ & - & - & - \\ 
4U 1538-522 & 0152780201 & I/E & $N_2$ & - & 1.44 & 5.65E-04 & $659.55_{-192.86}^{+227.25}$ & $80.48_{-13.17}^{+13.16}$ & $0.86_{-0.05}^{+0.05}$ & - & - & - \\ 
4U 1700-37 & 0083280101 & D & $N_2$ & - & 1.38 & 9.69E-03 & $161.58_{-52.25}^{+62.67}$ & $11.31_{-0.99}^{+1.49}$ & $1.20_{-0.11}^{+0.17}$ & - & - & - \\ 
4U 1700-37 & 0083280101 & F & $N_1$ & - & 0.97 & 5.73E-01 & $1790.09_{-545.05}^{+645.31}$ & $10.67_{-1.36}^{+0.46}$ & $1.10_{-0.09}^{+0.07}$ & - & - & - \\ 
4U 1700-37 & 0083280101 & Q & $N_1$ & - & 1.16 & 1.04E-01 & $624.17_{-190.74}^{+228.25}$ & 15.49 & 1.22 & - & - & - \\ 
4U 1700-37 & 0083280201 & F & $N_2$ & - & 1.28 & 1.23E-02 & $1952.17_{-589.18}^{+693.31}$ & $16.17_{-2.39}^{+0.54}$ & $1.12_{-0.03}^{+0.10}$ & - & - & - \\ 
4U 1700-37 & 0083280201 & Q & $TN_2$ & - & 0.90 & 7.98E-01 & $619.13_{-189.62}^{+224.34}$ & $6.02_{-0.79}^{+1.69}$ & $1.35_{-0.25}^{+0.54}$ & $2.59_{-0.26}^{+0.14}$ & - & - \\ 
4U 1700-37 & 0083280201 & Q & $N_2$ & - & 1.00 & 4.69E-01 & $623.31_{-190.77}^{+226.45}$ & $8.16_{-0.72}^{+2.55}$ & $1.03_{-0.04}^{+0.08}$ & - & - & - \\ 
4U 1700-37 & 0083280301 & Q & $N_2$ & - & 1.14 & 1.20E-01 & $718.06_{-217.98}^{+259.67}$ & $23.76_{-0.69}^{+0.65}$ & $1.00_{-0.05}^{+0.04}$ & - & - & - \\ 
4U 1700-37 & 0083280401 & EC & $N_1$ & - & 1.22 & 1.30E-01 & $24.12_{-8.93}^{+10.00}$ & 11.66 & 0.51 & - & - & - \\ 
4U 1700-37 & 0083280401 & I/E & $N_2$ & - & 1.13 & 2.01E-01 & $313.50_{-99.84}^{+116.85}$ & 97.57 & 0.58 & - & - & - \\ 
4U 1700-37 & 0600950101 & EC & $N_1$ & - & 1.18 & 6.97E-02 & $16.64_{-5.01}^{+5.93}$ & $11.36_{-0.16}^{+0.15}$ & 0.70 & - & - & - \\ 
4U 1700-37 & 0600950101 & EC & $N_2$ & - & 1.04 & 3.58E-01 & $16.77_{-5.05}^{+5.98}$ & $15.96_{-0.20}^{+0.22}$ & 0.70 & - & - & - \\ 
4U 1907+09 & 0555410101 & F & $TN_1$ & - & 1.10 & 1.89E-01 & $1288.31_{-4.92}^{+6.16}$ & $2.33_{-0.12}^{+0.12}$ & $0.75_{-0.09}^{+0.08}$ & $0.87_{-0.04}^{+0.03}$ & - & - \\ 
4U 1907+09 & 0555410101 & Q & $TN_2$ & - & 0.99 & 5.16E-01 & $721.87_{-3.28}^{+2.75}$ & $6.39_{-1.09}^{+0.52}$ & $0.45_{-0.10}^{+0.10}$ & $1.18_{-0.03}^{+0.07}$ & - & - \\ 
4U 1907+09 & 0555410101 & Q & $N_2$ & - & 1.13 & 1.30E-01 & $718.59_{-3.04}^{+3.08}$ & $9.10_{-1.21}^{+1.03}$ & $1.14_{-0.02}^{+0.02}$ & - & - & - \\ 
EXO1722-363 & 0206380401 & Q & $T_5$ & - & 1.33 & 9.84E-03 & $522.62_{-279.02}^{+389.71}$ & $17.65_{-0.68}^{+0.86}$ & - & $2.78_{-0.09}^{+0.10}$ & - & - \\ 
EXO1722-363 & 0206380401 & Q & $N_1$ & - & 1.28 & 2.30E-02 & $528.75_{-282.05}^{+393.88}$ & $24.14_{-0.95}^{+1.41}$ & $0.78_{-0.07}^{+0.09}$ & - & - & - \\ 
EXO1722-363 & 0405640201 & EC & $T_5$ & - & 0.71 & 7.47E-01 & $2.77_{-1.61}^{+2.35}$ & $17.78_{-3.72}^{+6.87}$ & - & 4.94 & - & - \\ 
EXO1722-363 & 0405640201 & EC & $N_1$ & - & 0.85 & 5.79E-01 & $2.82_{-1.72}^{+2.59}$ & $19.87_{-3.85}^{+21.04}$ & $-0.19_{-0.28}^{+0.75}$ & - & - & - \\ 
EXO1722-363 & 0405640301 & Q & $T_5$ & - & 0.91 & 7.42E-01 & $474.87_{-254.37}^{+357.48}$ & $13.10_{-0.66}^{+0.77}$ & - & $2.71_{-0.09}^{+0.11}$ & - & - \\ 
EXO1722-363 & 0405640301 & Q & $N_1$ & - & 1.04 & 3.67E-01 & $479.86_{-257.24}^{+360.57}$ & $18.55_{-0.88}^{+1.18}$ & $0.77_{-0.08}^{+0.09}$ & - & - & - \\ 
EXO1722-363 & 0405640401 & Q & $T_5$ & - & 1.09 & 2.69E-01 & $181.41_{-97.28}^{+136.89}$ & $36.68_{-2.02}^{+2.28}$ & - & $3.14_{-0.18}^{+0.21}$ & - & - \\ 
EXO1722-363 & 0405640401 & Q & $N_1$ & - & 1.00 & 4.77E-01 & $182.90_{-98.01}^{+137.96}$ & $44.56_{-2.00}^{+3.18}$ & $0.63_{-0.09}^{+0.14}$ & - & - & - \\ 
EXO1722-363 & 0405640901 & Q & $N_1$ & - & 1.14 & 1.50E-01 & $203.03_{-108.96}^{+150.31}$ & $30.84_{-2.01}^{+1.59}$ & $0.97_{-0.07}^{+0.18}$ & - & - & - \\ 
EXO1722-363 & 0405640801 & Q & $T_7$ & - & 1.15 & 1.44E-01 & $287.66_{-153.67}^{+215.12}$ & $31.45_{-3.11}^{+1.71}$ & - & $0.23_{-0.07}^{+0.00}$ & $2.62_{-0.09}^{+0.16}$ & - \\ 
GX 301-2 & 0555200401 & F & $N_3$ & slope & 1.29 & 1.21E-02 & $3412.87_{-5.37}^{+5.83}$ & $200.37_{-0.57}^{+0.53}$ & 0.81 & - & - & - \\ 
GX 301-2 & 0555200401 & Q & $N_3$ & slope & 1.82 & 9.70E-09 & $1334.78_{-2.47}^{+3.30}$ & $228.22_{-0.57}^{+0.49}$ & 1.04 & - & - & - \\ 
IGR J16207-5129 & 0402920201 & Q & $N_2$ & - & 1.17 & 8.16E-02 & $94.98_{--0.63}^{+2.93}$ & 16.27 & 1.27 & - & - & - \\ 
IGR J16318-4848 & 0154750401 & Q & $T_5$ & - & 1.12 & 2.31E-01 & $11.11_{-10.28}^{+22.83}$ & $227.17_{-13.52}^{+19.61}$ & - & $2.89_{-0.32}^{+0.43}$ & - & - \\ 
IGR J16318-4848 & 0154750401 & Q & $N_1$ & - & 1.05 & 3.63E-01 & $11.15_{-10.31}^{+22.66}$ & $232.66_{-8.47}^{+12.48}$ & $0.91_{-0.17}^{+0.19}$ & - & - & - \\ 
IGR J16318-4848 & 0201000201 & Q & $T_5$ & - & 0.89 & 7.12E-01 & $11.56_{-10.70}^{+23.87}$ & $206.14_{-13.46}^{+15.62}$ & - & $2.67_{-0.27}^{+0.39}$ & - & - \\ 
IGR J16318-4848 & 0201000201 & Q & $N_1$ & - & 0.92 & 6.36E-01 & $11.62_{-10.75}^{+23.87}$ & $214.22_{-9.48}^{+11.00}$ & $1.08_{-0.16}^{+0.20}$ & - & - & - \\ 
IGR J16318-4848 & 0201000301 & Q & $T_5$ & - & 0.73 & 6.83E-01 & $3.61_{-3.36}^{+8.26}$ & $309.57_{-76.63}^{+49.52}$ & - & $2.15_{-0.46}^{+2.90}$ & - & - \\ 
IGR J16318-4848 & 0201000301 & Q & $N_1$ & - & 0.76 & 6.55E-01 & $3.64_{-3.39}^{+7.93}$ & $317.51_{-14.71}^{+56.91}$ & $1.66_{-0.11}^{+0.66}$ & - & - & - \\ 
IGR J16318-4848 & 0201000401 & Q & $T_5$ & - & 1.12 & 2.57E-01 & $7.52_{-6.96}^{+15.62}$ & $182.75_{-18.20}^{+20.34}$ & - & $2.55_{-0.31}^{+0.45}$ & - & - \\ 
IGR J16318-4848 & 0201000401 & Q & $N_1$ & - & 1.16 & 2.12E-01 & $7.56_{-7.00}^{+15.54}$ & $191.92_{-8.77}^{+21.91}$ & $1.19_{-0.20}^{+0.45}$ & - & - & - \\ 
IGR J16320-4751 & 0128531101 & Q & $T_5$ & - & 0.99 & 4.94E-01 & $78.37_{-45.61}^{+69.98}$ & $24.97_{-4.87}^{+5.10}$ & - & $2.19_{-0.23}^{+0.33}$ & - & - \\ 
IGR J16320-4751 & 0128531101 & Q & $T_{11}$ & - & 1.03 & 4.23E-01 & $79.52_{-45.66}^{+70.79}$ & $34.91_{-4.88}^{+1.50}$ & - & 50.75 & - & - \\ 
IGR J16320-4751 & 0128531101 & Q & $N_1$ & - & 1.03 & 4.16E-01 & $79.73_{-46.82}^{+69.23}$ & $35.37_{-4.55}^{+8.66}$ & $1.41_{-0.28}^{+0.45}$ & - & - & - \\ 
IGR J16320-4751 & 0201700301 & F & $N_2$ & - & 0.89 & 8.13E-01 & $918.65_{-515.23}^{+724.71}$ & $20.53_{-1.57}^{+1.00}$ & $0.45_{-0.13}^{+0.12}$ & - & - & - \\ 
IGR J16320-4751 & 0201700301 & Q & $N_2$ & - & 0.93 & 6.96E-01 & $417.44_{-233.15}^{+328.31}$ & $27.23_{-0.70}^{+0.80}$ & $0.65_{-0.05}^{+0.07}$ & - & - & - \\ 
IGR J16320-4751 & 0556140101 & Q & $N_2$ & - & 0.99 & 5.15E-01 & $474.86_{-267.46}^{+382.82}$ & $45.79_{-1.53}^{+3.23}$ & $0.73_{-0.08}^{+0.13}$ & - & - & - \\ 
IGR J16320-4751 & 0556140201 & Q & $N_2$ & - & 1.05 & 3.47E-01 & $920.37_{-515.52}^{+729.24}$ & $32.21_{-1.01}^{+1.27}$ & $0.39_{-0.06}^{+0.07}$ & - & - & - \\ 
IGR J16320-4751 & 0556140301 & Q & $N_2$ & - & 1.09 & 2.28E-01 & $890.63_{-498.80}^{+705.54}$ & $32.11_{-0.92}^{+1.58}$ & $0.43_{-0.05}^{+0.09}$ & - & - & - \\ 
IGR J16320-4751 & 0556140401 & Q & $N_2$ & - & 1.21 & 6.25E-02 & $547.92_{-307.19}^{+436.81}$ & $38.80_{-1.26}^{+1.67}$ & $0.72_{-0.07}^{+0.08}$ & - & - & - \\ 
IGR J16320-4751 & 0556140501 & Q & $N_2$ & - & 0.84 & 8.57E-01 & $443.72_{-251.54}^{+363.15}$ & $37.64_{-2.65}^{+3.36}$ & $0.67_{-0.12}^{+0.17}$ & - & - & - \\ 
IGR J16320-4751 & 0556140601 & Q & $N_2$ & - & 1.04 & 3.73E-01 & $1145.80_{-640.15}^{+902.57}$ & $33.80_{-0.69}^{+0.92}$ & $0.32_{-0.03}^{+0.05}$ & - & - & - \\ 
IGR J16320-4751 & 0556140701 & Q & $N_2$ & - & 1.24 & 5.34E-02 & $418.19_{-234.89}^{+333.69}$ & $71.96_{-2.43}^{+3.22}$ & $0.17_{-0.07}^{+0.08}$ & - & - & - \\ 
IGR J16320-4751 & 0556140801 & Q & $N_2$ & - & 1.15 & 1.31E-01 & $533.78_{-299.24}^{+425.72}$ & $34.90_{-1.24}^{+1.67}$ & $0.53_{-0.06}^{+0.09}$ & - & - & - \\ 
IGR J16320-4751 & 0556141001 & Q & $N_2$ & - & 0.93 & 7.05E-01 & $666.20_{-373.27}^{+530.68}$ & $30.68_{-1.52}^{+2.11}$ & $0.52_{-0.08}^{+0.12}$ & - & - & - \\ 
IGR J16465-4507 & 0164561001 & Q & $N_1$ & - & 0.58 & 9.23E-01 & $74.04_{-8.28}^{+6.11}$ & $72.18_{-15.83}^{+24.40}$ & $0.42_{-0.50}^{+0.55}$ & - & - & - \\ 
SAX J1802.7-2017 & 0206380601 & Q & $N_1$ & - & 1.28 & 3.18E-02 & $624.71_{-20.14}^{+16.86}$ & $15.72_{-1.35}^{+1.99}$ & $0.96_{-0.13}^{+0.19}$ & - & - & - \\ 
Vela X-1 & 0111030101 & F & $N_2$ & - & 1.35 & 4.24E-03 & $547.45_{-112.92}^{+126.06}$ & $34.63_{-0.77}^{+1.64}$ & $1.66_{-0.07}^{+0.13}$ & - & - & - \\ 
Vela X-1 & 0111030101 & Q & $N_1$ & - & 0.87 & 8.65E-01 & $323.43_{-65.56}^{+73.19}$ & $39.68_{-0.82}^{+0.82}$ & $1.65_{-0.06}^{+0.05}$ & - & - & - \\ 
XTE J0421+560 & 0139760101 & Q & $N_1$ & - & 0.70 & 9.51E-01 & $3.99_{-3.84}^{+9.48}$ & $36.88_{-3.76}^{+13.28}$ & $0.57_{-0.26}^{+0.44}$ & - & - & - \\ 
\midrule
AXJ1841.0-0536 & 0604820301 & F & $N_3$ & - & 1.39 & 1.54E-03 & $501.72_{-94.20}^{+104.41}$ & $41.90_{-3.98}^{+8.22}$ & $1.27_{-0.05}^{+0.07}$ & - & - & - \\ 
IGR J00370+6122 & 0501450101 & Q & $T_{10}$ & - & 0.84 & 9.23E-01 & $47.94_{-0.60}^{+0.64}$ & $0.87_{-0.09}^{+0.13}$ & - & $1.16_{-0.08}^{+0.09}$ & $4.20_{-0.55}^{+0.69}$ & - \\ 
IGR J00370+6122 & 0501450101 & Q & $T_6$ & - & 0.91 & 7.71E-01 & $47.70_{-0.62}^{+0.75}$ & 0.65 & - & $0.76_{-0.05}^{+0.07}$ & $1.89_{-0.10}^{+0.18}$ & - \\ 
IGR J00370+6122 & 0501450101 & Q & $T_9$ & - & 0.97 & 5.92E-01 & $47.95_{-0.68}^{+0.59}$ & $1.17_{-0.05}^{+0.05}$ & - & $3.08_{-0.09}^{+0.09}$ & - & - \\ 
IGR J00370+6122 & 0501450101 & Q & $TN_1$ & - & 0.84 & 9.24E-01 & $48.01_{-0.74}^{+0.58}$ & $1.00_{-0.13}^{+0.17}$ & $1.16_{-0.14}^{+0.19}$ & $1.28_{-0.07}^{+0.11}$ & - & - \\ 
IGR J11215-5952 & 0405181901 & F & $TN_1$ & - & 1.04 & 3.64E-01 & $639.50_{-5.72}^{+5.64}$ & $0.89_{-0.06}^{+0.09}$ & $1.04_{-0.09}^{+0.15}$ & $1.87_{-0.23}^{+0.26}$ & - & - \\ 
IGR J11215-5952 & 0405181901 & Q & $TN_1$ & - & 0.98 & 5.39E-01 & $76.78_{-0.97}^{+0.87}$ & $1.02_{-0.12}^{+0.21}$ & $0.63_{-0.15}^{+0.21}$ & $1.22_{-0.06}^{+0.14}$ & - & - \\ 
IGR J16328-4726 & 0654190201 & Q & $T_5$ & - & 1.04 & 3.77E-01 & $32.07_{-25.42}^{+45.82}$ & $16.67_{-1.25}^{+1.32}$ & - & $2.14_{-0.10}^{+0.11}$ & - & - \\ 
IGR J16328-4726 & 0654190201 & Q & $T_9$ & - & 1.03 & 3.93E-01 & $32.35_{-25.65}^{+46.00}$ & $22.21_{-0.95}^{+1.58}$ & - & $4.48_{-0.59}^{+0.45}$ & - & - \\ 
IGR J16328-4726 & 0654190201 & Q & $N_1$ & - & 1.07 & 2.98E-01 & $32.54_{-25.80}^{+46.43}$ & $25.13_{-1.62}^{+2.26}$ & $1.41_{-0.13}^{+0.18}$ & - & - & - \\ 
IGR J16418-4532 & 0206380301 & Q & $T_5$ & - & 0.90 & 7.30E-01 & $122.35_{-4.50}^{+4.58}$ & $10.73_{-1.18}^{+1.42}$ & - & $2.11_{-0.13}^{+0.15}$ & - & - \\ 
IGR J16418-4532 & 0206380301 & Q & $N_1$ & - & 0.91 & 7.01E-01 & $124.60_{-5.00}^{+4.33}$ & $17.64_{-1.53}^{+2.35}$ & $1.35_{-0.15}^{+0.22}$ & - & - & - \\ 
IGR J16418-4532 & 0405180501 & Q & $T_5$ & - & 1.11 & 1.99E-01 & $78.92_{-2.38}^{+2.30}$ & $3.92_{-0.36}^{+0.42}$ & - & $1.77_{-0.07}^{+0.07}$ & - & - \\ 
IGR J16418-4532 & 0405180501 & Q & $N_1$ & - & 1.10 & 2.26E-01 & $82.17_{-2.32}^{+2.23}$ & $8.25_{-0.57}^{+0.77}$ & $1.49_{-0.10}^{+0.12}$ & - & - & - \\ 
IGRJ16479-4514 & 0512180101 & EC & $N_1$ & - & 0.95 & 5.92E-01 & $5.30_{-3.10}^{+4.63}$ & $8.21_{-1.14}^{+1.62}$ & $1.75_{-0.26}^{+0.32}$ & - & - & - \\ 
IGRJ16479-4514 & 0512180101 & I/E & $N_3$ & - & 0.74 & 9.67E-01 & $278.33_{-159.69}^{+225.37}$ & 69.51 & 1.12 & - & - & - \\ 
XTE J1739-302 & 0554720101 & F & $T_5$ & - & 1.25 & 5.66E-02 & $1.61_{-0.76}^{+1.05}$ & $1.51_{-0.18}^{+0.21}$ & - & $1.37_{-0.05}^{+0.05}$ & - & - \\ 
XTE J1739-302 & 0554720101 & F & $T_{11}$ & - & 1.08 & 2.88E-01 & $1.71_{-0.80}^{+1.11}$ & $3.81_{-0.32}^{+0.31}$ & - & $10.46_{-1.84}^{+3.51}$ & - & - \\ 
XTE J1739-302 & 0554720101 & F & $N_1$ & - & 1.20 & 9.43E-02 & $1.73_{-0.82}^{+1.11}$ & $4.29_{-0.33}^{+0.46}$ & $1.81_{-0.10}^{+0.13}$ & - & - & - \\ 
XTE J1739-302 & 0554720101 & Q & $T_5$ & - & 1.20 & 2.51E-01 & $0.16_{-0.08}^{+0.12}$ & $1.89_{-0.81}^{+1.12}$ & - & $1.00_{-0.12}^{+0.13}$ & - & - \\ 
XTE J1739-302 & 0554720101 & Q & $T_{11}$ & - & 1.39 & 1.26E-01 & $0.17_{-0.09}^{+0.13}$ & $4.15_{-1.08}^{+1.50}$ & - & $3.92_{-1.37}^{+2.55}$ & - & - \\ 
XTE J1739-302 & 0554720101 & Q & $N_1$ & - & 1.52 & 7.23E-02 & $0.18_{-0.09}^{+0.13}$ & $5.08_{-0.92}^{+1.59}$ & $2.49_{-0.30}^{+0.46}$ & - & - & - \\ 
XTE J1739-302 & 0561580101 & F & $T_5$ & - & 0.96 & 6.04E-01 & $5.68_{-2.67}^{+3.63}$ & $2.24_{-0.23}^{+0.26}$ & - & $1.67_{-0.06}^{+0.06}$ & - & - \\ 
XTE J1739-302 & 0561580101 & F & $T_{11}$ & - & 0.84 & 8.79E-01 & $5.96_{-2.79}^{+3.85}$ & $5.12_{-0.42}^{+0.29}$ & - & $34.27_{-8.96}^{+44.58}$ & - & - \\ 
XTE J1739-302 & 0561580101 & F & $N_1$ & - & 0.88 & 8.05E-01 & $6.00_{-2.82}^{+3.81}$ & $5.31_{-0.41}^{+0.50}$ & $1.43_{-0.09}^{+0.11}$ & - & - & - \\ 
XTE J1739-302 & 0561580101 & Q & $T_5$ & - & 1.24 & 6.93E-02 & $1.14_{-0.54}^{+0.74}$ & $2.28_{-0.25}^{+0.29}$ & - & $1.42_{-0.05}^{+0.06}$ & - & - \\ 
XTE J1739-302 & 0561580101 & Q & $T_{11}$ & - & 1.02 & 4.33E-01 & $1.20_{-0.57}^{+0.79}$ & $4.94_{-0.39}^{+0.38}$ & - & $10.85_{-1.85}^{+3.85}$ & - & - \\ 
XTE J1739-302 & 0561580101 & Q & $N_1$ & - & 1.09 & 2.77E-01 & $1.22_{-0.58}^{+0.78}$ & $5.52_{-0.42}^{+0.63}$ & $1.82_{-0.10}^{+0.15}$ & - & - & - \\ 
IGR J17544-2619 & 0148090501 & F & $T_5$ & - & 0.90 & 7.16E-01 & $16.58_{-9.11}^{+13.39}$ & $2.10_{-0.28}^{+0.34}$ & - & $1.32_{-0.06}^{+0.07}$ & - & - \\ 
IGR J17544-2619 & 0148090501 & F & $T_{11}$ & - & 0.81 & 8.91E-01 & $17.77_{-9.78}^{+14.39}$ & $4.58_{-0.43}^{+0.40}$ & - & $8.60_{-1.63}^{+3.08}$ & - & - \\ 
IGR J17544-2619 & 0148090501 & F & $N_1$ & - & 0.87 & 7.83E-01 & $18.12_{-10.01}^{+14.45}$ & $5.17_{-0.48}^{+0.75}$ & $1.92_{-0.14}^{+0.21}$ & - & - & - \\ 
IGR J18450-0435 & 0306170401 & F & $TN_1$ & - & 1.04 & 3.51E-01 & $131.25_{-1.55}^{+1.41}$ & $2.18_{-0.24}^{+0.31}$ & $0.84_{-0.22}^{+0.26}$ & $1.58_{-0.13}^{+0.26}$ & - & - \\ 
IGR J18450-0435 & 0306170401 & Q & $TN_1$ & - & 1.05 & 3.39E-01 & $22.69_{-0.46}^{+0.37}$ & $2.47_{-0.27}^{+0.58}$ & $0.79_{-0.19}^{+0.66}$ & $1.74_{-0.08}^{+0.23}$ & - & - \\ 
IGR J18483-0311 & 0406140201 & Q & $T_5$ & - & 1.17 & 1.98E-01 & $1.25_{-0.39}^{+0.48}$ & $4.64_{-0.75}^{+1.11}$ & - & $1.36_{-0.10}^{+0.09}$ & - & - \\ 
IGR J18483-0311 & 0406140201 & Q & $T_9$ & - & 1.28 & 9.12E-02 & $1.26_{-0.39}^{+0.49}$ & $7.42_{-0.83}^{+1.28}$ & - & $2.07_{-0.22}^{+0.25}$ & - & - \\ 
IGR J18483-0311 & 0406140201 & Q & $N_1$ & - & 1.39 & 4.16E-02 & $1.29_{-0.41}^{+0.48}$ & $11.40_{-0.33}^{+0.53}$ & $2.44_{-0.14}^{+0.15}$ & - & - & - \\ 
\midrule
$\gamma$~Cassiopeiae & 0201220101 & Q & $T_3$ & slope & 1.39 & 1.42E-03 & $0.24_{-0.05}^{+0.05}$ & 0.56 & - & 0.96 & 5.27 & 33.04 \\ 
$\gamma$~Cassiopeiae & 0651670201 & Q & $T_4$ & - & 1.22 & 3.62E-02 & $0.27_{-0.05}^{+0.06}$ & $0.23_{-0.02}^{+0.01}$ & - & $28.62_{-0.92}^{+1.14}$ & - & - \\ 
$\gamma$~Cassiopeiae & 0651670201 & Q & $T_3$ & - & 0.89 & 8.39E-01 & $0.27_{-0.05}^{+0.06}$ & $0.42_{-0.04}^{+0.05}$ & - & 0.74 & 3.42 & 31.13 \\ 
$\gamma$~Cassiopeiae & 0651670301 & Q & $T_3$ & - & 1.43 & 4.40E-04 & $0.24_{-0.05}^{+0.05}$ & 0.56 & - & 0.64 & 3.95 & 40.83 \\ 
$\gamma$~Cassiopeiae & 0651670401 & Q & $T_3$ & offset & 1.15 & 1.09E-01 & $0.33_{-0.07}^{+0.07}$ & 0.56 & - & 0.66 & 3.58 & 38.62 \\ 
$\gamma$~Cassiopeiae & 0651670501 & Q & $T_3$ & offset & 1.18 & 6.73E-02 & $0.23_{-0.05}^{+0.05}$ & 0.56 & - & 0.66 & 3.40 & 38.38 \\ 
HD 110432 & 0504730101 & Q & $N_1$ & - & 1.24 & 2.81E-02 & $0.21_{-0.06}^{+0.06}$ & $0.17_{-0.02}^{+0.02}$ & $1.52_{-0.02}^{+0.02}$ & - & - & - \\ 
HD 110432 & 0504730101 & Q & $TN_5$ & - & 1.03 & 3.78E-01 & $0.21_{-0.06}^{+0.06}$ & $0.11_{-0.02}^{+0.01}$ & $0.94_{-0.15}^{+0.18}$ & $8.69_{-1.14}^{+1.16}$ & - & - \\ 
HD 119682 & 0551000201 & Q & $T_2$ & - & 1.11 & 2.06E-01 & $0.16_{-0.01}^{+0.01}$ & 1.81 & - & $0.22_{-0.05}^{+0.00}$ & $14.53_{-3.50}^{+9.04}$ & - \\ 
HD 119682 & 0551000201 & Q & $TN_3$ & - & 1.18 & 1.06E-01 & $0.16_{-0.00}^{+0.01}$ & 1.81 & $1.63_{-0.16}^{+0.15}$ & $0.12_{-0.01}^{+0.00}$ & - & - \\ 
HD 157832 & 0551020101 & Q & $T_1$ & - & 1.07 & 3.12E-01 & $0.08_{-0.00}^{+0.00}$ & 0.34 & - & $6.50_{-0.73}^{+0.90}$ & - & - \\ 
HD 161103 & 0201200101 & Q & $T_2$ & - & 1.20 & 9.76E-02 & $0.41_{-0.23}^{+0.34}$ & 0.94 & - & 1.03 & 7.05 & - \\ 
HD 45314 & 0670080301 & Q & $T_2$ & - & 0.96 & 5.63E-01 & $0.30_{-0.10}^{+0.14}$ & 0.73 & - & 1.03 & 13.29 & - \\ 
SAO 49725 & 0201200201 & Q & $T_2$ & - & 1.31 & 7.39E-02 & $0.52_{-0.27}^{+0.39}$ & 0.77 & - & 0.82 & $18.86_{-2.80}^{+61.04}$ & - \\ 
SAO 49725 & 0201200201 & Q & $TN_5$ & - & 1.19 & 1.67E-01 & $0.55_{-0.29}^{+0.41}$ & 0.77 & $0.82_{-0.37}^{+0.40}$ & $1.97_{-0.53}^{+0.92}$ & - & - \\ 
SS397 & 0122700101 & Q & $T_1$ & - & 0.84 & 7.90E-01 & $0.17_{-0.05}^{+0.07}$ & 1.37 & - & $7.13_{-1.14}^{+2.49}$ & - & - \\ 
SS397 & 0122700101 & Q & $N_1$ & - & 1.00 & 4.67E-01 & $0.17_{-0.05}^{+0.06}$ & $1.45_{-0.27}^{+0.43}$ & $1.73_{-0.16}^{+0.24}$ & - & - & - \\ 
SS397 & 0122700201 & Q & $T_1$ & - & 0.89 & 6.88E-01 & $0.20_{-0.06}^{+0.08}$ & 1.37 & - & $10.03_{-2.09}^{+7.04}$ & - & - \\ 
SS397 & 0122700201 & Q & $N_1$ & - & 0.93 & 6.04E-01 & $0.20_{-0.06}^{+0.08}$ & 1.37 & $1.58_{-0.11}^{+0.12}$ & - & - & - \\ 
SS397 & 0122700301 & Q & $T_1$ & - & 1.03 & 4.21E-01 & $0.13_{-0.04}^{+0.06}$ & 1.37 & - & $6.92_{-1.23}^{+3.32}$ & - & - \\ 
SS397 & 0122700301 & Q & $N_1$ & - & 1.08 & 3.35E-01 & $0.14_{-0.05}^{+0.06}$ & 1.37 & $1.68_{-0.11}^{+0.13}$ & - & - & - \\ 
SS397 & 0122700501 & Q & $T_1$ & - & 1.01 & 4.49E-01 & $0.20_{-0.06}^{+0.09}$ & 1.37 & - & $28.01_{-11.74}^{+51.89}$ & - & - \\ 
SS397 & 0122700501 & Q & $N_1$ & - & 0.95 & 5.45E-01 & $0.19_{-0.07}^{+0.09}$ & 1.37 & $1.42_{-0.16}^{+0.20}$ & - & - & - \\ 
\midrule
LS I +61 303 & 0207260101 & Q & $TN_1$ & - & 1.09 & 2.25E-01 & $6.27_{-0.07}^{+0.08}$ & 0.90 & $1.98_{-0.08}^{+0.05}$ & $2.46_{-0.26}^{+1.65}$ & - & - \\ 
LS I +61 303 & 0505980801 & Q & $TN_1$ & - & 1.03 & 4.04E-01 & $5.89_{-0.12}^{+0.13}$ & 0.90 & $2.34_{-0.13}^{+0.12}$ & $1.78_{-0.15}^{+0.33}$ & - & - \\ 
LS I +61 303 & 0505980901 & Q & $TN_1$ & - & 1.19 & 8.86E-02 & $6.31_{-0.17}^{+0.20}$ & 0.90 & $2.47_{-0.24}^{+0.17}$ & $1.85_{-0.15}^{+0.34}$ & - & - \\ 
LS I +61 303 & 0505981001 & Q & $TN_1$ & - & 1.09 & 2.44E-01 & $6.00_{-0.16}^{+0.18}$ & 0.90 & $2.28_{-0.25}^{+0.16}$ & $1.79_{-0.16}^{+0.57}$ & - & - \\ 
LS I +61 303 & 0505981101 & Q & $TN_1$ & - & 0.92 & 7.24E-01 & $6.39_{-0.11}^{+0.16}$ & 0.90 & $2.08_{-0.16}^{+0.09}$ & $2.17_{-0.24}^{+2.48}$ & - & - \\ 
LS I +61 303 & 0505981201 & Q & $TN_1$ & - & 0.99 & 5.09E-01 & $11.94_{-0.18}^{+0.24}$ & 0.90 & $1.89_{-0.11}^{+0.07}$ & $2.58_{-0.31}^{+1.74}$ & - & - \\ 
LS I +61 303 & 0505981301 & Q & $TN_1$ & - & 1.06 & 3.03E-01 & $9.37_{-0.15}^{+0.18}$ & 0.90 & $2.07_{-0.12}^{+0.11}$ & $2.07_{-0.15}^{+0.34}$ & - & - \\ 
LS I +61 303 & 0505981401 & Q & $TN_1$ & - & 1.16 & 1.03E-01 & $6.22_{-0.12}^{+0.17}$ & 0.90 & $2.06_{-0.17}^{+0.11}$ & $2.12_{-0.22}^{+1.23}$ & - & - \\ 
LS 5039 & 0151160201 & Q & $T_{11}$ & - & 1.18 & 9.13E-02 & $5.00_{-0.52}^{+0.58}$ & 1.03 & - & $11.87_{-1.18}^{+2.26}$ & - & - \\ 
LS 5039 & 0151160201 & Q & $N_1$ & - & 0.97 & 5.76E-01 & $5.10_{-0.52}^{+0.56}$ & 1.03 & $1.60_{-0.04}^{+0.04}$ & - & - & - \\ 
LS 5039 & 0151160301 & Q & $T_{11}$ & - & 1.21 & 6.01E-02 & $5.07_{-0.53}^{+0.61}$ & 1.03 & - & $18.50_{-2.57}^{+4.55}$ & - & - \\ 
LS 5039 & 0151160301 & Q & $T_{10}$ & - & 1.20 & 6.97E-02 & $5.08_{-0.54}^{+0.64}$ & 1.03 & - & $2.01_{-0.16}^{+0.42}$ & $0.76_{-0.07}^{+0.16}$ & - \\ 
LS 5039 & 0151160301 & Q & $N_1$ & - & 1.11 & 1.99E-01 & $5.15_{-0.55}^{+0.59}$ & 1.03 & $1.48_{-0.04}^{+0.04}$ & - & - & - \\ 
LS 5039 & 0202950201 & Q & $T_{11}$ & - & 1.15 & 1.14E-01 & $5.93_{-0.58}^{+0.62}$ & 1.03 & - & $14.66_{-1.51}^{+1.68}$ & - & - \\ 
LS 5039 & 0202950201 & Q & $T_{10}$ & - & 1.10 & 2.09E-01 & $6.02_{-0.62}^{+0.65}$ & 1.03 & - & $2.14_{-0.16}^{+0.26}$ & $0.82_{-0.06}^{+0.08}$ & - \\ 
LS 5039 & 0202950201 & Q & $N_1$ & - & 0.93 & 6.99E-01 & $6.05_{-0.59}^{+0.63}$ & 1.03 & $1.54_{-0.03}^{+0.03}$ & - & - & - \\ 
LS 5039 & 0202950301 & Q & $T_{11}$ & - & 1.19 & 8.41E-02 & $3.50_{-0.38}^{+0.43}$ & 1.03 & - & $9.24_{-0.93}^{+1.38}$ & - & - \\ 
LS 5039 & 0202950301 & Q & $N_1$ & - & 0.92 & 7.12E-01 & $3.60_{-0.38}^{+0.41}$ & 1.03 & $1.69_{-0.04}^{+0.04}$ & - & - & - \\ 
\midrule
4U 2206+54 & 0650640101 & Q & $TN_1$ & - & 1.19 & 5.04E-02 & $294.46_{-39.89}^{+42.83}$ & $0.61_{-0.03}^{+0.03}$ & $1.01_{-0.04}^{+0.04}$ & $1.67_{-0.03}^{+0.04}$ & - & - \\ 
Cen X-3 & 0111010101 & I/E & $TN_4$ & - & 1.28 & 1.24E-02 & $2242.27_{-430.86}^{+477.52}$ & $5.96_{-0.05}^{+0.04}$ & 0.68 & 0.94 & - & - \\ 
Cen X-3 & 0111010101 & I/E & $N_2$ & - & 1.27 & 1.39E-02 & $2240.60_{-430.78}^{+477.65}$ & $6.07_{-0.09}^{+0.08}$ & 0.71 & - & - & - \\ 
Cygnus X-1 & 0202401101 & Q & $TN_4$ & - & 1.12 & 1.54E-01 & $10191.40_{-2289.25}^{+2586.02}$ & $3.85_{-0.51}^{+0.74}$ & $2.67_{-0.04}^{+0.06}$ & $0.46_{-0.01}^{+0.01}$ & - & - \\ 
Cygnus X-1 & 0202401201 & Q & $TN_3$ & - & 1.13 & 1.39E-01 & $8572.45_{-1929.60}^{+2176.95}$ & $0.77_{-0.01}^{+0.02}$ & $2.61_{-0.03}^{+0.05}$ & $0.43_{-0.01}^{+0.00}$ & - & - \\ 
Cygnus X-1 & 0202760201 & Q & $N_1$ & - & 1.49 & 2.94E-03 & $5061.89_{-1137.75}^{+1284.05}$ & 0.86 & $2.29_{-0.01}^{+0.03}$ & - & - & - \\ 
Cygnus X-1 & 0202760301 & Q & $N_1$ & - & 0.96 & 5.82E-01 & $3453.30_{-777.01}^{+877.54}$ & 0.86 & $2.07_{-0.01}^{+0.01}$ & - & - & - \\ 
Cygnus X-1 & 0202760401 & Q & $N_1$ & - & 1.72 & 6.18E-05 & $5009.50_{-1126.79}^{+1269.64}$ & 0.86 & $2.10_{-0.01}^{+0.02}$ & - & - & - \\ 
Cygnus X-1 & 0202760501 & Q & $N_1$ & - & 1.17 & 1.38E-01 & $4842.41_{-1091.06}^{+1232.95}$ & 0.86 & $2.18_{-0.02}^{+0.02}$ & - & - & - \\ 
Cygnus X-1 & 0500880201 & D & $TN_4$ & - & 1.19 & 6.15E-02 & $1899.40_{-430.07}^{+494.23}$ & $4.53_{-0.34}^{+0.26}$ & $1.43_{-0.03}^{+0.01}$ & $0.39_{-0.02}^{+0.02}$ & - & - \\ 
Cygnus X-1 & 0500880201 & D & $N_2$ & - & 1.19 & 5.54E-02 & $1899.66_{-434.10}^{+490.38}$ & $5.04_{-0.45}^{+0.35}$ & $1.42_{-0.03}^{+0.04}$ & - & - & - \\ 
Cygnus X-1 & 0500880201 & Q & $N_1$ & - & 1.46 & 3.73E-04 & $2679.69_{-600.04}^{+676.00}$ & 0.71 & 1.55 & 4.63 & - & - \\ 
Cygnus X-1 & 0610000401 & Q & $TN_4$ & - & 1.13 & 1.43E-01 & $3840.30_{-868.00}^{+977.45}$ & $6.18_{-0.07}^{+1.11}$ & $1.60_{-0.01}^{+0.03}$ & $0.57_{-0.01}^{+0.02}$ & - & - \\ 
AX J1749.1-2733 & 0510010401 & Q & $T_5$ & - & 0.92 & 6.80E-01 & $290.03_{-119.53}^{+156.80}$ & $22.81_{-2.25}^{+2.70}$ & - & $2.26_{-0.16}^{+0.18}$ & - & - \\ 
AX J1749.1-2733 & 0510010401 & Q & $N_1$ & - & 0.95 & 6.06E-01 & $293.29_{-121.29}^{+156.74}$ & $31.98_{-2.53}^{+3.03}$ & $1.32_{-0.16}^{+0.18}$ & - & - & - \\ 
\bottomrule
\label{tab:cont} 
\end{longtable}
\end{landscape}

\newpage
\topmargin=0.5cm
\begin{landscape}
\tabcolsep=0.11cm
\begin{longtable}{cccc|ccccc|ccccc}
\caption{FeK$\alpha$ parameters.} \\
\toprule
\multicolumn{3}{c}{ } & \multicolumn{5}{c}{Narrow FeK$\alpha$} & \multicolumn{5}{c}{Broad Fe feature} \\ 
Source & Obs ID & State & $P_0$ & Energy & Width & $L_{FeK\alpha}$ & EW & Signif. & Energy & Width & $L_{FeK\alpha}$ & EW & Signif. \\ 
 &  &  &  & (keV) & (eV) & ($10^{41} \, \gamma$/s) & (keV) & ($\sigma$) & (keV) & (eV) & ($10^{41} \, \gamma$/s) & (keV) & ($\sigma$) \\  
\endfirsthead
\caption{FeK$\alpha$ parameters (continued).} \label{tab:FeKa} \\
\endhead
\\
\caption{Fit parameters of FeK$\alpha$ and broad Fe feature detected. We present the centroid energy (keV), the width (eV), the luminosity (photons/s), EW (keV) and the statistical significance of the Gaussian lines. In the cases where we don't detect FeK$\alpha$ ($\sigma_{sign}<2$), we give the upper limits of EW and the luminosity. Parameters frozen or unbounded are included without an error estimation. Therefore they are not used in the plots and the subsequent discussion. The possible states are: quiescence(Q), flare(F), eclipse ingress/egress(I/E), eclipse(E) and dip(D). } \\
\endlastfoot
\midrule	
1A 0535+26 & 0674180101 & Q & 8.70E-01 & - & - & $<6.92E-02$ & $<0.06$ & - & - & - & - & - & - \\ 
2S 1845-024 & 0302970801 & Q & 7.67E-01 & - & - & $<1.22E+00$ & $<0.60$ & - & - & - & - & - & - \\ 
X Persei & 0151380101 & Q & 3.68E-03 & - & - & $<2.11E-01$ & $<0.03$ & - & - & - & - & - & - \\ 
X Persei & 0600980101 & Q & 1.06E-02 & - & - & $<1.76E-01$ & $<0.02$ & - & - & - & - & - & - \\ 
AX J1820.5-1434 & 0511010101 & Q & 9.92E-01 & - & - & $<4.03E-01$ & $<0.21$ & - & - & - & - & - & - \\ 
RX J0146.9+6121 & 0201160101 & Q & 8.58E-03 & - & - & $<4.25E-02$ & $<0.06$ & - & - & - & - & - & - \\ 
RX J0440.9+4431 & 0653660101 & Q & 8.55E-02 & - & - & $<4.73E-01$ & $<0.06$ & - & - & - & - & - & - \\ 
RX J1037.5-5647 & 0550560101 & Q & 4.16E-01 & - & - & $<1.72E-01$ & $<0.13$ & - & - & - & - & - & - \\ 
SAX J2103.5+4545 & 0149550401 & Q & 1.77E-02 & $6.42_{-0.04}^{+0.06}$ & $54.00$ & $3.87E+00_{-2.59E+00}^{+3.80E+00}$ & $0.04_{-0.01}^{+0.01}$ & 3.27 & $-$ & $-$ & $-$ & $-$ & - \\ 
Swift J045106.8-694803 & 0679381401 & Q & 3.28E-01 & - & - & $<1.80E+01$ & $<0.20$ & - & - & - & - & - & - \\ 
V0332+53 & 0506190101 & Q & 7.58E-01 & - & - & $<1.14E-01$ & $<2223.78$ & - & - & - & - & - & - \\ 
\midrule
4U 1538-522 & 0152780201 & EC & 7.12E-03 & $6.38_{-0.01}^{+0.02}$ & $0.00$ & $9.58E-01_{-3.93E-01}^{+5.48E-01}$ & $0.18_{-0.04}^{+0.04}$ & 8.70 & $6.54_{-0.02}^{+0.02}$ & $271.65_{-18.92}^{+20.73}$ & $3.15E+00_{-1.10E+00}^{+1.41E+00}$ & $0.92_{-0.13}^{+0.14}$ & 19.95 \\ 
4U 1538-522 & 0152780201 & I/E & 5.65E-04 & $6.40_{-0.03}^{+0.01}$ & $24.24$ & $3.42E+00_{-1.58E+00}^{+4.16E+00}$ & $0.03_{-0.01}^{+0.02}$ & 6.46 & $-$ & $-$ & $-$ & $-$ & - \\ 
4U 1700-37 & 0083280101 & D & 9.69E-03 & - & - & $<6.03E+00$ & $<0.27$ & - & - & - & - & - & - \\ 
4U 1700-37 & 0083280101 & F & 5.73E-01 & $6.48_{-0.16}^{+0.06}$ & $0.02$ & $6.86E+00_{-5.30E+00}^{+4.54E+00}$ & $0.03_{-0.02}^{+0.01}$ & 3.56 & $-$ & $-$ & $-$ & $-$ & - \\ 
4U 1700-37 & 0083280101 & Q & 1.04E-01 & - & - & $<1.23E+01$ & $<0.13$ & - & - & - & - & - & - \\ 
4U 1700-37 & 0083280201 & F & 1.23E-02 & $6.45_{-0.01}^{+0.01}$ & $90.60_{-33.19}^{+11.35}$ & $2.86E+01_{-1.31E+01}^{+1.13E+01}$ & $0.11_{-0.03}^{+0.00}$ & 19.07 & $-$ & $-$ & $-$ & $-$ & - \\ 
4U 1700-37 & 0083280201 & Q & 7.98E-01 & $6.44_{-0.04}^{+0.04}$ & $125.45_{-49.29}^{+48.71}$ & $9.07E+00_{-4.56E+00}^{+5.72E+00}$ & $0.11_{-0.03}^{+0.02}$ & 7.25 & $-$ & $-$ & $-$ & $-$ & - \\ 
4U 1700-37 & 0083280301 & Q & 1.20E-01 & $6.45_{-0.03}^{+0.03}$ & $87.78_{-46.84}^{+72.25}$ & $8.08E+00_{-3.82E+00}^{+6.26E+00}$ & $0.07_{-0.02}^{+0.02}$ & 8.15 & $-$ & $-$ & $-$ & $-$ & - \\ 
4U 1700-37 & 0083280401 & EC & 1.30E-01 & $6.49_{-0.02}^{+0.02}$ & $112.97_{-25.76}^{+23.82}$ & $3.54E+00_{-1.45E+00}^{+1.80E+00}$ & $1.25_{-0.28}^{+0.27}$ & 12.50 & $-$ & $-$ & $-$ & $-$ & - \\ 
4U 1700-37 & 0083280401 & I/E & 2.01E-01 & $6.48_{-0.03}^{+0.04}$ & $0.00$ & $7.28E+00_{-3.67E+00}^{+3.73E+00}$ & $0.13_{-0.04}^{+0.01}$ & 6.82 & $-$ & $-$ & $-$ & $-$ & - \\ 
4U 1700-37 & 0600950101 & EC & 6.97E-02 & $6.40_{-0.00}^{+0.00}$ & $38.57_{-4.67}^{+4.33}$ & $2.47E+00_{-7.77E-01}^{+9.42E-01}$ & $0.92_{-0.04}^{+0.05}$ & $>25.5$ & $6.54_{-0.03}^{+0.04}$ & $470.98_{-35.92}^{+47.86}$ & $1.07E+00_{-3.69E-01}^{+4.84E-01}$ & $0.19_{-0.01}^{+0.02}$ & 21.33 \\ 
4U 1907+09 & 0555410101 & F & 1.89E-01 & $6.43_{-0.02}^{+0.03}$ & $34.05$ & $6.04E+00_{-1.59E+00}^{+1.10E+00}$ & $0.04_{-0.01}^{+0.01}$ & 8.55 & $-$ & $-$ & $-$ & $-$ & - \\ 
4U 1907+09 & 0555410101 & Q & 5.16E-01 & $6.43_{-0.02}^{+0.02}$ & $56.22_{-46.41}^{+27.09}$ & $5.29E+00_{-8.56E-01}^{+7.56E-01}$ & $0.05_{-0.01}^{+0.01}$ & 11.60 & $-$ & $-$ & $-$ & $-$ & - \\ 
EXO1722-363 & 0206380401 & Q & 9.84E-03 & $6.40_{-0.01}^{+0.01}$ & $46.25$ & $1.80E+01_{-1.09E+01}^{+1.67E+01}$ & $0.15_{-0.03}^{+0.02}$ & 13.19 & $-$ & $-$ & $-$ & $-$ & - \\ 
EXO1722-363 & 0405640201 & EC & 7.47E-01 & $6.41_{-0.03}^{+0.05}$ & $124.04_{-39.53}^{+55.59}$ & $1.02E+00_{-6.50E-01}^{+1.23E+00}$ & $2.49_{-0.59}^{+0.71}$ & 7.92 & $-$ & $-$ & $-$ & $-$ & - \\ 
EXO1722-363 & 0405640301 & Q & 7.42E-01 & $6.42_{-0.02}^{+0.02}$ & $54.05$ & $1.36E+01_{-8.66E+00}^{+1.44E+01}$ & $0.14_{-0.03}^{+0.03}$ & 8.99 & $-$ & $-$ & $-$ & $-$ & - \\ 
EXO1722-363 & 0405640401 & Q & 2.69E-01 & $6.41_{-0.01}^{+0.01}$ & $45.19_{-43.01}^{+20.89}$ & $1.46E+01_{-8.60E+00}^{+1.33E+01}$ & $0.23_{-0.04}^{+0.03}$ & 14.26 & $-$ & $-$ & $-$ & $-$ & - \\ 
EXO1722-363 & 0405640901 & Q & 1.50E-01 & $6.42_{-0.48}^{+0.72}$ & $17.75$ & $2.99E+00_{-1.81E+00}^{+7.11E+01}$ & $0.05_{-0.01}^{+0.66}$ & 4.02 & $-$ & $-$ & $-$ & $-$ & - \\ 
EXO1722-363 & 0405640801 & Q & 1.44E-01 & $6.43_{-0.01}^{+0.01}$ & $50.70_{-37.21}^{+23.37}$ & $1.57E+01_{-9.36E+00}^{+1.49E+01}$ & $0.17_{-0.03}^{+0.03}$ & 14.20 & $-$ & $-$ & $-$ & $-$ & - \\ 
GX 301-2 & 0555200401 & F & 1.21E-02 & $6.37_{-0.00}^{+0.00}$ & $25.43_{-6.61}^{+1.57}$ & $2.79E+02_{-7.95E+00}^{+1.05E+00}$ & $0.46_{-0.02}^{+0.00}$ & $>25.5$ & $-$ & $-$ & $-$ & $-$ & - \\ 
GX 301-2 & 0555200401 & Q & 9.70E-09 & $6.38_{-0.01}^{+0.01}$ & $0.14$ & $1.41E+02_{-3.54E+00}^{+4.70E+00}$ & $0.44_{-0.02}^{+0.03}$ & $>25.5$ & $-$ & $-$ & $-$ & $-$ & - \\ 
IGR J16207-5129 & 0402920201 & Q & 8.16E-02 & $6.42_{-0.05}^{+0.06}$ & $0.00$ & $4.19E-01_{-1.77E-01}^{+3.09E-01}$ & $0.03_{-0.01}^{+0.02}$ & 2.71 & $-$ & $-$ & $-$ & $-$ & - \\ 
IGR J16318-4848 & 0154750401 & Q & 2.31E-01 & $6.40_{-0.00}^{+0.00}$ & $35.48_{-7.68}^{+6.79}$ & $2.55E+02_{-2.39E+02}^{+8.46E+02}$ & $2.47_{-0.21}^{+0.25}$ & $>25.5$ & $-$ & $-$ & $-$ & $-$ & - \\ 
IGR J16318-4848 & 0201000201 & Q & 7.12E-01 & $6.41_{-0.01}^{+0.01}$ & $26.59$ & $1.05E+02_{-9.88E+01}^{+3.07E+02}$ & $1.12_{-0.13}^{+0.13}$ & $>25.5$ & $-$ & $-$ & $-$ & $-$ & - \\ 
IGR J16318-4848 & 0201000301 & Q & 6.83E-01 & $6.42_{-0.01}^{+0.02}$ & $32.52$ & $3.96E+02_{-3.88E+02}^{+2.58E+03}$ & $2.28_{-0.68}^{+1.47}$ & 10.97 & $-$ & $-$ & $-$ & $-$ & - \\ 
IGR J16318-4848 & 0201000401 & Q & 2.57E-01 & $6.41_{-0.01}^{+0.01}$ & $27.11$ & $4.56E+01_{-4.31E+01}^{+1.51E+02}$ & $1.15_{-0.16}^{+0.15}$ & 22.68 & $-$ & $-$ & $-$ & $-$ & - \\ 
IGR J16320-4751 & 0128531101 & Q & 4.94E-01 & - & - & $<3.78E+00$ & $<0.28$ & - & - & - & - & - & - \\ 
IGR J16320-4751 & 0201700301 & F & 8.13E-01 & $6.41_{-0.01}^{+0.02}$ & $31.55$ & $1.92E+01_{-1.17E+01}^{+1.84E+01}$ & $0.13_{-0.02}^{+0.01}$ & 5.38 & $-$ & $-$ & $-$ & $-$ & - \\ 
IGR J16320-4751 & 0201700301 & Q & 6.96E-01 & $6.42_{-0.01}^{+0.00}$ & $0.00$ & $7.80E+00_{-4.60E+00}^{+7.17E+00}$ & $0.11_{-0.01}^{+0.01}$ & 9.58 & $-$ & $-$ & $-$ & $-$ & - \\ 
IGR J16320-4751 & 0556140101 & Q & 5.15E-01 & $6.44_{-0.07}^{+0.03}$ & $0.00$ & $1.83E+01_{-1.67E+01}^{+1.62E+01}$ & $0.22_{-0.19}^{+0.01}$ & 15.57 & $-$ & $-$ & $-$ & $-$ & - \\ 
IGR J16320-4751 & 0556140201 & Q & 3.47E-01 & $6.42_{-0.01}^{+0.01}$ & $26.65$ & $2.92E+01_{-1.80E+01}^{+2.56E+01}$ & $0.19_{-0.03}^{+0.01}$ & 6.60 & $-$ & $-$ & $-$ & $-$ & - \\ 
IGR J16320-4751 & 0556140301 & Q & 2.28E-01 & $6.42_{-0.01}^{+0.01}$ & $0.00$ & $2.23E+01_{-1.39E+01}^{+2.17E+01}$ & $0.15_{-0.03}^{+0.02}$ & 5.52 & $-$ & $-$ & $-$ & $-$ & - \\ 
IGR J16320-4751 & 0556140401 & Q & 6.25E-02 & $6.43_{-0.02}^{+0.01}$ & $68.31_{-23.31}^{+19.44}$ & $1.64E+01_{-9.98E+00}^{+1.60E+01}$ & $0.17_{-0.02}^{+0.02}$ & 17.17 & $-$ & $-$ & $-$ & $-$ & - \\ 
IGR J16320-4751 & 0556140501 & Q & 8.57E-01 & $6.41_{-0.02}^{+0.02}$ & $0.00$ & $1.13E+01_{-7.30E+00}^{+1.27E+01}$ & $0.15_{-0.03}^{+0.03}$ & 7.71 & $-$ & $-$ & $-$ & $-$ & - \\ 
IGR J16320-4751 & 0556140601 & Q & 3.73E-01 & $6.42_{-0.01}^{+0.01}$ & $34.48_{-31.19}^{+11.53}$ & $3.72E+01_{-2.29E+01}^{+3.25E+01}$ & $0.20_{-0.04}^{+0.01}$ & 15.70 & $-$ & $-$ & $-$ & $-$ & - \\ 
IGR J16320-4751 & 0556140701 & Q & 5.34E-02 & $6.42_{-0.01}^{+0.01}$ & $45.43_{-29.65}^{+8.55}$ & $2.83E+01_{-1.77E+01}^{+2.46E+01}$ & $0.41_{-0.09}^{+0.04}$ & 10.46 & $-$ & $-$ & $-$ & $-$ & - \\ 
IGR J16320-4751 & 0556140801 & Q & 1.31E-01 & $6.42_{-0.01}^{+0.01}$ & $52.95_{-29.37}^{+20.85}$ & $1.67E+01_{-1.01E+01}^{+1.61E+01}$ & $0.19_{-0.02}^{+0.02}$ & 19.00 & $-$ & $-$ & $-$ & $-$ & - \\ 
IGR J16320-4751 & 0556141001 & Q & 7.05E-01 & $6.42_{-0.01}^{+0.01}$ & $21.72$ & $1.72E+01_{-1.05E+01}^{+1.82E+01}$ & $0.15_{-0.02}^{+0.03}$ & 8.66 & $-$ & $-$ & $-$ & $-$ & - \\ 
IGR J16465-4507 & 0164561001 & Q & 9.23E-01 & - & - & $<5.27E+00$ & $<0.50$ & - & - & - & - & - & - \\ 
SAX J1802.7-2017 & 0206380601 & Q & 3.18E-02 & - & - & $<1.36E+01$ & $<0.14$ & - & - & - & - & - & - \\ 
Vela X-1 & 0111030101 & F & 4.24E-03 & $6.45_{-0.02}^{+0.02}$ & $55.91$ & $7.27E+00_{-2.56E+00}^{+3.78E+00}$ & $0.07_{-0.01}^{+0.02}$ & 9.75 & $-$ & $-$ & $-$ & $-$ & - \\ 
Vela X-1 & 0111030101 & Q & 8.65E-01 & $6.45_{-0.01}^{+0.01}$ & $45.04_{-25.41}^{+18.62}$ & $4.00E+00_{-1.15E+00}^{+1.50E+00}$ & $0.06_{-0.01}^{+0.01}$ & 19.86 & $6.62_{-0.11}^{+0.08}$ & $1220.66_{-76.84}^{+115.79}$ & $4.29E+01_{-1.24E+01}^{+1.91E+01}$ & $0.75_{-0.09}^{+0.15}$ & $>25.5$ \\ 
XTE J0421+560 & 0139760101 & Q & 9.51E-01 & $6.43_{-0.03}^{+0.03}$ & $0.00$ & $2.78E-01_{-2.69E-01}^{+1.61E+00}$ & $0.15_{-0.04}^{+0.21}$ & 2.65 & $6.56_{-2.68}^{+0.21}$ & $402.38_{-58.40}^{+4955.35}$ & $7.08E-01_{-6.86E-01}^{+2.67E+01}$ & $0.43_{-0.13}^{+4.28}$ & 3.51 \\ 
\midrule
AXJ1841.0-0536 & 0604820301 & F & 1.54E-03 & $6.46_{-0.06}^{+0.07}$ & $0.00$ & $1.67E+00_{-1.19E+00}^{+1.18E+00}$ & $0.02_{-0.01}^{+0.01}$ & 2.78 & $-$ & $-$ & $-$ & $-$ & - \\ 
IGR J00370+6122 & 0501450101 & Q & 9.23E-01 & - & - & $<2.97E-01$ & $<0.06$ & - & - & - & - & - & - \\ 
IGR J11215-5952 & 0405181901 & F & 3.64E-01 & $6.48_{-0.05}^{+0.04}$ & $62.01_{-45.59}^{+52.30}$ & $4.16E+00_{-1.45E+00}^{+1.38E+00}$ & $0.05_{-0.02}^{+0.02}$ & 3.49 & $-$ & $-$ & $-$ & $-$ & - \\ 
IGR J11215-5952 & 0405181901 & Q & 5.39E-01 & - & - & $<1.16E+00$ & $<0.13$ & - & - & - & - & - & - \\ 
IGR J16328-4726 & 0654190201 & Q & 3.77E-01 & - & - & $<4.94E-01$ & $<0.09$ & - & - & - & - & - & - \\ 
IGR J16418-4532 & 0206380301 & Q & 7.30E-01 & - & - & $<1.34E+00$ & $<0.07$ & - & - & - & - & - & - \\ 
IGR J16418-4532 & 0405180501 & Q & 1.99E-01 & - & - & $<6.39E-01$ & $<0.05$ & - & - & - & - & - & - \\ 
IGRJ16479-4514 & 0512180101 & EC & 5.92E-01 & $6.42_{-0.00}^{+0.02}$ & $0.00$ & $4.89E-01_{-3.34E-01}^{+5.75E-01}$ & $0.45_{-0.14}^{+0.11}$ & 6.36 & $-$ & $-$ & $-$ & $-$ & - \\ 
IGRJ16479-4514 & 0512180101 & I/E & 9.67E-01 & $6.40$ & $0.00$ & $1.72E+00$ & $0.03$ & 1.48 & $-$ & $-$ & $-$ & $-$ & - \\ 
XTE J1739-302 & 0554720101 & F & 5.66E-02 & - & - & $<3.81E-02$ & $<0.21$ & - & - & - & - & - & - \\ 
XTE J1739-302 & 0554720101 & Q & 2.51E-01 & - & - & $<1.83E-02$ & $<2.27$ & - & - & - & - & - & - \\ 
XTE J1739-302 & 0561580101 & F & 6.04E-01 & - & - & $<9.58E-02$ & $<0.12$ & - & - & - & - & - & - \\ 
XTE J1739-302 & 0561580101 & Q & 6.93E-02 & - & - & $<2.35E-02$ & $<0.17$ & - & - & - & - & - & - \\ 
IGR J17544-2619 & 0148090501 & F & 7.16E-01 & - & - & $<2.39E-01$ & $<0.12$ & - & - & - & - & - & - \\ 
IGR J18450-0435 & 0306170401 & F & 3.51E-01 & - & - & $<2.24E+00$ & $<0.13$ & - & - & - & - & - & - \\ 
IGR J18450-0435 & 0306170401 & Q & 3.39E-01 & - & - & $<4.27E-01$ & $<0.14$ & - & - & - & - & - & - \\ 
IGR J18483-0311 & 0406140201 & Q & 1.98E-01 & - & - & $<4.23E-02$ & $<0.27$ & - & - & - & - & - & - \\ 
\midrule
$\gamma$ Cassiopeiae & 0201220101 & Q & 1.42E-03 & $6.43_{-0.01}^{+0.01}$ & $0.00$ & $6.86E-04_{-1.95E-04}^{+2.62E-04}$ & $0.03_{-0.00}^{+0.00}$ & 13.80 & $-$ & $-$ & $-$ & $-$ & - \\ 
$\gamma$ Cassiopeiae & 0651670201 & Q & 3.62E-02 & $6.39_{-0.02}^{+0.03}$ & $23.94$ & $1.08E-03_{-4.24E-04}^{+5.03E-04}$ & $0.04_{-0.01}^{+0.01}$ & 7.82 & $-$ & $-$ & $-$ & $-$ & - \\ 
$\gamma$ Cassiopeiae & 0651670301 & Q & 4.40E-04 & $6.44_{-0.03}^{+0.26}$ & $6.39$ & $1.00E-03_{-3.77E-04}^{+2.86E-03}$ & $0.04_{-0.01}^{+0.11}$ & 7.41 & $-$ & $-$ & $-$ & $-$ & - \\ 
$\gamma$ Cassiopeiae & 0651670401 & Q & 1.09E-01 & $6.41_{-0.02}^{+0.02}$ & $1.50$ & $1.53E-03_{-4.17E-04}^{+6.94E-04}$ & $0.05_{-0.00}^{+0.01}$ & 10.26 & $-$ & $-$ & $-$ & $-$ & - \\ 
$\gamma$ Cassiopeiae & 0651670501 & Q & 6.73E-02 & $6.39_{-0.03}^{+0.02}$ & $0.00$ & $8.55E-04_{-3.18E-04}^{+3.70E-04}$ & $0.04_{-0.01}^{+0.01}$ & 7.91 & $-$ & $-$ & $-$ & $-$ & - \\ 
HD 110432 & 0504730101 & Q & 2.81E-02 & $6.44_{-0.04}^{+0.03}$ & $0.14$ & $1.21E-03_{-5.97E-04}^{+7.92E-04}$ & $0.05_{-0.02}^{+0.01}$ & 5.36 & $-$ & $-$ & $-$ & $-$ & - \\ 
HD 119682 & 0551000201 & Q & 2.06E-01 & - & - & $<7.53E-03$ & $<0.63$ & - & - & - & - & - & - \\ 
HD 157832 & 0551020101 & Q & 3.12E-01 & $6.40$ & $0.00$ & $1.32E-03_{-9.18E-04}^{+8.17E-04}$ & $0.10_{-0.07}^{+0.06}$ & 2.47 & $-$ & $-$ & $-$ & $-$ & - \\ 
HD 161103 & 0201200101 & Q & 9.76E-02 & $6.40$ & $0.00$ & $4.68E-03_{-4.19E-03}^{+9.96E-03}$ & $0.09_{-0.07}^{+0.07}$ & 2.18 & $-$ & $-$ & $-$ & $-$ & - \\ 
HD 45314 & 0670080301 & Q & 5.63E-01 & $6.39_{-0.65}^{+2.77}$ & $0.00$ & $5.45E-03$ & $0.17$ & 1.98 & $-$ & $-$ & $-$ & $-$ & - \\ 
SAO 49725 & 0201200201 & Q & 7.39E-02 & - & - & $<3.26E-02$ & $<0.71$ & - & - & - & - & - & - \\ 
SS397 & 0122700101 & Q & 7.90E-01 & - & - & $<1.00E-02$ & $<0.42$ & - & - & - & - & - & - \\ 
SS397 & 0122700201 & Q & 6.88E-01 & - & - & $<1.00E-02$ & $<0.43$ & - & - & - & - & - & - \\ 
SS397 & 0122700301 & Q & 4.21E-01 & - & - & $<1.10E-02$ & $<0.57$ & - & - & - & - & - & - \\ 
SS397 & 0122700501 & Q & 4.49E-01 & - & - & $<2.69E-02$ & $<1.31$ & - & - & - & - & - & - \\ 
\midrule
LS I +61 303 & 0207260101 & Q & 2.25E-01 & - & - & $<2.16E-02$ & $<0.04$ & - & - & - & - & - & - \\ 
LS I +61 303 & 0505980801 & Q & 4.04E-01 & - & - & $<5.80E-02$ & $<0.12$ & - & - & - & - & - & - \\ 
LS I +61 303 & 0505980901 & Q & 8.86E-02 & - & - & $<8.59E-02$ & $<0.14$ & - & - & - & - & - & - \\ 
LS I +61 303 & 0505981001 & Q & 2.44E-01 & - & - & $<9.20E-02$ & $<0.17$ & - & - & - & - & - & - \\ 
LS I +61 303 & 0505981101 & Q & 7.24E-01 & - & - & $<4.72E-02$ & $<0.08$ & - & - & - & - & - & - \\ 
LS I +61 303 & 0505981201 & Q & 5.09E-01 & - & - & $<1.16E-01$ & $<0.10$ & - & - & - & - & - & - \\ 
LS I +61 303 & 0505981301 & Q & 3.03E-01 & - & - & $<5.54E-02$ & $<0.06$ & - & - & - & - & - & - \\ 
LS I +61 303 & 0505981401 & Q & 1.03E-01 & - & - & $<6.98E-02$ & $<0.12$ & - & - & - & - & - & - \\ 
LS 5039 & 0151160201 & Q & 9.13E-02 & - & - & $<4.72E-02$ & $<0.10$ & - & - & - & - & - & - \\ 
LS 5039 & 0151160301 & Q & 6.01E-02 & - & - & $<1.06E-01$ & $<0.22$ & - & - & - & - & - & - \\ 
LS 5039 & 0202950201 & Q & 1.14E-01 & - & - & $<4.72E-02$ & $<0.08$ & - & - & - & - & - & - \\ 
LS 5039 & 0202950301 & Q & 8.41E-02 & - & - & $<8.39E-02$ & $<0.28$ & - & - & - & - & - & - \\ 
\midrule
4U 2206+54 & 0650640101 & Q & 5.04E-02 & - & - & $<1.21E+00$ & $<0.03$ & - & - & - & - & - & - \\ 
Cen X-3 & 0111010101 & I/E & 1.24E-02 & $6.41_{-0.01}^{+0.01}$ & $42.06_{-13.71}^{+17.30}$ & $5.77E+01_{-1.45E+01}^{+2.06E+01}$ & $0.14_{-0.01}^{+0.02}$ & 16.08 & $-$ & $-$ & $-$ & $-$ & - \\ 
Cygnus X-1 & 0202401101 & Q & 1.54E-01 & $6.59_{-0.04}^{+0.05}$ & $0.00$ & $6.22E+00_{-3.50E+00}^{+4.51E+00}$ & $0.01_{-0.01}^{+0.01}$ & 3.46 & $6.42_{-0.09}^{+0.06}$ & $873.82_{-110.40}^{+100.30}$ & $1.66E+02_{-4.86E+01}^{+8.33E+01}$ & $0.44_{-0.06}^{+0.09}$ & 18.93 \\ 
Cygnus X-1 & 0202401201 & Q & 1.39E-01 & $-$ & $-$ & $-$ & $-$ & - & $6.50_{-0.09}^{+0.07}$ & $944.58_{-84.66}^{+156.75}$ & $1.31E+02_{-4.24E+01}^{+7.54E+01}$ & $0.46_{-0.06}^{+0.13}$ & 19.49 \\ 
Cygnus X-1 & 0202760201 & Q & 2.94E-03 & $6.55_{-0.02}^{+0.01}$ & $0.00$ & $1.07E+01_{-3.49E+00}^{+4.57E+00}$ & $0.01_{-0.00}^{+0.00}$ & 5.82 & $6.55_{-0.03}^{+0.02}$ & $1101.52_{-35.58}^{+54.39}$ & $4.82E+02_{-1.26E+02}^{+1.80E+02}$ & $0.65_{-0.04}^{+0.07}$ & 5.82 \\ 
Cygnus X-1 & 0202760301 & Q & 5.82E-01 & $6.54_{-0.02}^{+0.02}$ & $0.00$ & $5.54E+00_{-2.11E+00}^{+2.72E+00}$ & $0.01_{-0.00}^{+0.00}$ & 7.72 & $6.41_{-0.04}^{+0.04}$ & $1243.17_{-52.76}^{+66.00}$ & $3.29E+02_{-9.09E+01}^{+1.19E+02}$ & $0.61_{-0.04}^{+0.05}$ & $>25.5$ \\ 
Cygnus X-1 & 0202760401 & Q & 6.18E-05 & $6.48_{-0.01}^{+0.03}$ & $0.00$ & $6.01E+00_{-2.43E+00}^{+3.04E+00}$ & $0.01_{-0.00}^{+0.00}$ & 7.47 & $6.47_{-0.03}^{+0.03}$ & $1199.67_{-37.68}^{+57.79}$ & $4.64E+02_{-1.18E+02}^{+1.65E+02}$ & $0.61_{-0.02}^{+0.06}$ & 4.02 \\ 
Cygnus X-1 & 0202760501 & Q & 1.38E-01 & $6.53_{-0.02}^{+0.02}$ & $0.00$ & $1.12E+01_{-3.99E+00}^{+5.53E+00}$ & $0.01_{-0.00}^{+0.00}$ & 8.97 & $6.46_{-0.06}^{+0.03}$ & $1297.32_{-52.59}^{+91.10}$ & $6.03E+02_{-1.67E+02}^{+2.45E+02}$ & $0.83_{-0.06}^{+0.11}$ & $>25.5$ \\ 
Cygnus X-1 & 0500880201 & D & 6.15E-02 & $-$ & $-$ & $-$ & $-$ & - & $6.36_{-0.22}^{+0.21}$ & $595.01_{-181.09}^{+365.26}$ & $2.13E+01_{-1.10E+01}^{+1.59E+01}$ & $0.09_{-0.03}^{+0.03}$ & 4.99 \\ 
Cygnus X-1 & 0500880201 & Q & 3.73E-04 & $6.41$ & $0.00$ & $4.94E+00_{-1.11E+00}^{+1.25E+00}$ & $0.02$ & 6.95 & $-$ & $-$ & $-$ & $-$ & - \\ 
Cygnus X-1 & 0610000401 & Q & 1.43E-01 & $6.46_{-0.09}^{+0.03}$ & $82.68_{-34.87}^{+142.20}$ & $7.05E+00_{-2.98E+00}^{+6.63E+00}$ & $0.02_{-0.01}^{+0.01}$ & 5.09 & $-$ & $-$ & $-$ & $-$ & - \\ 
AX J1749.1-2733 & 0510010401 & Q & 6.80E-01 & - & - & $<6.11E+00$ & $<0.12$ & - & - & - & - & - & - \\ 
\bottomrule
\end{longtable}
\end{landscape}

\newpage
\section{Spectral atlas \label{Atlas}}

\begin{figure}[ht]
\centering
\subsection*{\Large \textbf{BeXB}}
\includegraphics[angle=-90, width=0.237\textwidth]{./Figures/plot0.ps}
\includegraphics[angle=-90, width=0.23\textwidth]{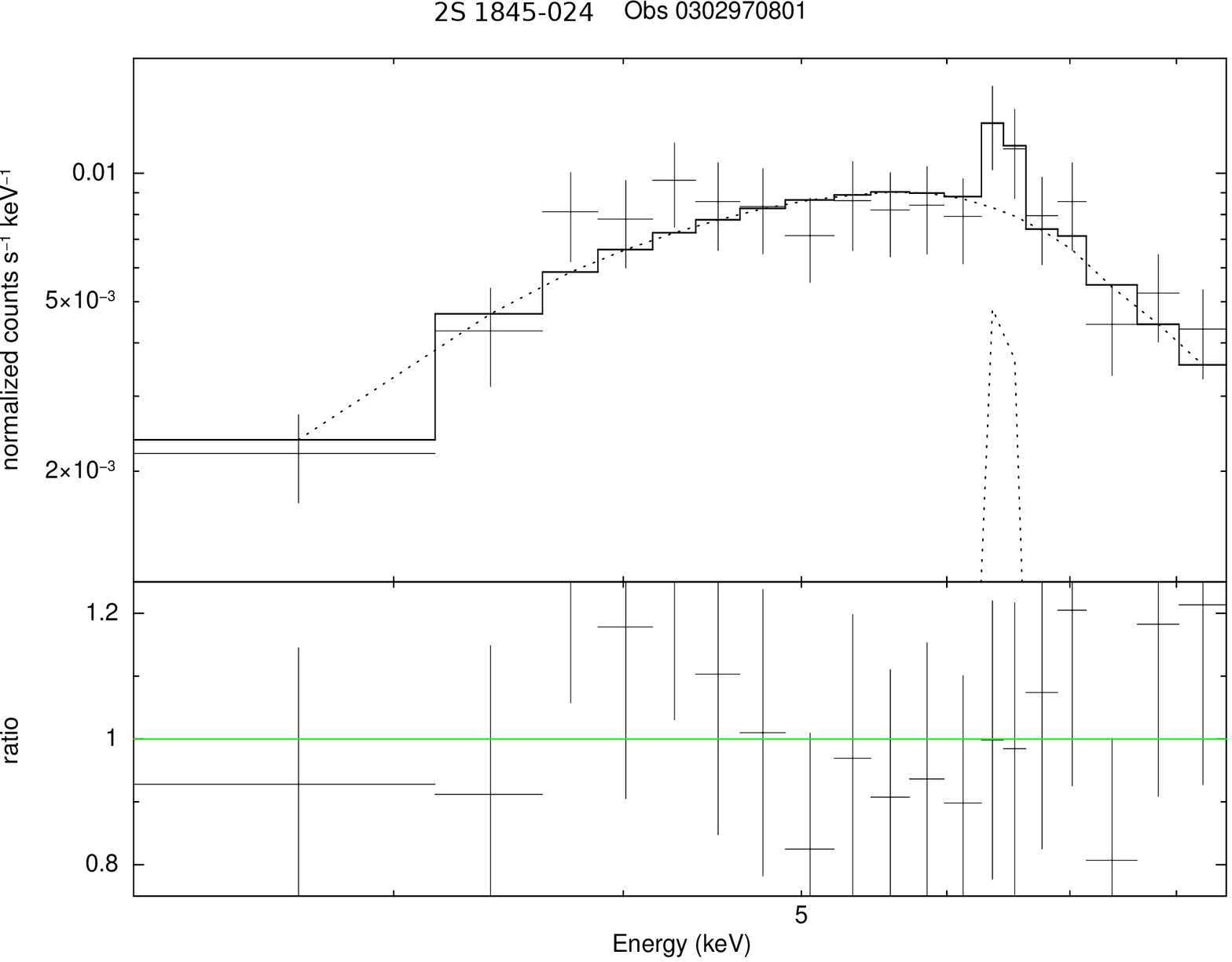}
\includegraphics[angle=-90, width=0.237\textwidth]{./Figures/plot2.ps}
\includegraphics[angle=-90, width=0.237\textwidth]{./Figures/plot3.ps}
\medskip
\includegraphics[angle=-90, width=0.237\textwidth]{./Figures/plot4.ps}
\includegraphics[angle=-90, width=0.21\textwidth]{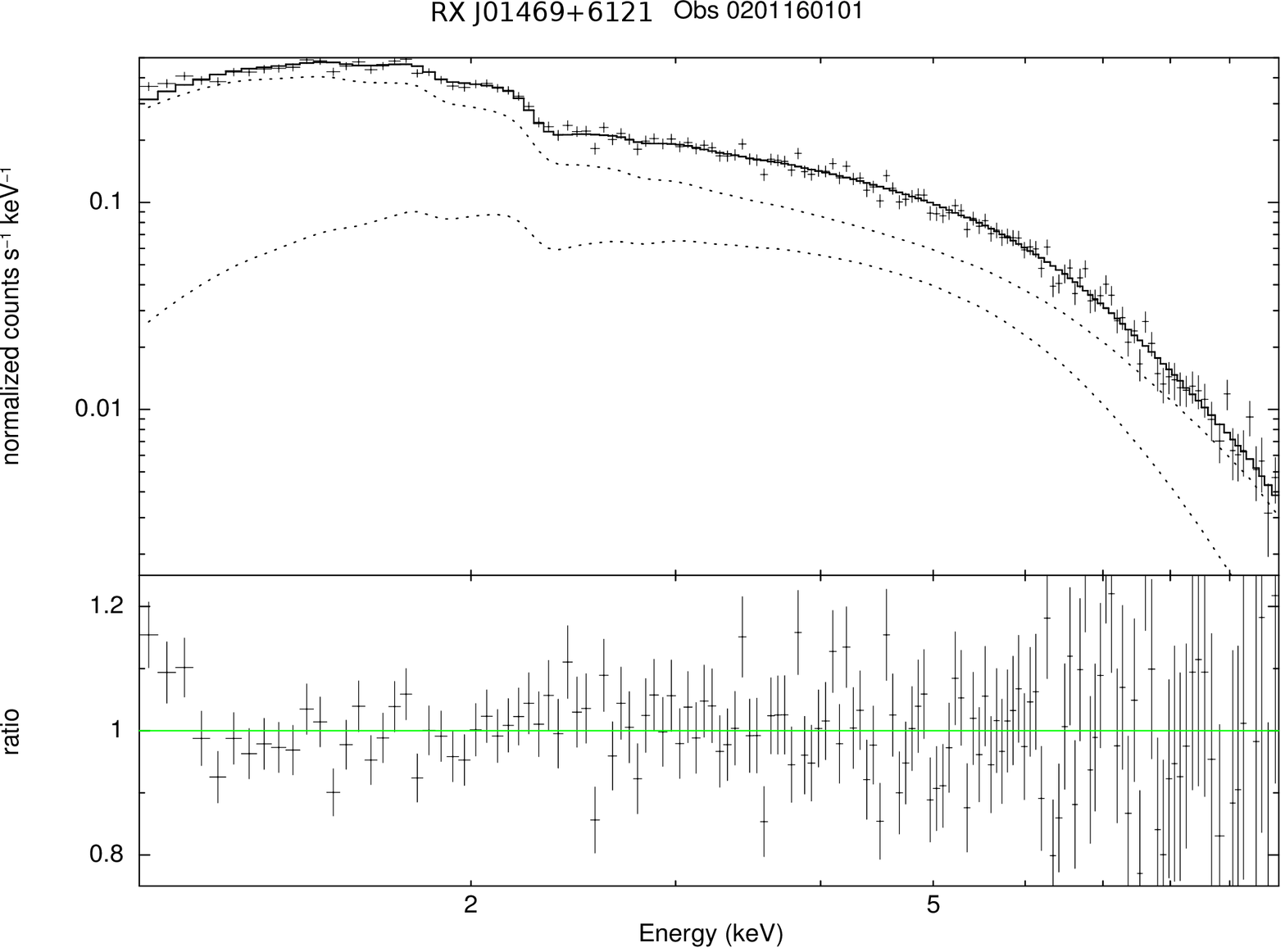}
\includegraphics[angle=-90, width=0.237\textwidth]{./Figures/plot6.ps}
\includegraphics[angle=-90, width=0.237\textwidth]{./Figures/plot7.ps}
\medskip
\includegraphics[angle=-90, width=0.237\textwidth]{./Figures/plot8.ps}
\includegraphics[angle=-90, width=0.237\textwidth]{./Figures/plot9.ps}
\includegraphics[angle=-90, width=0.237\textwidth]{./Figures/plot10.ps}
\caption{BeXBs data, model, model components and ratio data/model. The spectra are typically soft, with no Fe emission lines.}
\label{Atlas BeXB}
\end{figure}

\begin{figure}[ht]
\centering
\subsection*{\begin{Large}
\textbf{SGXB}
\end{Large}}
\includegraphics[angle=-90, width=0.237\textwidth]{./Figures/plot11.ps}
\includegraphics[angle=-90, width=0.237\textwidth]{./Figures/plot12.ps}
\includegraphics[angle=-90, width=0.237\textwidth]{./Figures/plot13.ps}
\includegraphics[angle=-90, width=0.237\textwidth]{./Figures/plot14.ps}
\medskip
\includegraphics[angle=-90, width=0.237\textwidth]{./Figures/plot15.ps}
\includegraphics[angle=-90, width=0.237\textwidth]{./Figures/plot16.ps}
\includegraphics[angle=-90, width=0.237\textwidth]{./Figures/plot17.ps}
\includegraphics[angle=-90, width=0.237\textwidth]{./Figures/plot18.ps}
\medskip
\includegraphics[angle=-90, width=0.237\textwidth]{./Figures/plot19.ps}
\includegraphics[angle=-90, width=0.237\textwidth]{./Figures/plot20.ps}
\includegraphics[angle=-90, width=0.237\textwidth]{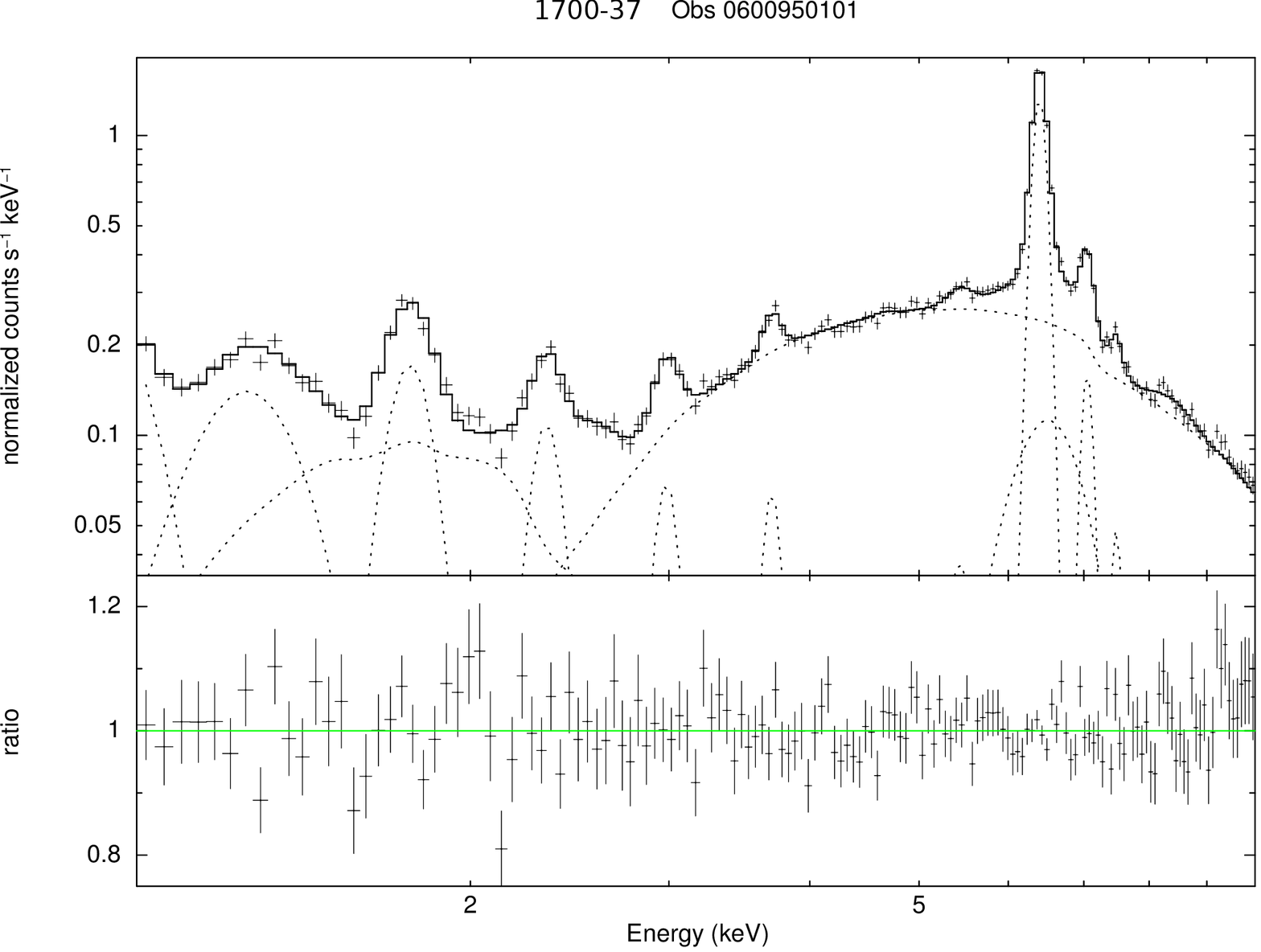}
\includegraphics[angle=-90, width=0.237\textwidth]{./Figures/plot22.ps}
\medskip
\includegraphics[angle=-90, width=0.237\textwidth]{./Figures/plot23.ps}
\includegraphics[angle=-90, width=0.237\textwidth]{./Figures/plot24.ps}
\includegraphics[angle=-90, width=0.237\textwidth]{./Figures/plot25.ps}
\includegraphics[angle=-90, width=0.237\textwidth]{./Figures/plot26.ps}
\medskip
\includegraphics[angle=-90, width=0.237\textwidth]{./Figures/plot27.ps}
\includegraphics[angle=-90, width=0.237\textwidth]{./Figures/plot28.ps}
\includegraphics[angle=-90, width=0.237\textwidth]{./Figures/plot29.ps}
\includegraphics[angle=-90, width=0.237\textwidth]{./Figures/plot30.ps}
\medskip
\includegraphics[angle=-90, width=0.237\textwidth]{./Figures/plot31.ps}
\includegraphics[angle=-90, width=0.237\textwidth]{./Figures/plot32.ps}
\includegraphics[angle=-90, width=0.237\textwidth]{./Figures/plot33.ps}
\includegraphics[angle=-90, width=0.237\textwidth]{./Figures/plot34.ps}
\caption{continued.}
\label{}
\end{figure}

\begin{figure}[ht]
\centering
\ContinuedFloat
\includegraphics[angle=-90, width=0.237\textwidth]{./Figures/plot35.ps}
\includegraphics[angle=-90, width=0.237\textwidth]{./Figures/plot36.ps}
\includegraphics[angle=-90, width=0.237\textwidth]{./Figures/plot37.ps}
\includegraphics[angle=-90, width=0.237\textwidth]{./Figures/plot38.ps}
\medskip
\includegraphics[angle=-90, width=0.237\textwidth]{./Figures/plot39.ps}
\includegraphics[angle=-90, width=0.237\textwidth]{./Figures/plot40.ps}
\includegraphics[angle=-90, width=0.237\textwidth]{./Figures/plot41.ps}
\includegraphics[angle=-90, width=0.237\textwidth]{./Figures/plot42.ps}
\medskip
\includegraphics[angle=-90, width=0.237\textwidth]{./Figures/plot43.ps}
\includegraphics[angle=-90, width=0.237\textwidth]{./Figures/plot44.ps}
\includegraphics[angle=-90, width=0.237\textwidth]{./Figures/plot45.ps}
\includegraphics[angle=-90, width=0.237\textwidth]{./Figures/plot46.ps}
\medskip
\includegraphics[angle=-90, width=0.237\textwidth]{./Figures/plot47.ps}
\includegraphics[angle=-90, width=0.237\textwidth]{./Figures/plot48.ps}
\includegraphics[angle=-90, width=0.237\textwidth]{./Figures/plot49.ps}
\includegraphics[angle=-90, width=0.237\textwidth]{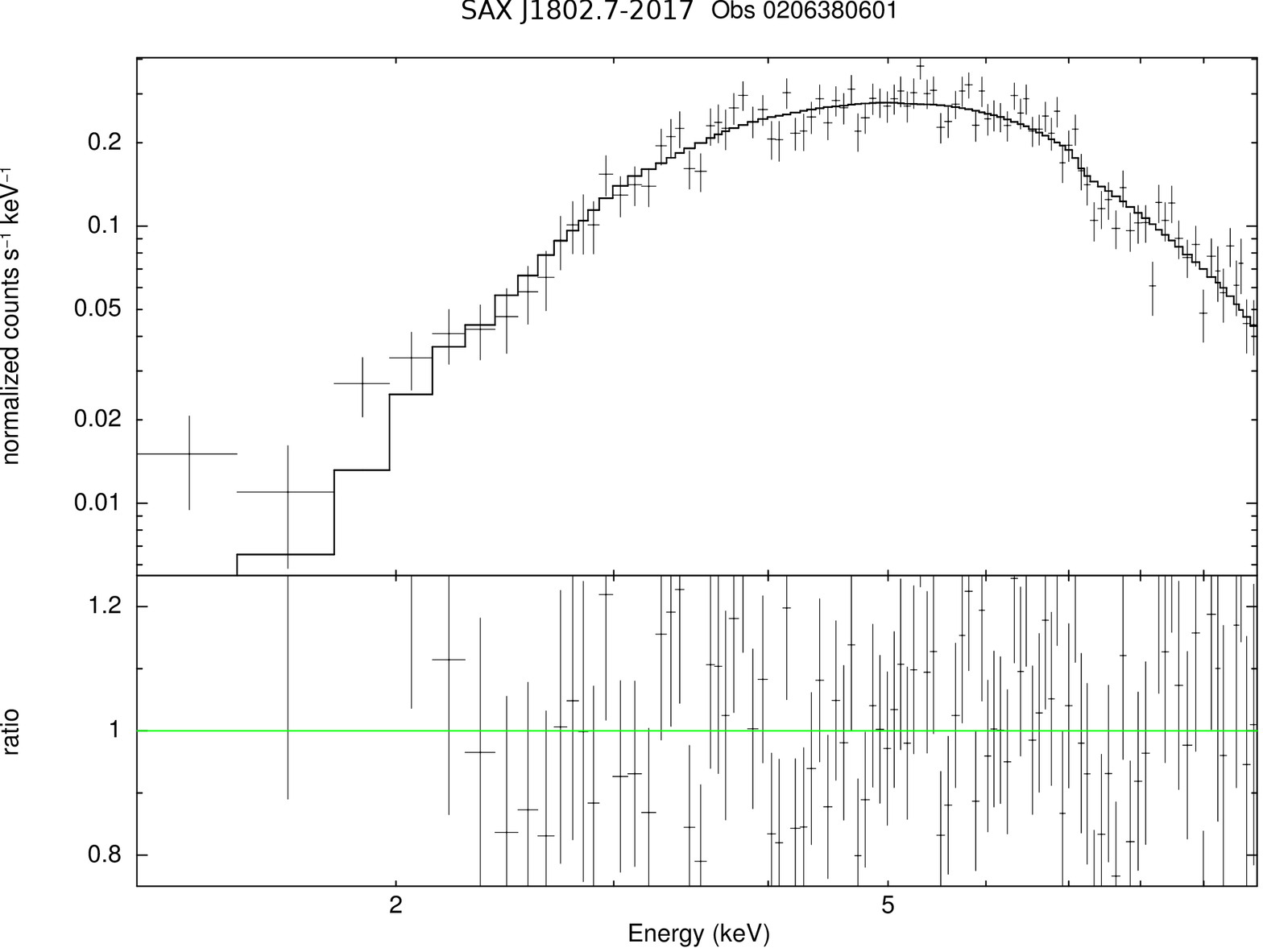}
\medskip
\includegraphics[angle=-90, width=0.237\textwidth]{./Figures/plot51.ps}
\includegraphics[angle=-90, width=0.237\textwidth]{./Figures/plot52.ps}
\includegraphics[angle=-90, width=0.237\textwidth]{./Figures/plot53.ps}
\caption{SGXBs data, model, model components and ratio data/model. The spectra are characteristically affected by high absorption, with Fe fluorescent lines.}
\label{}
\end{figure}

\begin{figure}[ht]
\centering
\subsection*{\begin{Large}
\textbf{SFXT}
\end{Large}}
\includegraphics[angle=-90, width=0.237\textwidth]{./Figures/plot54.ps}
\includegraphics[angle=-90, width=0.237\textwidth]{./Figures/plot55.ps}
\includegraphics[angle=-90, width=0.237\textwidth]{./Figures/plot56.ps}
\includegraphics[angle=-90, width=0.237\textwidth]{./Figures/plot57.ps}
\medskip
\includegraphics[angle=-90, width=0.237\textwidth]{./Figures/plot58.ps}
\includegraphics[angle=-90, width=0.237\textwidth]{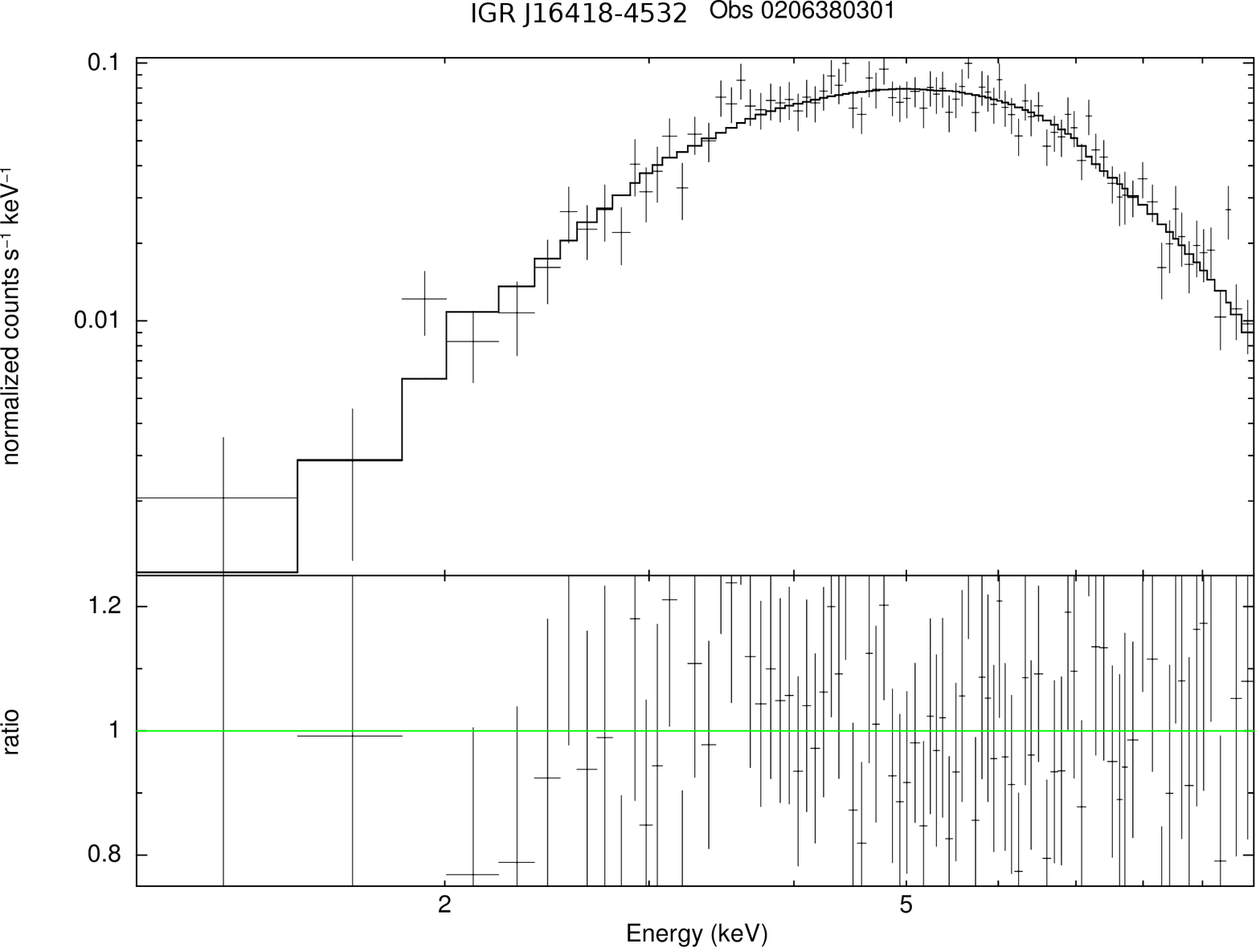}
\includegraphics[angle=-90, width=0.237\textwidth]{./Figures/plot60.ps}
\includegraphics[angle=-90, width=0.237\textwidth]{./Figures/plot61.ps}
\medskip
\includegraphics[angle=-90, width=0.237\textwidth]{./Figures/plot62.ps}
\includegraphics[angle=-90, width=0.237\textwidth]{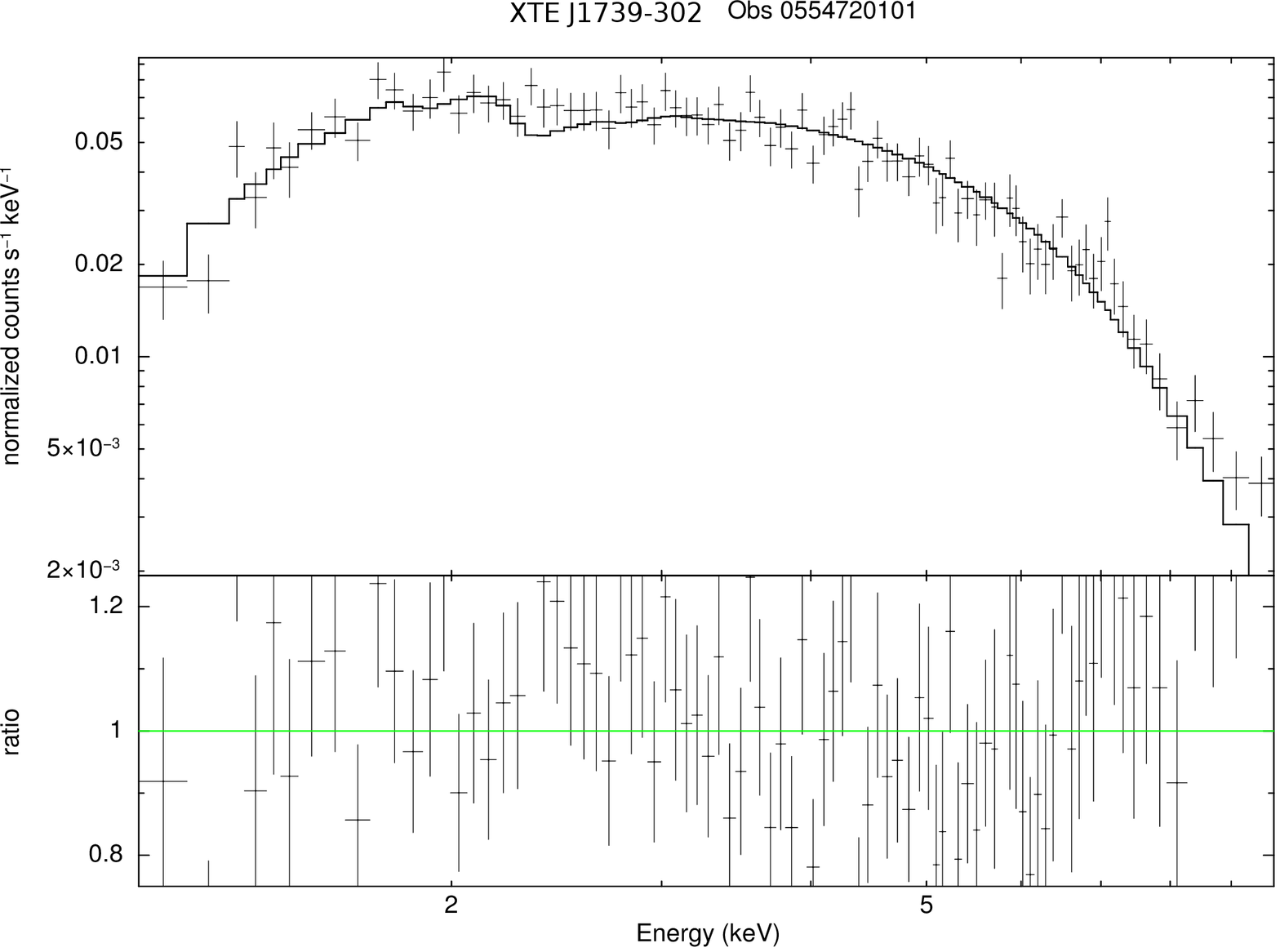}
\includegraphics[angle=-90, width=0.237\textwidth]{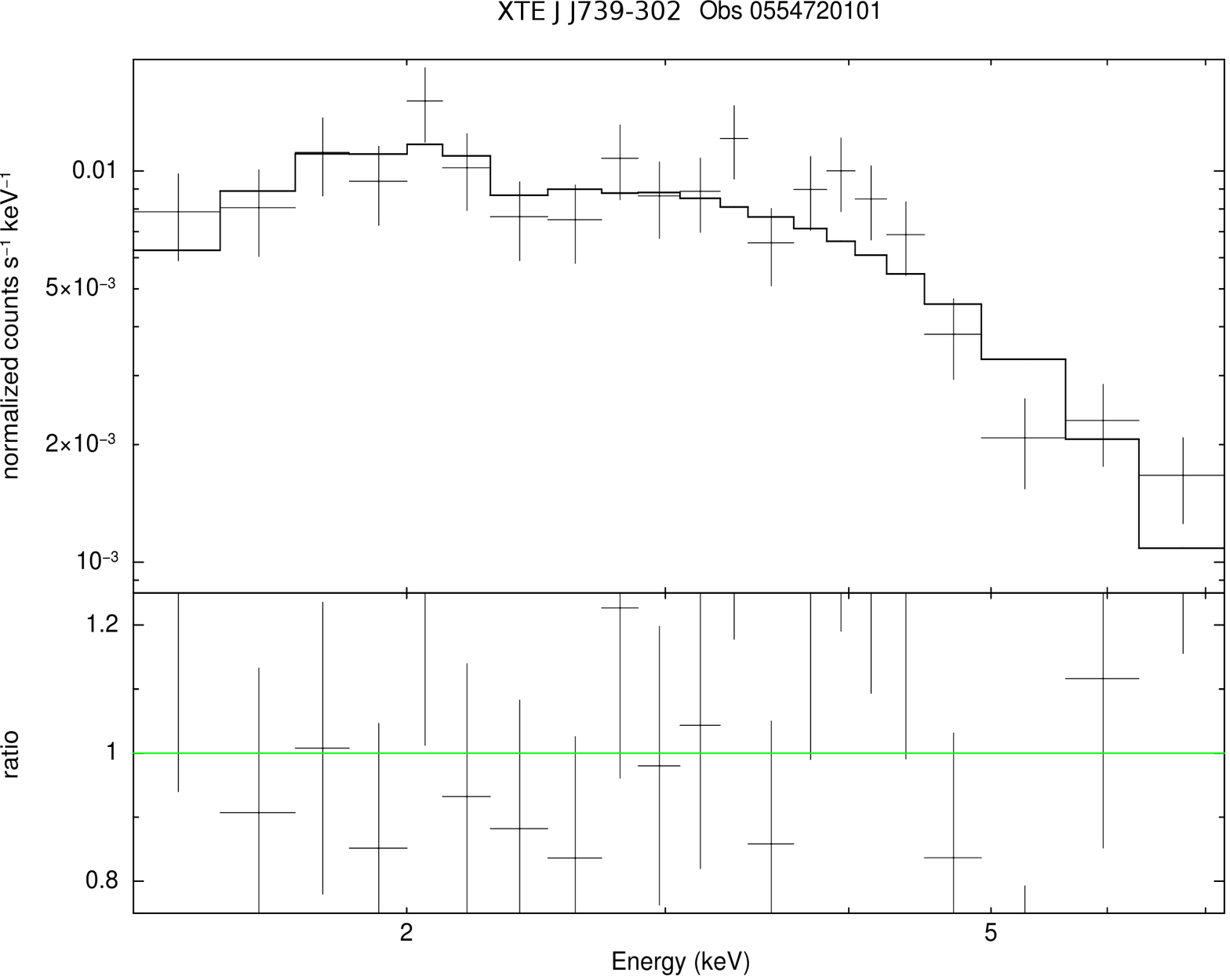}
\includegraphics[angle=-90, width=0.237\textwidth]{./Figures/plot65.ps}
\medskip
\includegraphics[angle=-90, width=0.237\textwidth]{./Figures/plot66.ps}
\includegraphics[angle=-90, width=0.237\textwidth]{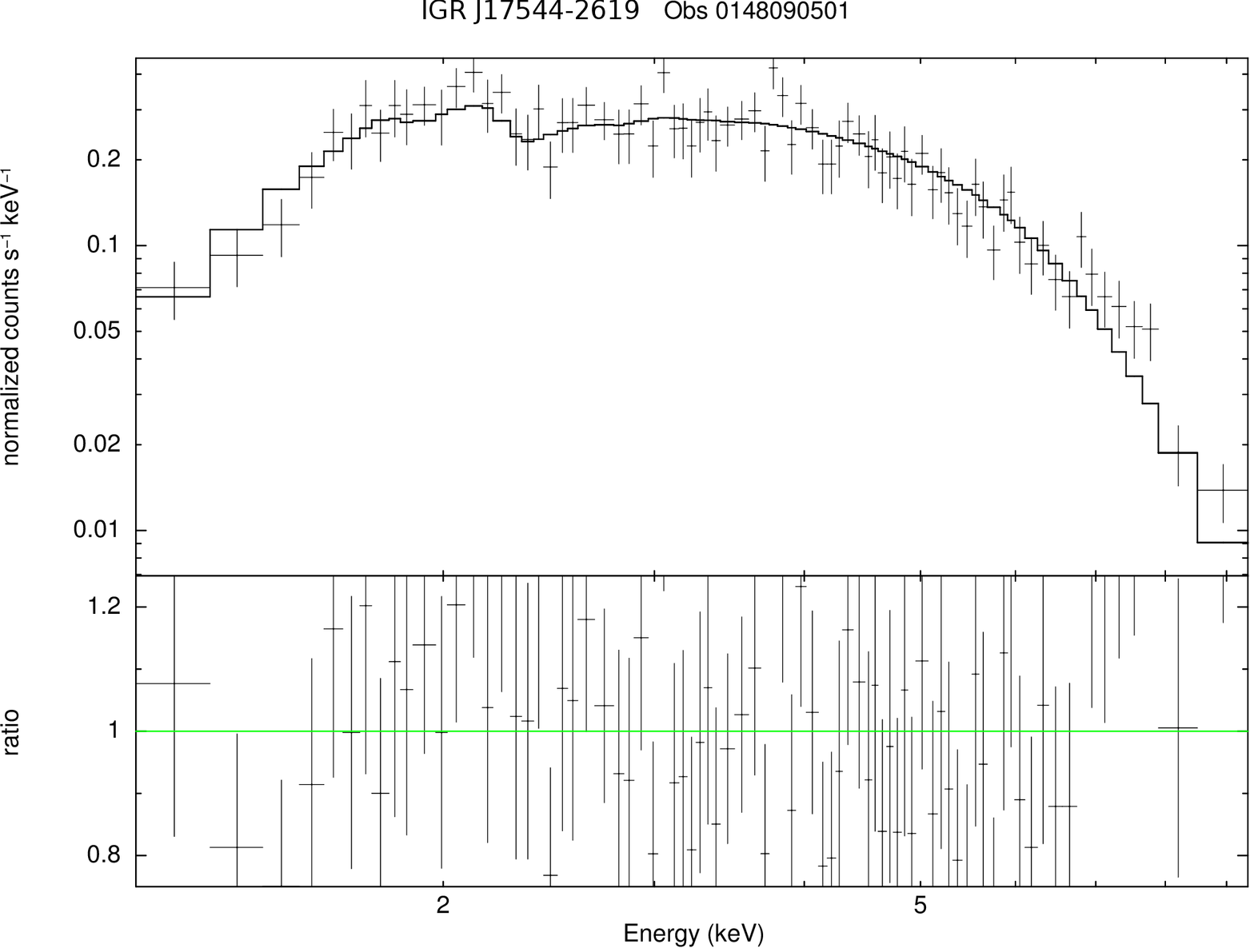}
\includegraphics[angle=-90, width=0.237\textwidth]{./Figures/plot68.ps}
\includegraphics[angle=-90, width=0.237\textwidth]{./Figures/plot69.ps}
\medskip
\includegraphics[angle=-90, width=0.237\textwidth]{./Figures/plot70.ps}
\caption{SFXTs data, model, model components and ratio data/model. }
\label{}
\end{figure}

\begin{figure}[ht]
\centering
\subsection*{\begin{Large}
$\gamma$~\textbf{Cassiopeae-like}
\end{Large}}
\includegraphics[angle=-90, width=0.237\textwidth]{./Figures/plot71.ps}
\includegraphics[angle=-90, width=0.237\textwidth]{./Figures/plot72.ps}
\includegraphics[angle=-90, width=0.237\textwidth]{./Figures/plot73.ps}
\includegraphics[angle=-90, width=0.237\textwidth]{./Figures/plot74.ps}
\medskip
\includegraphics[angle=-90, width=0.237\textwidth]{./Figures/plot75.ps}
\includegraphics[angle=-90, width=0.237\textwidth]{./Figures/plot76.ps}
\includegraphics[angle=-90, width=0.237\textwidth]{./Figures/plot77.ps}
\includegraphics[angle=-90, width=0.237\textwidth]{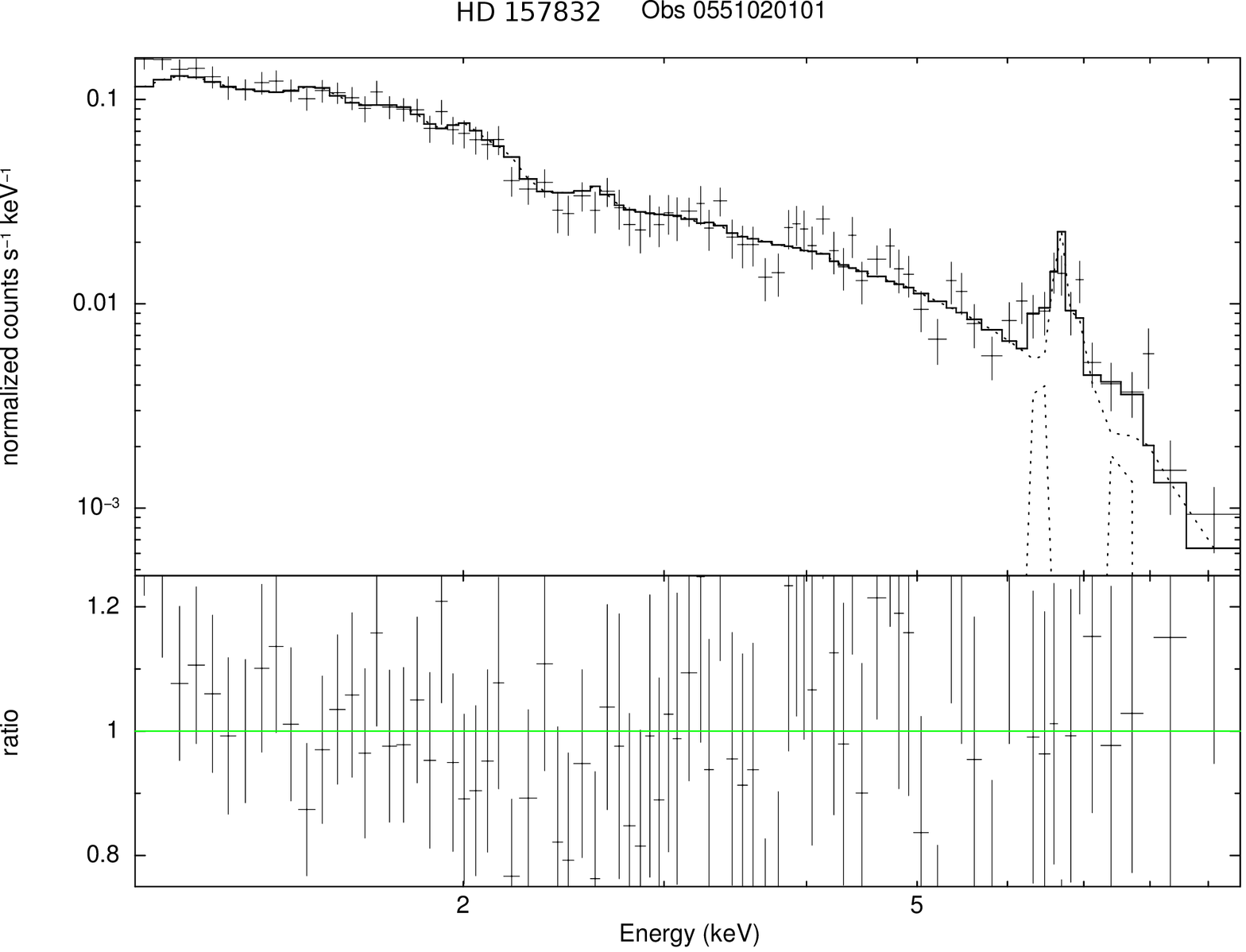}
\medskip
\includegraphics[angle=-90, width=0.237\textwidth]{./Figures/plot79.ps}
\includegraphics[angle=-90, width=0.237\textwidth]{./Figures/plot80.ps}
\includegraphics[angle=-90, width=0.237\textwidth]{./Figures/plot81.ps}
\includegraphics[angle=-90, width=0.237\textwidth]{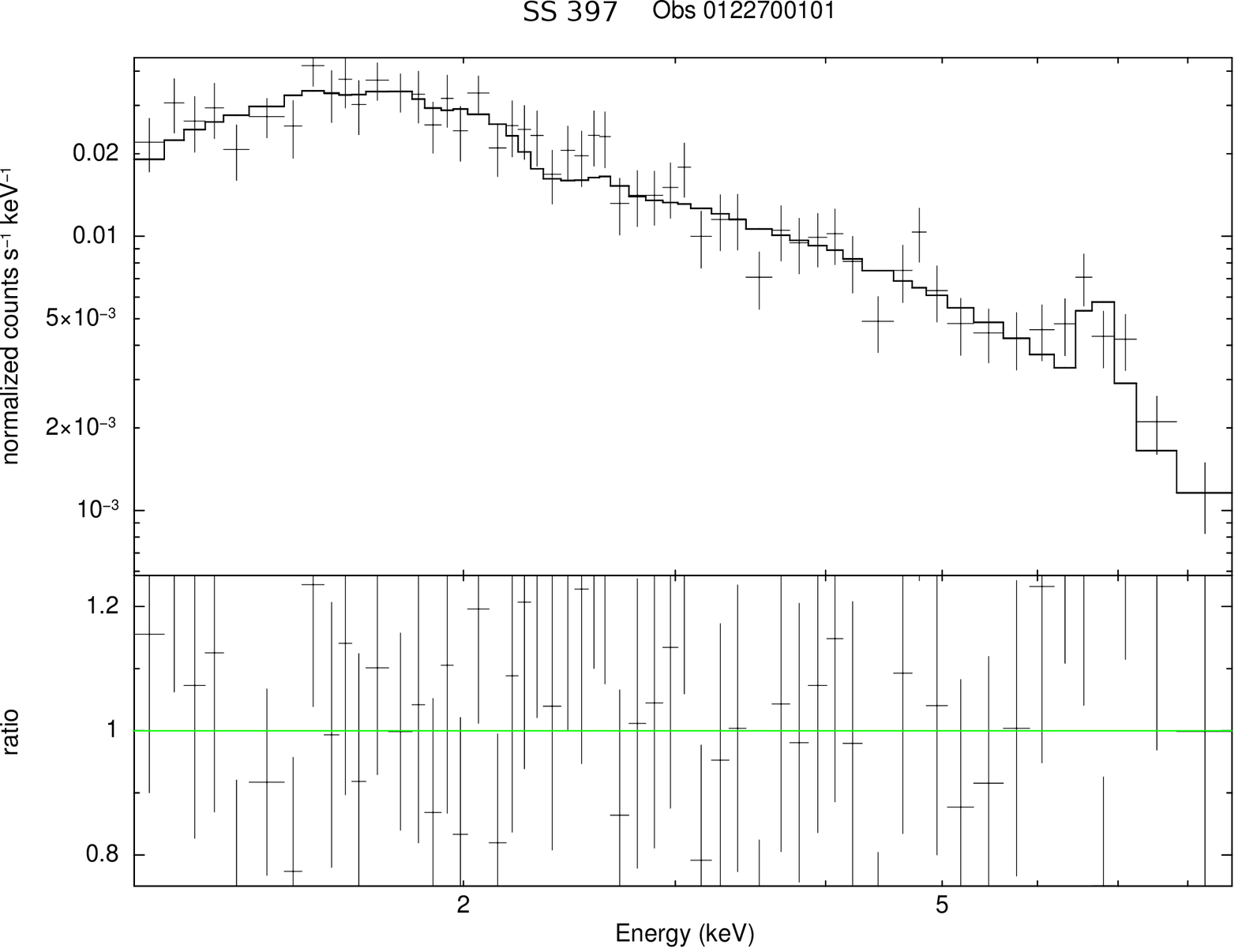}
\medskip
\includegraphics[angle=-90, width=0.237\textwidth]{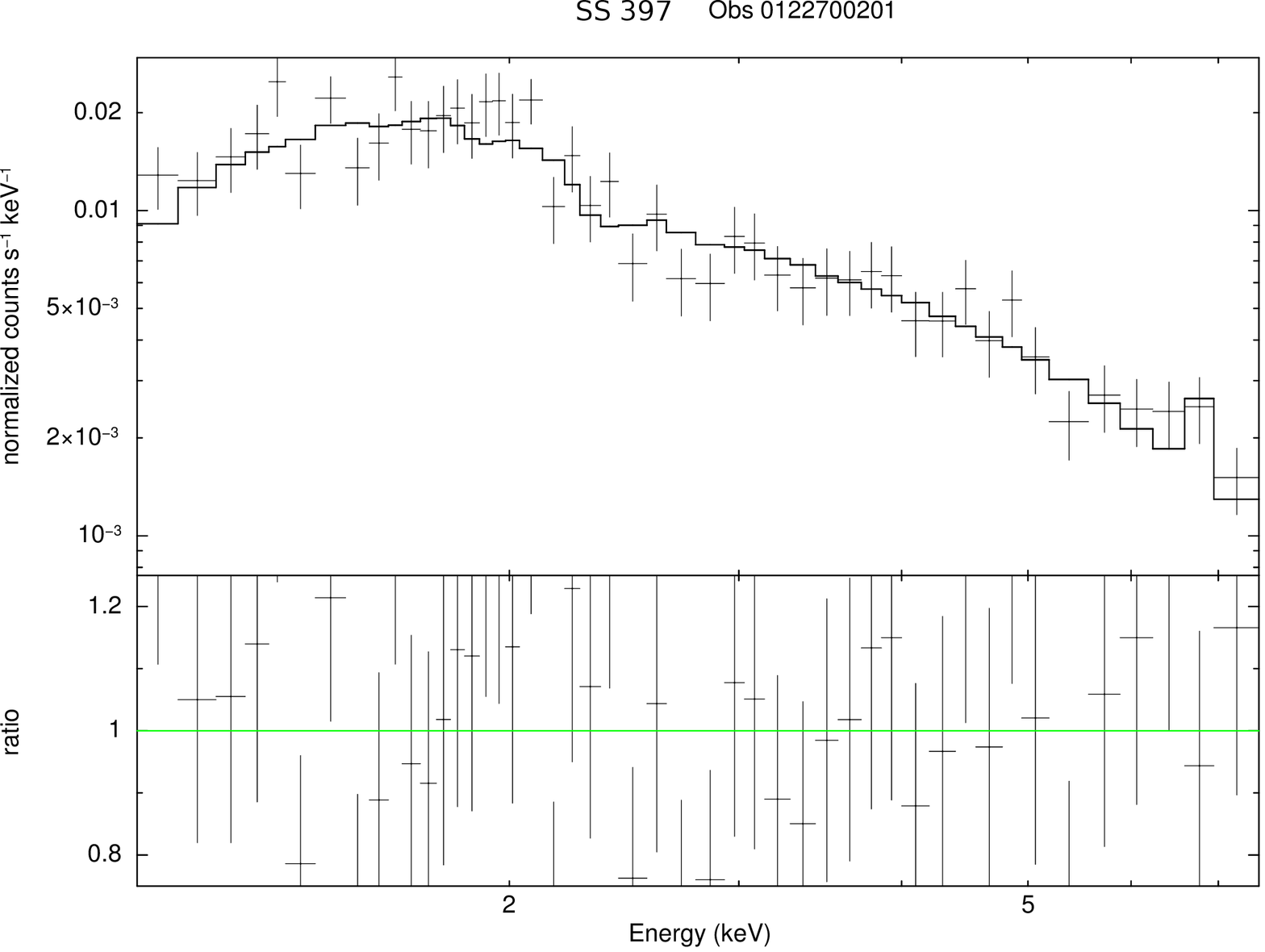}
\includegraphics[angle=-90, width=0.237\textwidth]{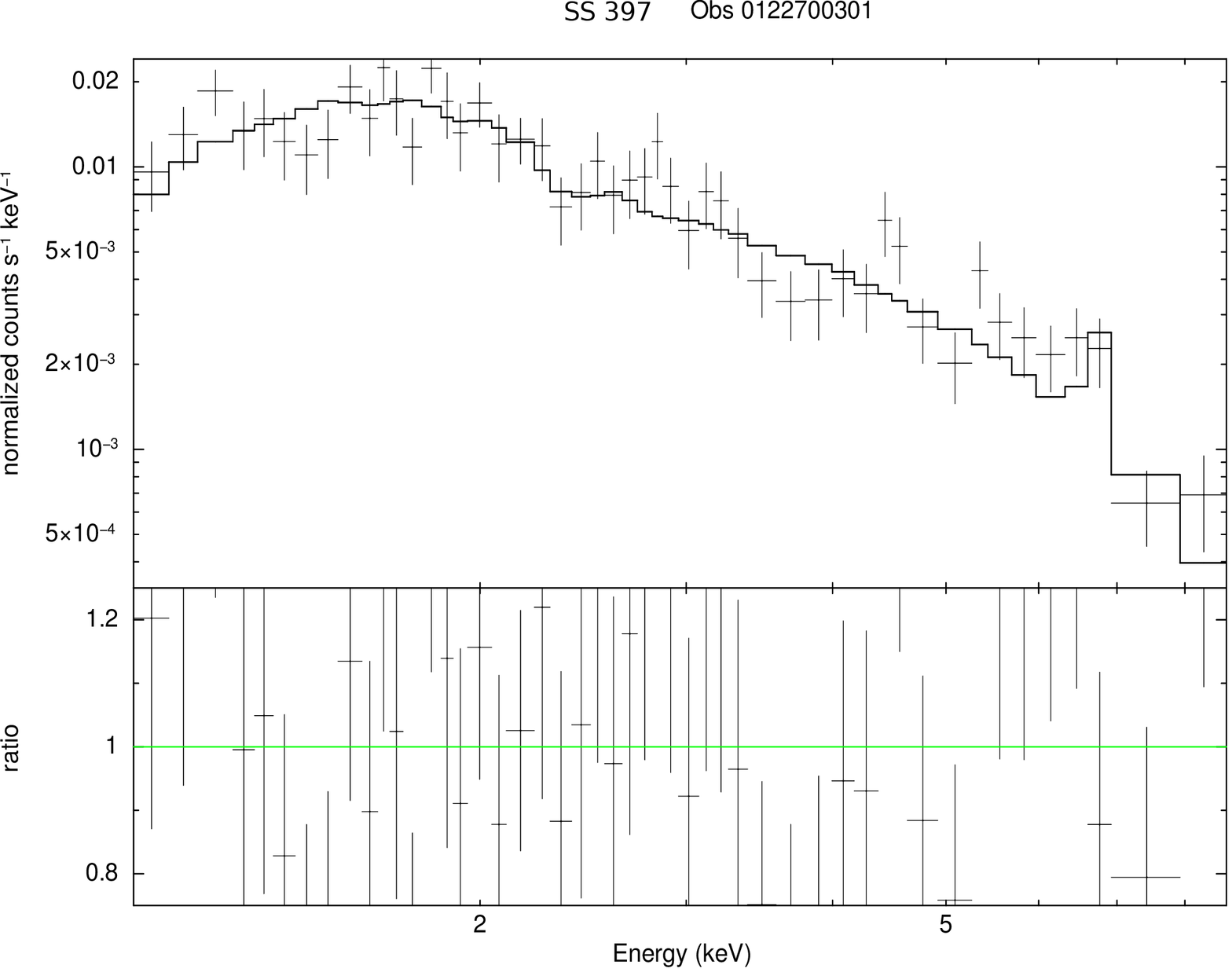}
\includegraphics[angle=-90, width=0.237\textwidth]{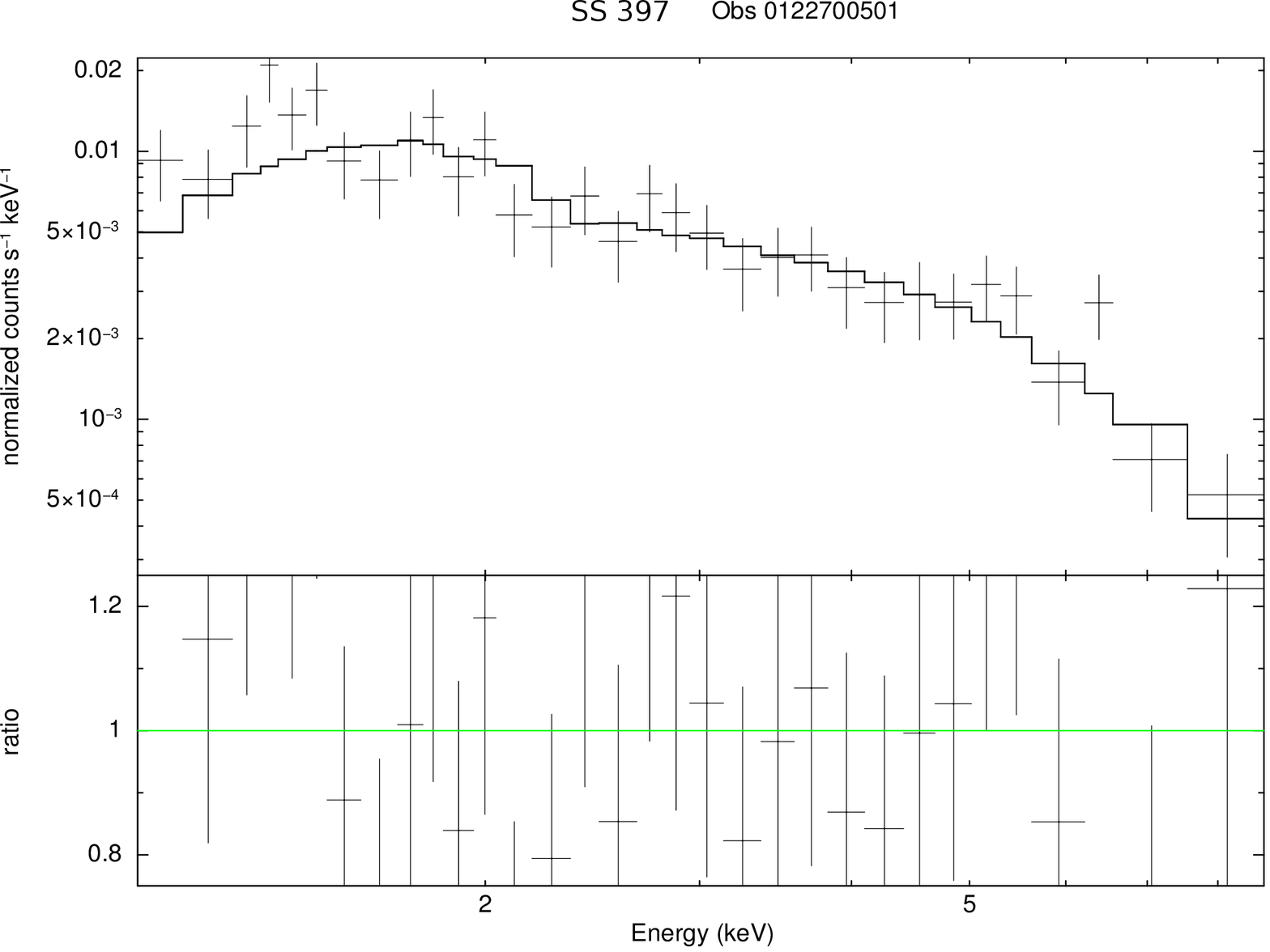}
\caption{$\gamma$~Cass-like systems data, model, model components and ratio data/model. The data is usually fitted using thermal models, including Fe recombination lines. FeK$\alpha$ is also usually visible. }
\label{} 
\end{figure}

\begin{figure}[ht]
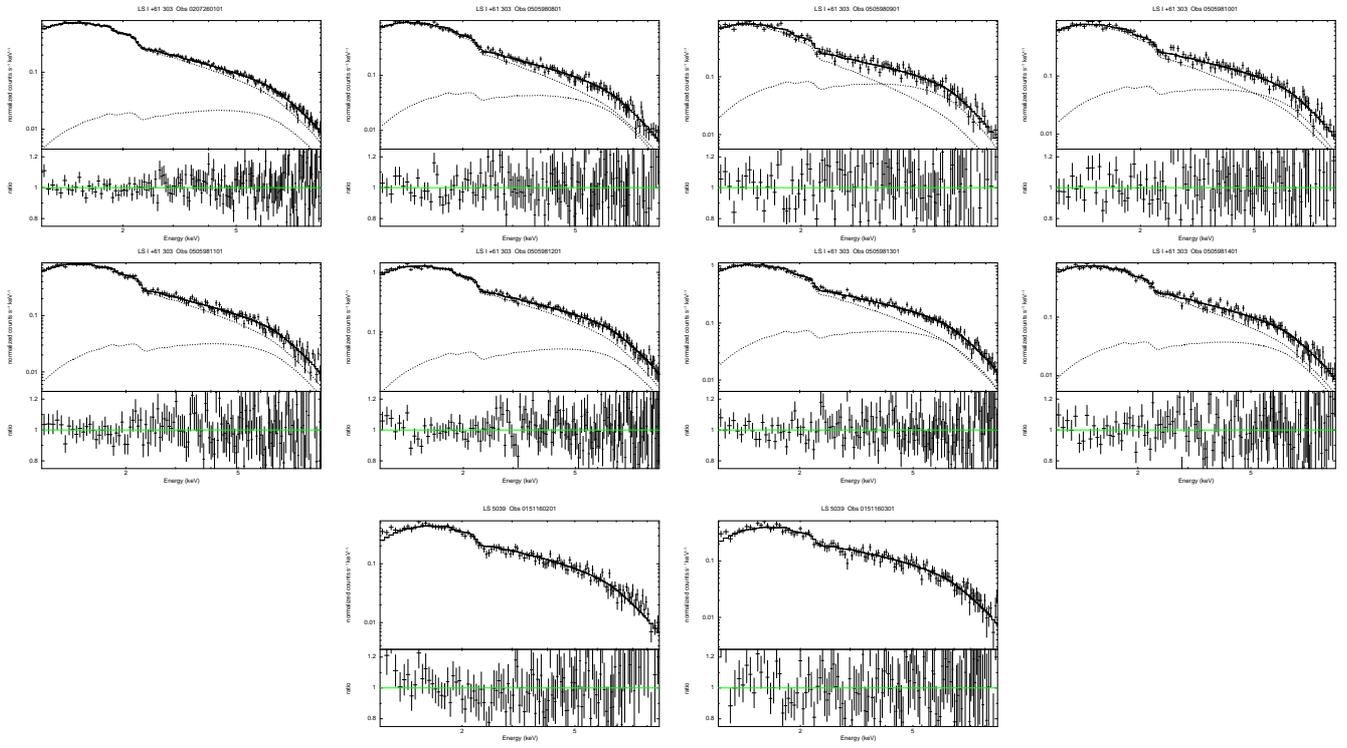

\centering
\subsection*{\begin{Large}
\textbf{HMGB}
\end{Large}}
\includegraphics[angle=-90, width=0.237\textwidth]{./Figures/plot86.ps}
\includegraphics[angle=-90, width=0.237\textwidth]{./Figures/plot87.ps}
\includegraphics[angle=-90, width=0.237\textwidth]{./Figures/plot88.ps}
\includegraphics[angle=-90, width=0.237\textwidth]{./Figures/plot89.ps}
\medskip
\includegraphics[angle=-90, width=0.237\textwidth]{./Figures/plot90.ps}
\includegraphics[angle=-90, width=0.237\textwidth]{./Figures/plot91.ps}
\includegraphics[angle=-90, width=0.237\textwidth]{./Figures/plot92.ps}
\includegraphics[angle=-90, width=0.237\textwidth]{./Figures/plot93.ps}
\medskip
\includegraphics[angle=-90, width=0.237\textwidth]{./Figures/plot94.ps}
\includegraphics[angle=-90, width=0.237\textwidth]{./Figures/plot95.ps}
\caption{HMGBs data, model, model components and ratio data/model. Soft spectra with no sign of Fe lines. }
\label{}
\end{figure}

\begin{figure}[ht]
\centering
\vspace{2cm}
\subsection*{{\Large \textbf{Non Classified}}}
\includegraphics[angle=-90, width=0.237\textwidth]{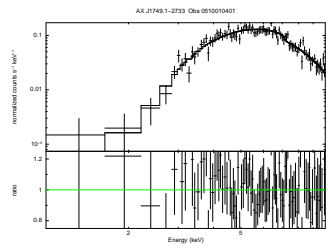}
\caption{AX~J1749.1-2733 data, model, model components and ratio data/model. We have no references for the luminosity class of the optical start, although the high absorption observed points to a supergiant companion. }
\label{}
\end{figure}

\begin{figure}[ht]
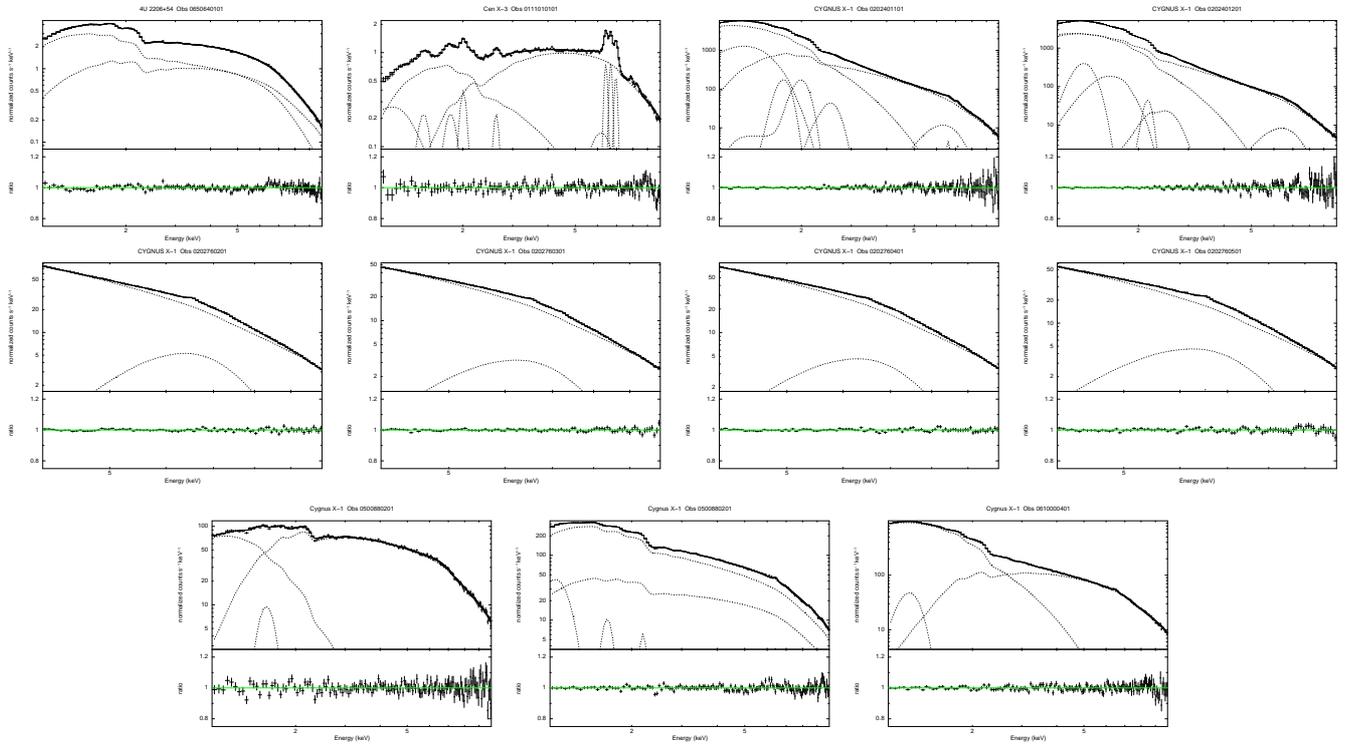

\centering
\subsection*{\begin{Large}
\textbf{\textit{Peculiars}}
\end{Large}}
\includegraphics[angle=-90, width=0.237\textwidth]{./Figures/plot97.ps}
\includegraphics[angle=-90, width=0.237\textwidth]{./Figures/plot98.ps}
\includegraphics[angle=-90, width=0.237\textwidth]{./Figures/plot99.ps}
\includegraphics[angle=-90, width=0.237\textwidth]{./Figures/plot100.ps}
\medskip
\includegraphics[angle=-90, width=0.237\textwidth]{./Figures/plot101.ps}
\includegraphics[angle=-90, width=0.237\textwidth]{./Figures/plot102.ps}
\includegraphics[angle=-90, width=0.237\textwidth]{./Figures/plot103.ps}
\includegraphics[angle=-90, width=0.237\textwidth]{./Figures/plot104.ps}
\medskip
\includegraphics[angle=-90, width=0.237\textwidth]{./Figures/plot105.ps}
\includegraphics[angle=-90, width=0.237\textwidth]{./Figures/plot106.ps}
\includegraphics[angle=-90, width=0.237\textwidth]{./Figures/plot107.ps}
\caption{Peculiar sources data, model, model components and ratio data/model. These sources can be hardly categorized in any of the described HMXBs standard groups. }
\label{}
\end{figure}

\end{document}